\documentclass[a4paper,11pt]{article}
\pdfoutput=1 

\usepackage{jheppub} 

\usepackage{latexsym,amssymb,amstext,amsmath}
\usepackage{hyperref}
\usepackage{graphicx} 
\usepackage{subcaption}
\usepackage{amsthm}
\usepackage{esint}
\usepackage{float}
\usepackage{color}
\newcommand{\ft}[2]{{\textstyle\frac{#1}{#2}}}

\begin{document}
%
\begin{titlepage}

\begin{center}
 {\LARGE\bfseries An approach to BPS black hole  \vskip 3mm
 microstate counting in an  $N=2$ STU model}
 \\[10mm]

\textbf{G. L.~Cardoso, S. Nampuri  and D. Polini}

\vskip 6mm
{\em  Center for Mathematical Analysis, Geometry and Dynamical Systems,\\
  Department of Mathematics, 
  Instituto Superior T\'ecnico,\\ Universidade de Lisboa,
  Av. Rovisco Pais, 1049-001 Lisboa, Portugal}\\[.5ex]

{\tt gcardoso@math.tecnico.ulisboa.pt},\;\,{\tt  nampuri@gmail.com},\;\,
\\
{\tt
 davide.polini@tecnico.ulisboa.pt}
\end{center}

\vskip .2in
\begin{center} {\bf ABSTRACT } \end{center}
\begin{quotation}\noindent

\vskip 3mm
\noindent
We consider four-dimensional dyonic single-center BPS black holes in the $N=2$ STU model of Sen and Vafa.
By working in a region of moduli space where the real part of two of the three complex scalars $S, T, U$ are 
taken to be large, we evaluate the quantum entropy function for these BPS black holes.
In this regime, 
the subleading corrections point to a microstate counting formula partly based on a Siegel modular
form of weight two. This is supplemented by another modular object that takes into account the dependence
on $Y^0$, 
a complex scalar field belonging to one of the four off-shell vector multiplets of the underlying supergravity
theory.
We also observe interesting connections to the rational Calogero model and  to formal deformation of a Poisson algebra, and 
suggest a string web picture
of our counting proposal.

\end{quotation}
\vfill
\today
\end{titlepage}

\tableofcontents

\section{Introduction }

A highly active area of focused research in string theory aims at a formulation of the microscopic 
degeneracy of states of black holes to arrive at a statistical description of black hole thermodynamics. This is a critical first step in gleaning insights into the organization of the degrees of freedom in any purported theory of quantum gravity. A series of steady definitive advancements has been accomplished in the last decades in this field, specifically w.r.t the exact counting of microstates of 
$\frac{1}{4}$ BPS and $\frac{1}{8}$ 
BPS black holes in four-dimensional $N=4$ and $N=8$ string theories,
respectively 
\cite{Dijkgraaf:1996it,Shih:2005uc,Shih:2005qf,Gaiotto:2005hc,Jatkar:2005bh,David:2006ji,Dabholkar:2006xa,
David:2006yn,David:2006ru,David:2006ud,Dabholkar:2006bj,Sen:2007vb,Dabholkar:2007vk,
Banerjee:2007ub,Sen:2007pg,Banerjee:2008pv,Banerjee:2008pu,Sen:2008ta,Dabholkar:2008zy}. In this paper, we address the problem of black hole microstate counting for $\frac{1}{2}$ BPS asymptotically flat black hole backgrounds.
We do this in a specific four-dimensional $N=2$ string theory model, namely the STU model of Sen and Vafa 
(example D in \cite{Sen:1995ff}), for which the holomorphic function $F$, which encodes (part of)
the Wilsonian effective action, has been determined recently \cite{CdWM}.  The function $F$ encodes an infinite set of gravitational coupling functions 
$\omega^{(n)}$ ($n \geq 1$).
A proposal for the counting of microstates in this model has been made in \cite{David:2007tq}, by only
taking into account
the gravitational coupling function $\omega^{(1)}$.
In this work, 
by using the detailed knowledge about $F$
given in \cite{CdWM}, we arrive at a different proposal.
While doing
so, we will work in a certain region of moduli space, as we will explain below.

In four-dimensional $N=2$ string models, extremal black hole backgrounds are characterized by
 electric and magnetic charges $(q_I, p^I)$ w.r.t  the various $U(1)$ gauge fields in these models. These backgrounds, 
 which at weak string coupling 
 correspond to BPS states, possess a 
 near-horizon $AdS_2$ geometry \cite{Strominger:1996sh}. This geometry is an exact solution of the equations of motions that is
decoupled from the asymptotics of the background.
The decoupling ensures that the values of scalar fields $\{\phi\}$  in the near-horizon black hole background as well as all relevant length scales, such as the $AdS_2$ radius, are solely fixed in terms of the charges and are
 independent of the asymptotic data. The near-horizon geometry, in effect, acts as an attractor \cite{Ferrara:1995ih,Ferrara:1996dd,Ferrara:1996um} for the scalar  fields 
$\{ \phi\}$  
 in the black hole background: the attractor values of these scalars determine
   the near-horizon length scale, and hence the black hole entropy ${\cal S}_{\rm BH}$. This, in turn, implies that the quantum gravity partition function 
 $Z_{AdS_2}(\phi,p)$ in  this attractor 
  background,
 evaluated in a canonical ensemble defined at fixed magnetic charges, is the generator of the dimension 
 of the black hole microstate Hilbert space, i.e. of the black hole microstate degeneracy 
 $d_{\rm BH}(q,p)$, graded by electric and magnetic charges. Hence, formally, one has
\begin{equation}
Z_{AdS_2} (\phi,p) = \sum_{q} d_{\rm BH} (q,p) e^{\pi q \cdot \phi} \:.
\label{zpf}
\end{equation}
The $SL(2, \mathbb{R})$-isometries of $AdS_2$ indicate that the generator of degeneracies has modular properties, 
an expectation  that has been extensively borne out in all cases where the exact microscopic degeneracy is known, such as in 
four-dimensional
$N=4$ and $N=8$ string theories.
Here the microstate degeneracy  $d_{\rm BH}(q,p)$ is related to the macroscopic entropy of the black hole by the Boltzmann relation
\begin{equation}
{\cal S}_{\rm BH}= \ln\, d_{\rm BH}(q,p) \;.
\end{equation}

The macroscopic entropy ${\cal S}_{\rm BH} $
has an expansion in terms of the dimensionless black hole area $A_{\rm BH}$ as\footnote{Additionally,
a constant contribution may be present. We will neglect such a contribution
when computing the quantum entropy function. For a recent development in this regard
see \cite{Chattopadhyaya:2018xvg}.}
\begin{equation}
{\cal S}_{\rm BH}  = \frac{A_{\rm BH}}{4} + c_1 \, \ln A_{\rm BH} + \frac{c_2}{A_{\rm BH}}+ \dots \;.
\end{equation}
For extremal black holes with a near-horizon $AdS_2$ geometry, a proposal
for their exact macroscopic entropy was put forward in \cite{Sen:2008yk,Sen:2008vm}.
This proposal is called the quantum entropy function, and is defined in terms of 
a regulated path integral over $AdS_2$ space-times with an insertion of a Wilson line corresponding to electric charges.
This in turn provides an operational definition for \eqref{zpf}.

For BPS black holes in four-dimensional $N=2$ string theories, the quantum entropy function 
$W(q,p)$
can be expressed as \cite{Dabholkar:2010uh} (when strictly restricted to smooth configurations)
\begin{equation}
W(q,p) = \int _{\mathcal{C}} 
d \phi \, \mu(\phi) \,  Z_{\rm 1-loop}(\phi)  \; e^{\pi [ 4 \, {\rm Im} F(\phi + i p) -  q \cdot \phi] } \, 
\;.
\label{qentr}
\end{equation}
The various ingredients that go into this formula will be reviewed in 
section \ref{sec:QuantumentropyfunctionS}.
Here we note that the holomorphic function $F$ appearing in the exponent is
the holomorphic function mentioned above that encodes (part of) the Wilsonian effective action.
The function $F$ can be decomposed as $F = F^{(0)} + 2 i \Omega$, where $F^{(0)}$ denotes the prepotential of the model,
whereas $\Omega$ encodes all the gravitational coupling functions $\omega^{(n)}$, as a series expansion in powers of
$\Upsilon/(Y^0)^2$. Here, $(\Upsilon, Y^0)$ denote rescaled supergravity fields $(A, X^0)$, where $A$ denotes the lowest
component of the square of a chiral superfield that describes the Weyl multiplet, and where $X^0$ denotes a complex scalar field
belonging to one of the off-shell vector multiplets in the associated supergravity theory.

The STU model 
of Sen and Vafa (example D in \cite{Sen:1995ff})
is based on four off-shell vector multiplets, and $\Omega$ will therefore depend on four complex scalar fields $Y^I$
that reside in these multiplets. 
Introducing projective coordinates $S = -i Y^1/Y^0, T= -i Y^2/Y^0, U = -i Y^3/Y^0$,
 $\Omega$ has the series expansion
\begin{equation}
\Omega(Y, \Upsilon) = \Upsilon \, \sum_{n=0}^{\infty} \left(\frac{\Upsilon}{(Y^0)^2} \right)^n \, \omega^{(n+1)} (S, T, U) \;.
\label{omWils}
\end{equation}
This STU model possesses duality symmetries, namely $[ \Gamma_0(2)]^3$ symmetry and triality symmetry.
The former refers to 
  the duality symmetry
  $\Gamma_0(2)_S \times
\Gamma_0(2)_T \times \Gamma_0(2)_U$, associated with each of the moduli $S, T, U$. 
Here, $\Gamma_0(2)$ denotes a certain congruence subgroup of
$SL(2, \mathbb{Z})$.
In addition, the model possesses
triality symmetry, i.e. invariance under the exchange of $S, T$ and $U$. 
Using all these symmetries, it was demonstrated in \cite{CdWM} that the gravitational coupling functions 
$ \omega^{(n+1)} (S, T, U)$ are expressed in terms of one single function $\omega$, and derivatives thereof with respect to $S, T, U$.
The duality symmetries were implemented by adding to $\Omega$ in \eqref{omWils} a term proportional to $ \Upsilon \, \ln Y^0$.
However, such a term is, strictly speaking, not part of the Wilsonian effective action. Since we will take $\Omega$ as an input in \eqref{qentr},
we will refrain from adding this term to $\Omega$. This term will be effectively generated from $Z_{\rm 1-loop}$ in \eqref{qentr},
as already noted in \cite{Sen:2011ba}.

The explicit expressions for the $ \omega^{(n+1)} (S, T, U)$ in the STU model are complicated. However, in \cite{CdWM} it was observed that they dramatically simplify
in the limit where the real part of two of the three moduli $S, T, U$ are taken to be large.
This is the limit in which we will work in this paper. We will
thus fix two of the moduli, say $T$ and $U$, to large values, which we will denote by $T_0, U_0$. In this limit, $\Omega$ becomes effectively replaced by
\begin{equation}
{\widehat \Omega} (Y^0, S, \Upsilon) = \Upsilon \left( \, \omega (S) 
+ \frac{\Upsilon}{(Y^0)^2} \, 
\frac{\alpha}{\gamma} \, \frac{\partial \omega (S)}{\partial S} + 
\Xi (Y^0, S, \Upsilon) \right) \;,
\label{hatOmI}
\end{equation}
where $\alpha, \gamma$ denotes constants, and 
where $\Xi$ encodes 
the higher gravitational couplings 
$\omega^{(n+1)}$ with $n \geq 2$. The latter are expressed in terms of modular forms $I_n (S)$ and products thereof. 
The $I_n$ are themselves products of Eisenstein series for $\Gamma_0(2)$, see \eqref{Ikeps2eps4}.
We are thus led to an expansion of $\Xi$ in terms of $I_n$ as
\begin{eqnarray}
\label{sigIn}
\Xi (Y^0, S, \Upsilon) &=& \sum_{n=2}^{\infty} \left(\frac{\Upsilon}{(Y^0)^2} \right)^n \, \omega^{(n+1)} (S) \\
&=& \sum_{n = 2}^{\infty} \alpha_n \; \left( \frac{ \Upsilon}{(Y^0)^2} \right)^n
 \; I_n(S) + \sum_{m, n=2}^{\infty}
\alpha_{m,n} \;  \left( \frac{ \Upsilon}{(Y^0)^2} \right)^{m+n} \; I_m(S) \; I_n(S) + \dots \;. \nonumber
\end{eqnarray}
Here, the right hand side is organised in powers of $I_n$. We call the first sector (which is linear in $I_n$) the monomial sector, the
second sector (which is quadratic in $I_n$) the binomial sector, and so on. In \cite{CdWM}, explicit expressions were only given for the coefficients
of the first two sectors: the $\alpha_n$ were determined fully,  and partial expressions for the $\alpha_{m,n}$ were given.
Notwithstanding this limited knowledge about the expansion coefficients,
there are several lessons that one can draw from the power series expansion \eqref{hatOmI},  \eqref{sigIn}. Firstly, under $\Gamma_0(2)$-transformations,
${\widehat \Omega}$ transforms in the same manner as the logarithm of 
$\vartheta_2^8 (\tau,z)$, where $\vartheta_2^8 (\tau,z)$ denotes one of the Jacobi theta functions, and where $\tau = i S, \, z = 1/Y^0$.
Moreover, 
the series expansion \eqref{hatOmI},  \eqref{sigIn}
is reminiscent of the Taylor series expansion around $z=0$ of 
$\ln \vartheta_2^8 (\tau,z)$, see \eqref{fImonbin}. 
In this expansion, the coefficients $\alpha_n$  in the monomial sector behave as $1/(2n)!$, whereas in \eqref{sigIn} they behave as $1/n!$.  Secondly,
each of the $I_n$-sectors in the expansion of $\ln \vartheta_2^8 (\tau,z)$ defines a function that is given as a series expansion in $z$,
and the sum over all these functions gives rise to $\ln \vartheta_2^8 (\tau,z)$, which is naturally expressed in terms of
variables $q = \exp ( 2 \pi i \tau) $ and $y = \exp (2 \pi i z)$. It is tempting
to conclude that a similar story may apply to
${\widehat \Omega} (Y^0, S, \Upsilon)$. Thirdly, $\ln \vartheta_2^8 (\tau,z)$ is a solution
to a non-linear PDE derived from the heat equation. So we may similarly 
ask whether $\Xi$ given in \eqref{sigIn}
arises as solution to a non-linear PDE. The answer is affirmative, and we give a PDE
whose solution yields the monomial and binomial sectors displayed in 
\eqref{sigIn} for the expressions of the coefficients given in \cite{CdWM}
(\cite{CdWM} obtained
exact expressions for $\alpha_n$ and partial expressions for $\alpha_{m,n}$).
The higher sectors in \eqref{sigIn}, whose expansion coefficients
have not been determined in \cite{CdWM}, will lead to modifications of this PDE,
and we give a candidate for the modified PDE. Note that $\Xi$ has weight $0$
under $\Gamma_0(2)$-transformations. Finally, we note that the monomial sector
exhibits a connection with the two-particle rational Calogero model
\cite{Avan:2012dd,Lechtenfeld:2015wka,Benincasa:2017hcj}, as well as a relation 
with Serre-Rankin-Cohen brackets which is suggestive of formal deformation \cite{cdmr}.

In this paper we will compute the 
quantum entropy function \eqref{qentr} for  large single-center BPS black holes in the
STU model.  
We follow \cite{Gomes:2015xcf,Murthy:2015zzy} and add a boundary term to the 
quantum entropy function,
so as to bring it into a manifestly duality invariant form. 
This modified form is what we thenceforth call the quantum entropy function.
It requires specifying an integration contour ${\cal C}$, which we take
to be the one constructed in \cite{Sen:2007qy} in the context of microstate counting for
$N=4$ BPS black holes.
The quantum entropy function
will depend on the three ingredients displayed in
\eqref{hatOmI}, namely $\omega (S)$, $I_1 (S) \equiv \partial \omega / \partial S$ and
$\Xi (Y^0,S)$. Since ${\widehat \Omega} (Y^0, S, \Upsilon) $ depends on $Y^0$,
so will the quantum entropy function.  The dependence on $Y^0$ brings in
an explicit dependence on the magnetic charges $(p^0, p^1)$, as we will see.
A microstate counting formula that reproduces the quantum entropy function
will have to take this dependence on $Y^0$ into account.
Microstate counting formulae for BPS black holes in four-dimensional $N=4,8$ string
theories \cite{Dijkgraaf:1996it,Shih:2005qf,Jatkar:2005bh,David:2006ru} 
are based on modular objects.
In the $N=4$ context, these are Siegel modular forms, and \cite{David:2007tq,Huang:2015sta} indicates
that they should also play a role in $N=2$ theories. However, in the latter case,
microstate counting formulae cannot be solely based on Siegel modular forms
due to the dependence of the quantum entropy function on 
the additional modulus $Y^0$.

Let us briefly describe our microstate proposal.  We find that $\omega (S)$,
which is proportional to $\ln \vartheta_2^8 (S)$,
 together with
subleading corrections in the quantum entropy function, point
to a dependence of the microstate counting formula on a Siegel modular
form $\Phi_2$ of weight $2$.  This Siegel modular form can be constructed by applying
a Hecke lift \cite{Jatkar:2005bh} to a specific Jacobi form constructed from the seed
$\vartheta_2^8$, and it
differs from the Siegel modular form proposed in 
\cite{David:2007tq}, which did not take into account the subleading corrections
just mentioned.  Due to the dependence on $Y^0$, this needs to be 
supplemented by a modular object that depends on $Y^0$. By focusing attention onto
the leading divisor of the Siegel modular form $\Phi_2$, we give an expression
for this modular object: one of the ingredients that goes into it  is $\Xi$, which we view as a solution to the aforementioned
non-linear PDE.  We then verify that on the leading divisor, the proposed microstate
counting formula is invariant under the subgroup $H \subset Sp (4, \mathbb{Z})$
that 
acts on the Siegel upper half plane and implements 
the $\Gamma_0(2)_S$-symmetry of the STU model.
We stress that the various approximations that we have implemented
(such as working in a certain region of moduli space, or focussing attention on the
leading divisor of $\Phi_2$), render the microstate proposal to be
only an approximate one.

The paper is structured as follows.  In section \ref{sec:QuantumentropyfunctionS}, we give a brief review of the quantum entropy function
for BPS black holes in $N=2$ supergravity theories in four dimensions. In section  \ref{quantumSTU}  we evaluate the quantum entropy function 
for large single-center BPS black holes in the STU model. 
While doing so, we work in a certain region of moduli space
in which the  function $F = F^{(0)} + 2 i \Omega$ that encodes the Wilsonian effective action simplifies: 
$\Omega$ becomes effectively replaced by \eqref{hatOmI}. The modular invariant 
function $\Xi$ in \eqref{hatOmI} 
arises as a solution to a non-linear PDE, and 
 exhibits a relation with the two-particle rational Calogero model,
  as well as with Serre-Rankin-Cohen brackets and formal deformation. We
  discuss our integration contour $\cal C$, which we take to be the one constructed in 
   \cite{Sen:2007qy}.
   The result for the quantum
 entropy function is based on several assumptions and approximations which we explain.
 In section \ref{sec:microstatesec}, we propose a microstate counting formula that reproduces this result for the quantum entropy function.  It is based
 on the Siegel modular form $\Phi_2$
  as well as on another modular object that captures the dependence on the complex scalar $Y^0$.
 Given the various approximations that went into computing the quantum entropy function, the proposed counting function will be an approximation
 to the (as yet unkown) exact counting function.
 In section \ref{sec:concl},
we conclude with a brief summary and a few observations, and we suggest a string web picture
of our counting proposal.
In appendices \ref{sec:modsl} - \ref{sec:Heckelift}, we collect results 
 about
 modular forms for $SL(2, \mathbb{Z})$ and $\Gamma_0 (2)$, Jacobi forms, Siegel modular forms, Rankin-Cohen brackets and Hecke lifts.

\section{Quantum entropy function for BPS black holes in $N=2$ supergravity theories}
\label{sec:QuantumentropyfunctionS}
\setcounter{equation}{0}

\subsection{Generic structure}

The equations of motion of $N=2$ supergravity coupled to $n_V$ Abelian vector multiplets
 in four dimensions admit dyonic
single-center BPS black hole
solutions \cite{Ferrara:1995ih}. These are static, 
extremal black hole solutions that carry electric/magnetic charges $(q_I,p^I)$, where the index $I = 0, \dots, n_V$ 
labels the Maxwell fields in the theory. The
near-horizon geometry of these solutions contains an $AdS_2$ factor. These BPS black holes are supported
by complex scalar fields $Y^I$ that reside in the vector multiplets. At the horizon,
these scalar fields are fixed in terms of specific values that are entirely specified by the charges carried
by the black hole.

In the presence of higher-derivative terms proportional to the
square of the Weyl tensor, the associated $N=2$ Wilsonian effective action is encoded
in a holomorphic function $F(Y, \Upsilon)$ \cite{deWit:1996gjy}, where the complex field $\Upsilon$
is related to the lowest component of the square of the Weyl superfield. At the horizon of a BPS
black hole, the field $\Upsilon$ takes the real value $\Upsilon = - 64$, and  Wald's entropy
of the BPS black hole \cite{LopesCardoso:1998tkj} can be written as \cite{Ooguri:2004zv}
\begin{equation}
{\cal S}_{\rm BH}
 (q,p) = \pi \left[ 4 \, {\rm Im} F (Y, \Upsilon) - q_I (Y^I + {\bar Y}^I )  \right]  \;,
\label{wald_entro}
\end{equation}
where
\begin{equation}
Y^I = \tfrac12 \left( \phi^I + i p^I \right) \;\;\;,\;\;\; \Upsilon = -64 \;
\label{Xphi}
\end{equation}
are evaluated at the horizon. The horizon value of 
$\phi^I$  is determined
by extremizing
the right hand side of \eqref{wald_entro} with respect to $\phi^I$,
\begin{equation}
4 \, \frac{\partial}{\partial \phi^I} \, 
 {\rm Im}  F(\phi + i p, -64)
  =  q_I \;.
  \label{attracq}
\end{equation}

Wald's entropy  \eqref{wald_entro} constitutes the semi-classical approximation to the
exact macroscopic entropy of a BPS black hole. For extremal black holes whose near-horizon geometry 
contains an $AdS_2$ factor, a proposal 
for the  exact macroscopic entropy of  extremal black holes has been put forward in 
\cite{Sen:2008yk,Sen:2008vm}. This proposal is called the
quantum entropy function, and it is a regulated
partition function for quantum gravity in a near-horizon $AdS_2$ space-time.

The quantum
entropy function is a functional integral, which for the subclass of BPS black holes
can be defined by means of equivariant localization techniques
\cite{Dabholkar:2010uh,Murthy:2015yfa,Gomes:2015xcf,Murthy:2015zzy,Gupta:2015gga,deWit:2018dix,Jeon:2018kec}
that reduce the infinite dimensional functional integral to a finite dimensional integral
over a bosonic localization manifold. When restricting to smooth field configurations in near-horizon $AdS_2$ backgrounds,  
the resulting expression for the quantum entropy function
$W(q,p)$ for BPS black holes 
takes the form
\begin{equation}
W(q,p) = \int _{\mathcal{C}} 
d \phi \, \mu(\phi) \,  Z_{\rm 1-loop}(\phi)  \; e^{\pi [ 4 \, {\rm Im} F(\phi + i p) -  q \cdot \phi] } \, 
\;,
\label{quantumentro}
\end{equation}
where $q \cdot \phi = q_I \, \phi^I$, with $I = 0, \dots, n_V$. 
The localization manifold is labelled by $n_V + 1$ parameters $\{ \phi^ I \}$.
Note that the integral in \eqref{quantumentro} requires specifying a contour $\mathcal{C}$.
The measure $\mu(\phi)$
arises as a result of the localization procedure implemented
on the field configuration space.
The term $Z_{\rm 1-loop} (\phi)$ describes the semi-classical correction,
giving rise to a super determinant, that arises when 
performing the Gaussian integration over terms quadratic in 
quantum fluctuations around the localization manifold. 
 The function $F$ entering in 
\eqref{quantumentro} is the holomorphic function that defines the $N=2$ Wilsonian effective action 
(with  $\Upsilon =-64$).

Let us describe the quantities that enter in \eqref{quantumentro} in more detail.
The Wilsonian function $F$  can be decomposed into
\begin{equation}
F(Y, \Upsilon) = F^{(0)}(Y) + 2 i \, \Omega(Y, \Upsilon) \;,
\label{FWilson}
\end{equation}
where the dependence on the field $\Upsilon$ is solely contained in $\Omega$. The first term
$F^{(0)}(Y)$ is the so-called prepotential of the $N=2$ model.
Local supersymmetry requires both
$F^{(0)}$ and $\Omega$ to be homogeneous of degree $2$ under complex rescalings \cite{deWit:1984wbb}, 
\begin{eqnarray}
F^{(0)} (\lambda Y) &=& \lambda^2 \, F^{(0)}(Y) \;, \nonumber\\
\Omega (\lambda Y, \lambda^2 \Upsilon) &=& \lambda^2 \, \Omega (Y, \Upsilon) \;\;\;,\;\;\; \lambda \in \mathbb{C}\backslash \{0\} \;,
\label{rescaling}
\end{eqnarray}
and hence they
satisfy the homogeneity relations
\begin{eqnarray}
2 F^{(0)} &=& Y^I \, F_I^{(0)} \;, \nonumber\\
2 \Omega &=& Y^I\, \Omega_I + 2 \Upsilon \, \Omega_{\Upsilon} \;.
\label{homog}
\end{eqnarray}
Here, we have introduced the notation $F_I = \partial F/\partial Y^I, \; \Omega_{\Upsilon} = \partial \Omega /
\partial \Upsilon$.

The Wilsonian effective action is based on a function $F$ where
$\Omega(Y, \Upsilon)$ is 
given in terms of a power series expansion in $\Upsilon$.
Choosing $\lambda = 1/Y^0$ in the homogeneity relation \eqref{homog} implies
that $\Omega(Y, \Upsilon)$ takes the form
\begin{equation}
\Omega(Y^0, z^A, \Upsilon) = \Upsilon\, \sum_{n=0}^{\infty} \left( \frac{\Upsilon}{(Y^0)^2} \right)^n \, 
\omega^{(n+1)} (z^A) \;.
\label{Omomexp}
\end{equation}
 Here, the gravitational coupling 
functions $\omega^{(n+1)}$ only depend on the projective coordinates 
$z^A = Y^A / Y^0$ ($A=1, \dots, n_V)$.
In a generic $N=2$ model, this perturbative series
in $\Upsilon/(Y^0)^2$ may not be convergent.

Next, let us discuss the measure factor $\mu$. While this measure factor
has not yet been worked out from first principles\footnote{A formalism
for carrying out localization calculations in the presence of a supersymmetric
background has recently been given in \cite{deWit:2018dix,Jeon:2018kec}.}, an approximate expression for it can be
obtained by demanding consistency under electric-magnetic duality transformations
as well as consistency with semi-classical results for BPS entropy, as follows.

Electric-magnetic duality is a characteristic
feature of systems with $N=2$ Abelian vector multiplets in four dimensions, also
in the presence of a chiral background field \cite{deWit:1996gjy}.
Duality transformations act as symplectic transformations on the Abelian field strengths, and hence, on the associated
charge vector $(p^I, q_I)$, as well as on the vector $(Y^I, F_I)$ \cite{deWit:1984wbb}.
The semi-classical entropy \eqref{wald_entro}
of a BPS black hole transforms as a function under symplectic
transformations, that is, under changes of the duality frame,
and this should also be the case for the exact macroscopic entropy $W(q,p)$ of the 
BPS black hole. 
To ensure that \eqref{quantumentro} transforms 
as a function under symplectic transformations,
the measure factor has to be proportional to \cite{LopesCardoso:2006ugz,Murthy:2015yfa}
\begin{equation}
\sqrt{ \det {\rm Im} F_{KL}} \;,
\end{equation}
as we will review in the next subsection.
Here, $F_{KL}$ denotes the second derivative of the Wilsonian function $F$ \eqref{FWilson} with respect to the $Y^I$.

In addition, the measure factor $\mu$ should also include non-holomorphic terms,
which we will denote by $\Sigma$, that
have their origin in the holomorphic anomaly equations for the free energies
of perturbative
topological string theory, and are needed for consistency with semi-classical results \cite{Cardoso:2008fr}. Keeping
only non-holomorphic terms associated with the topological
free energy $F^{(1)}$, $\Sigma$ takes the form
\begin{equation}
\Sigma = - \tfrac12 \ln \vert \det {\rm Im} F^{(0)}_{IJ} \vert + \left( \frac{\chi}{24} -1 \right) \ln
\left( \frac{ e^{- {\cal K}^{(0)}} }{|Y^0|^2} \right) \;,
\end{equation}
where 
 $\chi = 2 (n_V - n_H + 1 )$, which is determined in terms of the number of vector and
 hyper multiplets of the $N=2$ model ($n_V$ and $n_H$, respectively),
denotes the Euler number of the Calabi-Yau threefold underlying the model,
and
\begin{equation}
e^{- {\cal K}^{(0)}} = i \left( {\bar Y}^I \, F_I^{(0)} - Y^I \,{\bar F}_I^{(0)} \right) \;.
\label{KF0}
\end{equation}
We are thus led to consider a measure factor of the form
\begin{equation}
\mu = \sqrt{ \det {\rm Im} F_{KL}} \; e^{\Sigma} \;,
\label{measDel}
\end{equation}
which, when approximating 
$F$ by $F^{(0)}$ 
 takes the form
(c.f. eq. (4.21) in \cite{Cardoso:2008fr})
\begin{equation}
\mu
\approx
\left( \frac{ e^{- {\cal K}^{(0)}} }{|Y^0|^2} \right)^{\chi/24-1} \:.
\label{exp-mu}
\end{equation}
The expressions \eqref{measDel} and \eqref{exp-mu} are evaluated at \eqref{Xphi}.
We stress that the measure factor given in \eqref{exp-mu} is an approximate measure factor that will
receive further corrections stemming from $\Omega$ in \eqref{FWilson}. We will, in due course, make use of this
observation. Note that in this approximation, $\mu$ only depends on the projective coordinates $z^A,
{\bar z}^A$.

Now let us turn to the 1-loop determinant  $Z_{\rm 1-loop}$, which reads \cite{Murthy:2015yfa,Gupta:2015gga,Jeon:2018kec}
\begin{equation}
 Z_{\rm 1-loop}  = e^{- (2 - \chi/24) \, {\cal K}} \;.
 \label{Z1loop}
\end{equation}
Here, the quantity
\begin{equation}
e^{- {\cal K}} = i \left( {\bar Y}^I F_I - Y^I {\bar F}_I \right) 
\end{equation}
is computed from the Wilsonian function $F(Y, \Upsilon)$, and not just from the prepotential $F^{(0)}(Y) $.
The factor $2$ in the exponent of \eqref{Z1loop} denotes the contribution from fluctuations of the Weyl multiplet. Note that
the expression for $Z_{\rm 1-loop}$ depends on $Y^0$ and $z^A$ (and their complex conjugates), and
that it is a symplectic function. $Z_{\rm 1-loop}$ is again evaluated at \eqref{Xphi}.

 The quantum entropy function \eqref{quantumentro}
 requires a choice of integration contour ${\cal C}$, which we will specify in
 subsection  \ref{sec:contour} for the specific 
   $N=2$ model under consideration.

Let us close this subsection by mentioning 
three checks that one can perform on the proposed approximate measure factor $\mu$
and on $Z_{\rm 1-loop}$. 
Firstly, when
replacing ${\cal K}$ by ${\cal K}^{(0)}$ in $Z_{\rm 1-loop}$, one infers \cite{Murthy:2015yfa,Gupta:2015gga} that the macroscopic
entropy ${\cal S}_{\rm BH}
(q,p) = \ln {W}(q,p)$ receives a logarithmic correction given by
\begin{equation}
 \Delta {\cal S}_{\rm BH} = \ln Z_{\rm 1-loop}\vert_*  =  (2 - \chi/24) \, \ln e^{- {\cal K}^{(0)}}\vert_* \;.
 \label{delBPSentro}
\end{equation}
Here, $*$ indicates that the expression is evaluated at the attractor values
\eqref{attracq}.
For a large supersymmetric black hole,
$ \pi \, e^{- {\cal K}^{(0)}}\vert_*$ equals the area $A_{\rm BH}$ of the event horizon at the two-derivative level
\cite{Behrndt:1996jn}, and hence
\begin{equation}
 \Delta {\cal S}_{\rm BH}
 =  (2 - \chi/24) \, \ln A_{\rm BH}\;,
\end{equation}
which reproduces the  logarithmic area correction to the BPS entropy of a large black hole
computed in  \cite{Sen:2011ba}.

Secondly, approximating ${\cal K}$ by ${\cal K}^{(0)}$ in \eqref{Z1loop} results in \cite{Sen:2011ba}
\begin{equation}
\mu \,  Z_{\rm 1-loop} = |Y^0|^{(2 - \chi/12)} \, e^{- {\cal K}^{(0)}} \;,
\label{approxmeas}
\end{equation}
which is the combination that plays 
the role of a measure factor in the study \cite{Denef:2007vg} of the OSV conjecture \cite{Ooguri:2004zv}.

Thirdly,
as mentioned above, the measure factor $\mu$ is required to ensure that the quantum entropy function
is a symplectic function.  To verify this, we will now compute ${W}(q,p)$ in saddle-point approximation.

\subsection{Saddle-point approximation}

Let us denote the exponent 
in \eqref{quantumentro} by
\begin{equation}
H(\phi, p, q) =  4 \, {\rm Im} F(\phi + i p) -  q \cdot \phi \;.
\label{Hexp}
\end{equation}
Then,
equations \eqref{attracq} follow by extremizing $H(\phi, p, q)$,
\begin{equation}
\frac{\partial H(\phi, p, q) }{\partial \phi^I} = 0 \;.
\label{saddleeqH}
\end{equation}
Let us assume that for a given set of black hole charges $(q_I, p^I)$
there exists only one non-inflective 
critical  point $\phi_*$, corresponding
to BPS attractor values such that $H(\phi_*, p, q) > 0$.
Expanding $H(\phi, p, q) $ around the 
critical point $\phi_*$, we obtain
\begin{eqnarray}
H(\phi, p, q) &=& H(\phi_*, p, q)  + \tfrac12 \frac{\partial^2 H}{\partial \phi^I \partial \phi^J}{|_{\phi_*}} \,
\delta \phi^I \, \delta \phi^J + {\cal O} ( (\delta \phi)^3) \;, \nonumber\\
&=& H(\phi_*, p, q)  + \tfrac12  \frac{\partial^2 (4 \,  {\rm Im} F)}{\partial \phi^I \partial \phi^J}{|_{\phi_*}} \,
\delta \phi^I \, \delta \phi^J + {\cal O} ( (\delta \phi)^3) \;.
\label{Hexpsadd}
\end{eqnarray}
Using \eqref{Xphi}, this equals
\begin{eqnarray}
H(\phi, p, q)
&=& H(\phi_*, p, q)  + \tfrac12  \, {\rm Im} F_{IJ}|_{\phi_*} \,
\delta \phi^I \, \delta \phi^J  + {\cal O} ( (\delta \phi)^3) \;.
\end{eqnarray}
Next, we evaluate 
 \eqref{quantumentro} in saddle point approximation by
 taking $\mu \,  Z_{\rm 1-loop}$ at the attractor point, 
\begin{equation}
\mu \,  Z_{\rm 1-loop}\vert_{\phi_*} \;,
\end{equation}
 and by
   integrating
 over the fluctuations
$\delta \phi^I \in \mathbb{C}$
in Gaussian approximation,
\begin{eqnarray}
{W}(q,p) &\approx&
\left(\frac{\mu \,  \, Z_{\rm 1-loop} }{\sqrt{ \det {\rm Im} F_{KL}} }\right)\vert_{\phi_*} \, e^{\pi 
H(\phi_*, p, q)} \nonumber\\
&=& \left(e^{\Sigma}   \, Z_{\rm 1-loop} \right)\vert_{\phi_*} \, e^{\pi 
H(\phi_*, p, q)} \;.
\label{Wapproxsymp}
\end{eqnarray}
We note that the factor $\sqrt{\det {\rm Im} F_{KL} }$ has cancelled out against the corresponding factor in \eqref{measDel}.

In the absence of non-holomorphic corrections ($\Sigma =0$), $H$ is a symplectic function \cite{LopesCardoso:2006ugz},
and hence, the result \eqref{Wapproxsymp} is a symplectic function.  
This justifies the presence of the factor $\sqrt{\det {\rm Im} F_{KL}} $ 
in \eqref{measDel}.
In the presence of non-holomorphic corrections
($\Sigma \neq 0$), the discussion of 
symplectic covariance is more involved 
\cite{LopesCardoso:2006ugz,Cardoso:2008fr,Cardoso:2014kwa}, but 
it can be shown that 
the combination $\Sigma + \pi H$ 
in \eqref{Wapproxsymp} is a symplectic function.

In this paper, we will focus on a specific $N=2$ model, namely the 
STU model of Sen and Vafa (example D in \cite{Sen:1995ff}).
This is a model for which the duality symmetries are known exactly.
We will
analyze the expression \eqref{Wapproxsymp}
for this model in subsection \ref{sec:spH}, 
and verify that it is consistent
with the duality invariance of the model.

Finally, we recall that the attractor equations \eqref{saddleeqH} take the form
\begin{equation}
F_I - {\bar F}_{I} = i q_I \;\;\;,\;\; I = 0, \dots, n_V \;,
\end{equation}
with $Y^I = \tfrac12 (\phi^I + i p^I)$. 
The attractor value $H(\phi_*, p, q)$ can be expressed as \cite{LopesCardoso:2006ugz}
\begin{eqnarray}
 H(\phi_*, p, q) = \left[ i \left( {\bar Y}^I F_I - Y^I {\bar F}_I \right)
- 2i \left( \Upsilon \, F_{\Upsilon} -
 \bar{\Upsilon} \, {\bar F}_{\Upsilon}
 \right) \right]\vert_{Y = \tfrac12 (\phi_* + i p)} 
 \label{Hsymp}
\end{eqnarray}
by means of homogeneity relations \eqref{homog}.

\section{Quantum entropy function for the $N=2$ STU model of Sen and Vafa   \label{quantumSTU}  }
\setcounter{equation}{0}

Next, we specialize to the $N=2$ STU model of
Sen and Vafa (example D in 
\cite{Sen:1995ff}). This is a model with duality symmetries which are so restrictive that they 
have recently 
led to the determination \cite{CdWM} of $\Omega$ in \eqref{FWilson}.
We will consider large BPS black holes in this model, and we will
evaluate
the quantum entropy function \eqref{quantumentro} for these black holes.
In doing so, we will work 
 in a certain region of moduli space of the STU model
in which the function $F$ simplifies.
We begin by reviewing the form of $F$ for
this model. 

\subsection{The $N=2$ STU model with $\chi =0$ \label{sec-stu}}

The $N=2$ STU model of \cite{Sen:1995ff}  (example D), obtained
by a $\mathbb{Z}_2 \times \mathbb{Z}_2$ orbifold compactification of
type II string theory,
is a model with $n_V =3$ vector multiplets and $n_H=4$ hyper multiplets,
and hence vanishing Euler number $\chi = 2 (n_V - n_H + 1 )$.
The complex scalar fields residing in the three vector multiplets are denoted by
$S, T$ and $U$,
\begin{equation}
S = -i z^1 = -i \frac{Y^1}{Y^0} \;\;\;,\;\;\; T = -i z^2 = -i \frac{Y^2}{Y^0} \;\;\;,\;\;\;
U = -i z^3 = -i \frac{Y^3}{Y^0} \;.
\label{STUzA}
\end{equation}
The prepotential $F^{(0)} (Y)$ is exact and given by
\begin{equation}
F^{(0)} (Y) = - \frac{Y^1 Y^2 Y^3}{Y^0} \;.
\end{equation}
The model possesses symmetries, in particular a duality symmetry
$\Gamma_0(2)_S \times \Gamma_0(2)_T \times
\Gamma_0(2)_U$, where $\Gamma_0(2)$ denotes a congruence subgroup of the
group $\mathrm{SL}(2, \mathbb{Z})$,
\begin{eqnarray}
\Gamma_0 (2) = \left\{\begin{pmatrix}
a & b\\
c & d\\
\end{pmatrix} \in  \mathrm{SL}(2, \mathbb{Z}): 
\begin{pmatrix}
a & b\\
c & d\\
\end{pmatrix} \equiv \begin{pmatrix}
1 & *\\
0 & 1\\
\end{pmatrix} \; ({\rm mod} \;2)
\right\} \;,
\end{eqnarray}
where $*$ can take any value in $\mathbb{Z}$.

Duality transformations act as symplectic transformations
on the vector $(Y^I, F_I)$.
Under $\Gamma_0(2)_S$ transformations, the complex scalars $S$ and $Y^0$ transform as 
\begin{equation}
S \rightarrow \frac{ a S  - i b}{i c S  + d } \;\;\;,\,\,\, Y^0 \rightarrow (i c S + d) \, Y^0 \;,
\label{SY0transf}
\end{equation}
while the scalars $T, U$ transform as \cite{Cardoso:2008fr}
\begin{eqnarray}
T \rightarrow T + \frac{2i c}{(i c S + d) (Y^0)^2} \frac{\partial \Omega}{\partial U} \;, \nonumber\\
U\rightarrow U + \frac{2i c}{(i c S + d) (Y^0)^2} \frac{\partial \Omega}{\partial T} \;,
\end{eqnarray}
and hence are not inert in the presence of $\Omega$.  Similar transformation rules apply under 
$\Gamma_0(2)_{T, U}$ transformations.

The 
symmetries of the model, namely $\Gamma_0(2)_S \times \Gamma_0(2)_T \times
\Gamma_0(2)_U$ and triality symmetry under exchange of $S, T$ and $U$, are very restrictive, and 
have recently been used \cite{CdWM} to determine the coupling functions 
$\omega^{(n+1)} (z^A)$ in $\Omega$,
c.f. \eqref{Omomexp}. 
This was achieved by adding to $\Omega$ the term $2 \Upsilon \, \gamma \ln Y^0$, with the constant $\gamma$
given  by \eqref{omegam}. This term, which is not of Wilsonian type, since it is not of the power law type,
was crucial to implement the duality symmetries of the STU model while maintaining holomorphy.
It was found that the higher gravitational coupling functions $\omega^{(n+1)} (z^A)$ (with $n \geq 1$)
are determined in terms of the first gravitational coupling function  $\omega^{(1)} (z^A)$. The latter
takes the form \cite{Gregori:1999ns}
\begin{equation}
\omega^{(1)} (S, T, U) = \omega (S) + \omega (T ) + \omega (U) \;, 
\label{om1}
\end{equation}
where 
\begin{equation}
\omega (S) 
= - \tfrac12 \gamma \ln \vartheta^8_2 (S)
\;\;\;,\;\;\; \gamma = - \frac{1}{256 \pi} \;,
\label{omegam}
\end{equation}
with $\vartheta_2(S) = 2 \eta^2 (2 S)/\eta(S)$, and likewise for $\omega (T)$ and $\omega (U)$.
Here, $\vartheta^8_2 (S)$ denotes a modular form of weight $4$ under $\Gamma_0(2)_S$
(with trivial multiplier system \cite{farkas}), 
\begin{equation}
\vartheta^8_2 \left(\frac{a S - i b}{i c S + d} \right) = \Delta^4 (S) \, \vartheta^8_2 (S) \;,
\end{equation} 
where
\begin{equation}
\Delta (S) = i c S + d \;.
\label{deltaS}
\end{equation}
We refer to appendices \ref{sec:modsl} and \ref{sec:G02}
for a brief review of modular forms.
We  pick a single-valued analytic branch of $\ln \vartheta^8_2 (S)$ such that 
under $\Gamma_{0} (2)_S$, $\omega (S)$ transforms into
\begin{equation}
\omega (S) \longrightarrow \omega (S) - 2 \gamma \, \ln \Delta (S) \;.
\label{omSGam02}
\end{equation}

Now let us describe the region in moduli space in which we will work.
It was observed in \cite{CdWM} that the coupling functions $\omega^{(n+1)} (z^A)$ in $\Omega$ simplify 
dramatically when working 
in a regime where two of the three moduli $S, T, U$ (say $T$ and $U$)
are taken to be large, i.e. ${\rm Re} \, T, \; {\rm Re} \, U \gg 1 $.
In this limit,
\begin{equation}
\omega(T) \propto T \;\;\;,\;\;\; \omega(U)  \propto U\;,
\label{wTU}
\end{equation}
that is,
\begin{equation}
\frac{\partial \omega (T) }{\partial T} = - \frac{1}{256} \;\;\;,\;\;\; \frac{\partial \omega (U)}{\partial U} 
= - \frac{1}{256}  \;.
\end{equation}
We introduce the combination
\begin{equation}
\alpha \equiv \frac{\partial \omega(T) }{\partial T} \, \frac{\partial \omega (U) }{\partial U} = \frac{1}{(256)^2}\;,
\label{alpha}
\end{equation}
which is constant in this limit. In this region of moduli space,
the coupling functions $\omega^{(n+1)}$ with $n\geq 1$ in \eqref{Omomexp}
are functions of $S$ only, and are moreover expressed in terms of derivatives of $\omega (S)$,
as we now describe.

In this limit, the coupling function $\omega^{(2)}$ is given by
\begin{equation}
\omega^{(2)} (S) = \frac{\alpha}{\gamma} \, \frac{\partial \omega (S)}{\partial S} \;,
\end{equation}
 and
 the Wilsonian function $\Omega (Y, \Upsilon)$ in \eqref{Omomexp} takes the form\footnote{Note that the Wilsonian function $\Omega$
 used here differs from the function $\Omega$ used in \cite{CdWM} by a term $ \Upsilon \, \gamma \, \ln [(Y^0)^2/\Upsilon]$.}
\begin{eqnarray}
\Omega (Y^0, S, T, U, \Upsilon) &=& \Upsilon \left( \omega (T) + \omega (U) \right) + {\widehat \Omega} (Y^0, S, \Upsilon)
\;, \nonumber\\
{\widehat \Omega} (Y^0, S, \Upsilon) &=& \Upsilon \left[ \omega (S) + \frac{\Upsilon}{(Y^0)^2} \, 
\frac{\alpha}{\gamma} \, \frac{\partial \omega (S)}{\partial S} + 
 \Xi (Y^0, S, \Upsilon) \right] \;,
\label{Sig}
\end{eqnarray}
where we used \eqref{om1}. Here, $\Xi$ 
contains
the coupling functions $\omega^{(n+1)} (S)$ with $n \geq 2$, and it 
denotes a function that is invariant under 
$\Gamma_0(2)_S$ transformations.
This can be established as follows.

Instead of working with variables $S, Y^0$, we 
find it convenient to work with variables $S, z$, where\footnote{The variable
$z \equiv 1/Y^0$ should not be confused with the projective coordinates $z^A$ defined in
\eqref{STUzA}.}
\begin{equation}
z \equiv 1/Y^0 \;.
\end{equation}
Under  $\Gamma_0(2)_S$ transformations, these variables transform according to \eqref{SY0transf}, and hence,
$z \mapsto z/\Delta$. 
The derivatives of ${\widehat \Omega} (Y^0, S, \Upsilon)$ are required \cite{CdWM} to transform according to 
\begin{eqnarray}
  \bigg(\frac{\partial \widehat \Omega}{\partial S}\bigg)^\prime -
  \Delta{\!}^2\,\frac{\partial\widehat \Omega}{\partial S} &=&
  \frac{\partial\Delta}{\partial{S}} 
  \bigg[  \Delta \, z \, \left(  \frac{\partial  \widehat \Omega}{\partial z}  - \frac{2 \Upsilon \, \gamma }{ z} \right)
  -2 \, \Upsilon^2 \,  \alpha \, z^2 \,
  \frac{\partial\Delta}{\partial{S}} 
  \bigg] \;, \nonumber\\
\left(  \frac{\partial  \widehat \Omega}{\partial z}  - \frac{2 \Upsilon \, \gamma }{ z} \right)^\prime 
  &=&   \Delta \,  
  \left(  \frac{\partial  \widehat \Omega}{\partial z}  - \frac{2 \Upsilon \, \gamma }{ z} \right)
     - 4 \,  \Upsilon^2  \, \alpha \,  z 
  \,\frac{\partial\Delta}{\partial{S}} \;,
  \label{hatOmecons}
\end{eqnarray}
where $^\prime$ denotes the transformed quantity. 
Inserting the expression for $\widehat \Omega$ given in \eqref{Sig} into 
\eqref{hatOmecons} results in 
\begin{eqnarray}
\left( \frac{\partial \Xi}{\partial S} \right)^{\prime} &=& \Delta^2 \, \frac{\partial \Xi}{\partial S} 
+ i c \,  \Delta \, z \frac{\partial \Xi}{\partial z} 
\;, \nonumber\\
\left( \frac{\partial \Xi}{\partial z} \right)^{\prime} &=& \Delta \, \frac{\partial \Xi}{\partial z} \;,
\label{transfom2}
\end{eqnarray}
where we used \eqref{omSGam02}. 
Note that \eqref{transfom2}
is linear in $\Xi$. Expanding $\Xi$ in powers of $\Upsilon \, z^2$ \cite{CdWM}, 
\begin{equation}
\Xi (z, S, \Upsilon) = \sum_{n=2}^{\infty} \Upsilon^n \, z^{2n} \, \omega^{(n+1)} (S) \;,
\label{sigmz}
\end{equation}
we find\footnote{Here we assume that we may differentiate \eqref{sigmz}
 term by term, with respect to both $z$ and $S$. } 
that \eqref{sigmz} solves \eqref{transfom2}, provided 
the coupling
functions $\omega^{(n+1)} (S)$ in \eqref{sigmz} are modular forms of weight 
$2n$. 
The solution \eqref{sigmz} is therefore invariant
under $\Gamma_0(2)_S$ transformations. 
The  coupling
functions $\omega^{(n+1)} (S)$ are
expressed in terms of modular forms $I_n$ introduced in \cite{CdWM}, as follows.

Consider the combinations 
\begin{eqnarray}
I_1(S) &=& \frac{\partial \omega (S)}{\partial S} \;, \nonumber\\
I_2 (S) &=&   \partial_S I_1 + \frac{1}{2 \gamma} I_1^2 = 
 \frac{\partial^2\omega (S)}{\partial S^2} +\frac1{2\gamma} \,
  \Big(\frac{\partial\omega (S) }{\partial S}\Big)^2\,.
  \label{I1I2}
\end{eqnarray}
$I_2(S)$ transforms as a modular form under $\Gamma_0(2)_S$ transformations, whereas $I_1(S)$ does not in view of
\eqref{omSGam02}. Therefore, and for latter use, we introduce the combination 
\begin{equation}
\hat{I}_1 (S, \bar S) =   \frac{\partial \omega}{\partial S} 
- \frac{2 \gamma}{ (S + \bar S)} \;,
\label{I1hat}
\end{equation}
which transforms covariantly (with weight 2) under $\Gamma_0(2)_S$ transformations
by virtue of 
\begin{eqnarray}
\frac{1}{S + \bar S} \longrightarrow  \frac{\Delta^2(S)}{S + \bar S} - i c \, \Delta (S) \;.
\end{eqnarray}
Next, define higher $I_n(S)$  by
\begin{eqnarray}
 I_{n+1}(S)  &=& \mathcal{D}_S  I_{n}(S) \;\;\;,\;\;\; n \geq 2 \;.
    \label{eq:DI-I}
\end{eqnarray}
Here, $\mathcal{D}_S$ denotes a holomorphic covariant derivative \cite{hahn}, 
an analogue of Serre derivative,
which acts as follows on $\Gamma_0(2)_S$ modular forms $f_n (S)$ of
weight $2n$,
\begin{equation}
  \label{eq:serre-derivative}
  \mathcal{D}_S\, f_n (S)  = \Big(\frac{\partial}{\partial S} +
  \frac{n}{\gamma}\, \frac{\partial \omega(S)}{\partial S}\Big)
  \,  f_n(S) \,\;\;,\;\;\; n \geq 2 \;.
\end{equation}
Under $\Gamma_0(2)_S$ transformations, the $I_n$ (with $n \geq 2$) transform as
modular forms of weight $2n$, i.e. $I_n (S) \rightarrow \Delta^{2n} \, I_n (S) $.
We refer to appendix \ref{sec:G02} for further properties of the $I_n$, in particular their
relation with Eisenstein series of $\Gamma_0(2)$ \cite{hahn}.

The explicit form of $\Xi (z,S, \Upsilon)$, which was determined in 
\cite{CdWM}, is in terms of a power series
in $I_n (S)$ with $n \geq 2$, 
\begin{equation}
\Xi (z, S,  \Upsilon) = \sum_{n = 2}^{\infty} \alpha_n \; \Upsilon^n \;  z^{2n} \; I_n(S) + \sum_{m, n=2}^{\infty}
\alpha_{m,n} \; \Upsilon^{m+n} \;  z^{2(m+n)} \; I_m(S) \; I_n(S) + \dots \;.
\label{Inexp}
\end{equation}
In this expansion, each summand is invariant under  $\Gamma_0(2)_S$ transformations.
This expansion contains an infinite number of different sectors, characterized by
different powers of $I_n$.
The expansion coefficients in \eqref{Inexp} can, in principle, be determined 
by following the rather laborious procedure described in \cite{CdWM}, which consists
in working at a generic point in moduli space, solving the associated conditions on $\Omega$ imposed
by duality, and only then taking $T$ and $U$ to be large. 
In \cite{CdWM}, only expressions for the coefficients $\alpha_n$ and $\alpha_{m,n}$ of the two first sectors were given:
the $\alpha_n$ were determined fully, and partial expressions for the $\alpha_{m,n}$ were given.
Below we write down explicit expressions for these coefficients (with $m,n \geq 2$):
\begin{eqnarray}
\label{coefficientsalphanmn}
\alpha_n &=& \frac{1}{n!} \left( \frac{\alpha}{\gamma} \right)^n  \;, \\
\alpha_{m,n} &=& \beta_{m,n} \, \left( \frac{\alpha}{\gamma} \right)^{m +n}  \;, \nonumber\\
\beta_{m,n} &=& - \frac{1}{2 \gamma} \, 
 \frac{1}{
  (m-1)!\, (n-1)!} \, \bigg[ \frac{1}{m(m+1)}
  +\frac{1}{n(n+1)}  - \frac{(m+n-2)!}{(m+n)!} +\frac{1}{2 } \,
  \bigg] \;. \nonumber
 \end{eqnarray}
The coefficients  $\beta_{m,n} $ satisfy the following recursion relation (with $m,n \geq 2$),
\begin{eqnarray}
(m + n) \, \beta_{m,n} &= &\beta_{m,n-1} + \beta_{m-1,n} - \frac{1}{2 \gamma}   \frac{1}{
  (m-1)!\, (n-1)!} \nonumber\\
  && -  \frac{1}{2 \gamma} \;   \frac{1}{ 
  (m-1)! } \;  \delta_{n,2} - \frac{1}{2 \gamma}  \;  \frac{1}{ (n-1)!  } \;\delta_{m,2} \;,
  \label{recursion_betamn}
    \end{eqnarray}
with $\beta_{1,n} = \beta_{m,1} \equiv 0$. Note that the terms in the bracket of $\beta_{m,n}$
are related to triangular numbers.

In this paper we will focus on the two first sectors displayed in \eqref{Inexp}, with the coefficients given by 
\eqref{coefficientsalphanmn}.
Both series are convergent, as we will show in appendix \ref{sec:jacfor}.
The quantity $\Xi (z, S, \Upsilon)$
will play a role in the microstate counting
proposal for large BPS black holes in section \ref{sec:microstatesec}.

The expansion \eqref{Inexp} in powers of $I_n$ may look unfamiliar. 
In appendix \ref{sec:jacfor}, we discuss another example of such an
expansion, namely the expansion of $\ln \vartheta_2^8 (S,z)$.

 \subsection{Properties of $\Xi (z, S, \Upsilon)$}
 
 The expansion \eqref{Inexp} with coefficients \eqref{coefficientsalphanmn}
 satisfies various interesting relations, which we now describe. 
First, we present a non-linear PDE that is satisfied by $\Xi$ in \eqref{Inexp}.
We also display a candidate for a non-linear PDE governing the all-order completion of 
 \eqref{Inexp}. Next, we relate the monomial sector in \eqref{Inexp} to (Serre-) Rankin-Cohen
 brackets at level $1$ and comment on its relation with formal deformation. And finally,
 we deduce a Hamilton-Jacobi equation for $\widehat{\Omega}$ in \eqref{Sig}, whose Hamiltonian
 describes
 a time-dependent deformation of a 
 rational Calogero model.

 \subsubsection{Non-linear PDE}
 
 Let $S = - i \tau$, with $\tau$ taking values in the complex upper half plane $ \mathcal{H}$,
 and $z \in \mathbb{C}$. Consider the following 
  non-linear PDE for a complex function $\Xi$ depending on $S$ and $z^2$,
 \begin{equation}
\nu \, {D}_S \Xi = \frac{\partial \Xi}{\partial u} + \frac{1}{2 \gamma} \,  u  \left(
\frac{\partial \Xi}{\partial u}  \right)^2 + \frac{\nu^2}{\gamma} \, I_2 \, u^2 \, 
\frac{\partial \Xi}{\partial u}
- \nu^2 \, I_2 \, u  \;,
\label{PDEXi}
\end{equation}
where we set $u \equiv z^2$ for ease of notation in this subsection, and where 
$\nu = \alpha \Upsilon/\gamma$. 
Here we defined
\begin{equation}
{D}_S \Xi \equiv 
\partial_S \Xi + \frac{1}{\gamma} \, I_1 \,  u \, \partial_{u}  \Xi \;.
\end{equation}
$I_1$ and $I_2$ denote the combinations 
given in \eqref{I1I2}.\\

\noindent
{\bf Proposition:}
The non-linear PDE \eqref{PDEXi}, subject to $\Xi \vert_{u=0} = 0$ and $(\partial_{u} \Xi) \vert_{u=0} = 0$, admits a 
 $\Gamma_0(2)_S$ invariant 
solution that is analytic in $u$ in an open neighbourhood of $u=0$.
This solution is given by
the series \eqref{Inexp} with coefficients  \eqref{coefficientsalphanmn}, up to terms that involve products of three $I_n$ or higher.

\begin{proof}

We begin by showing that if a solution $\Xi$ to \eqref{PDEXi} exists that
satisfies $\Xi \vert_{u=0} = 0$ and $(\partial_{u} \Xi) \vert_{u=0} = 0$, 
and that is analytic in $u$ in an open neighbourhood of $u=0$, then it is unique.
Such a solution will be given by the series 
 $\Xi (u, S) = \sum_{n = 2}^{\infty}  u^{n} \; h_n (S ) $, with $h_2(S) =  \frac{\nu^2}{2} \, I_2(S)$.
 The latter follows immediately by inserting this series into \eqref{PDEXi} and assuming
 that one can differentiate it on a term by term basis. Now 
suppose that there are two such solutions,
namely $\Xi_1 (u, S) = \sum_{n = 2}^{\infty}  u^{n} \; h_n (S ) $ and 
 $\Xi_2 (u, S) = \sum_{n = 2}^{\infty}  u^{n} \; g_n (S ) $, such that $h_2 (S) = g_2(S)$. Then, if the $n$-th coefficient functions are the same, i.e. $h_n (S) = g_n(S)$,
 also the $(n+1)$-st coefficient functions will agree, i.e. $h_{n+1} (S) = g_{n+1}(S)$.
This follows by direct inspection of the differential equation  \eqref{PDEXi}, 
which shows that the $(n+1)$-st coefficient function is determined in
terms of the lower coefficient functions. This also shows that the coefficient functions $h_n (S ) $ 
are modular forms of weight $2n$ under  $\Gamma_0(2)_S$ transformations, and hence
$\Xi (u, S)$ is a  $\Gamma_0(2)_S$ invariant solution.

Next, let us construct this $\Gamma_0(2)_S$ invariant 
solution $\Xi (u, S) = \sum_{n = 2}^{\infty}  u^{n} \; h_n (S ) $.
This solution will be given in terms of an expansion in powers
of $I_n(S)$ (with $n \geq 2$).  To distinguish between these various
powers, we rescale $\Xi$ and $I_2$ in \eqref{PDEXi} by a real parameter
$\lambda \in \mathbb{R} \backslash \{0\}$, to obtain
\begin{equation}
\nu \, {D}_S \Xi = \frac{\partial \Xi}{\partial u} + \frac{\lambda}{2 \gamma} \,  u
\left(
\frac{\partial \Xi}{\partial u}  \right)^2 + \lambda \,\frac{\nu^2}{\gamma} \, I_2 \, u^2 \, 
\frac{\partial \Xi}{\partial u}
- \nu^2 \, I_2 \, u  \;.
\label{PDEXilamb}
\end{equation}
We then 
construct a solution to 
\eqref{PDEXilamb} order by order in $\lambda$. Since 
$\lambda$ is a dummy variable, we set $\lambda =1$ at the end,
thereby arriving at a $\Gamma_0(2)_S$ invariant 
solution $\Xi (u, S)$ that is organized in powers of $ I_n(S)$, as in \eqref{Inexp}.

To lowest order
in $\lambda$, \eqref{PDEXilamb} reduces to 
\begin{equation}
\nu \, {D}_S \Xi = \frac{\partial \Xi}{\partial u} 
- \nu^2 \, I_2 \, u  \;.
\label{PDEXilow}
\end{equation}
We seek a solution to this differential equation of the form
\begin{equation}
\Xi (u, S) = \sum_{n = 2}^{\infty}  u^{n} \; f_n(S )\:.
\label{mod_inv_sol}
\end{equation}
Inserting \eqref{mod_inv_sol} into  \eqref{PDEXilow}, and
assuming that we can 
differentiate \eqref{mod_inv_sol} term by 
term\footnote{This will be shown to be the case in appendix \ref{sec:jacfor}, at least so long
as ${\rm Re} \, S $ is taken to be large.},
one immediately infers that 
$f_n(S) = \frac{\nu^n}{n!} \, I_n(S)$ for $n \geq 2$
by virtue of \eqref{eq:serre-derivative}.
This reproduces the first (monomial) sector in the expansion
\eqref{Inexp}. 

At the next order in $\lambda$, we return to \eqref{PDEXilamb} and substitute
\eqref{mod_inv_sol} into the term $\left( \partial \Xi / \partial u  \right)^2$.
At this order, \eqref{PDEXilamb} is solved by the second (binomial) sector 
in the series 
\eqref{Inexp}, with the coefficients 
satisfying 
the recursion relation \eqref{recursion_betamn}, thus reproducing the result \eqref{coefficientsalphanmn}.

Setting $\lambda =1$ we
conclude that \eqref{Inexp} solves the non-linear PDE \eqref{PDEXi}, 
up to terms that involve products of three $I_n$ or higher.

\end{proof}

The expansion $\Xi$ in  \eqref{Inexp} receives corrections that are of higher order (higher than two) in the $I_n$.
These will in turn lead to a modification of the PDE \eqref{PDEXi}.  The coefficients of these
higher order terms were not determined in \cite{CdWM}. Inspection of \eqref{PDEXi}
suggests the following all-order completion of  \eqref{Inexp} (under the assumption that the expressions for the coefficients $\alpha_{m,n}$ 
displayed in \eqref{coefficientsalphanmn} are actually the exact expressions).\\

\noindent
{\bf Conjecture:}
The non-linear PDE governing the all-order completion of \eqref{Inexp} is given by
\begin{equation}
\nu \, {D}_S \left( \Xi/\gamma\right) = \frac{1}{u} \left(
e^{u \partial \left( \Xi/\gamma\right)/\partial  u}
-1 
- \frac{ \nu^2 \,  u^2 \,  I_2/\gamma }{ 1 + u \, \frac{\partial \left( \Xi/\gamma\right)}{\partial 
u }
}
\right) \;.
\label{PDEXiall}
\end{equation}
However, at this stage, we cannot verify this, since we do not have at our disposal the coefficients
of the higher-order terms.\\

Note that
$D_S$ acts as follows on a given summand of $\Xi$ in  \eqref{Inexp}. Consider the summand
\begin{equation}
\alpha_{n_1, n_2, \dots, n_k} \, z^{2(n_1 + n_2 + \dots n_k)} \, I_{n_1} \, I_{n_2} \dots I_{n_k} \;,
\end{equation}
with $\alpha_{n_1, n_2, \dots, n_k}$ constant coefficients.
Then,
\begin{equation}
D_S \left( \alpha_{n_1, n_2, \dots, n_k} \, z^{2(n_1 + n_2 + \dots n_k)} \, I_{n_1} \, I_{n_2} \dots I_{n_k} \right) = \alpha_{n_1, n_2, \dots, n_k} \, z^{2(n_1 + n_2 + \dots n_k)} \, 
{\cal D}_S \left( I_{n_1} \, I_{n_2} \dots I_{n_k} \right)\;,
\end{equation}
where ${\cal D}_S $
denotes the covariant derivative introduced in \eqref{eq:serre-derivative}.

\subsubsection{$I_n$ and Rankin-Cohen brackets }

Second, let us focus on the terms linear in $I_n$ in the expansion \eqref{Inexp},
\begin{equation}
\Xi_1 (z, S, \Upsilon) =  \sum_{n = 2}^{\infty}  \frac{\nu^n}{n!} \, 
\;  z^{2n} \; I_n(S) \;.
\label{xi1}
\end{equation}
Note that the coefficients have a $1/n!$-suppression. This series is convergent, 
at least so long
as ${\rm Re} \, S $  and $|Y^0|$ are
taken to be large,
as we show
in appendix \ref{sec:jacfor}.

Using the property $I_{n+1} (S) = {\cal D}_S
I_n (S)$ for $n\geq 2$, together with $I_2(S) = \frac12 \pi \gamma \, {\cal D}_S {\tilde {\cal E}}_2 (S)$,
where  ${\tilde {\cal E}}_2 (S)$ denotes the basis vector of the vector space ${\cal M}_2 (\Gamma_0(2))$, see \eqref{I2E2}, we obtain
\begin{equation}
\Xi_1 (z, S, \Upsilon) =  \frac12 \pi \gamma \,  \sum_{n = 1}^{\infty}  \frac{1}{(n+1)!} \, \nu^{n+1} 
\;  z^{2(n+1)} \; {\cal D}_S^{n} \, {\tilde {\cal E}}_2 (S) =  \frac12 \pi \gamma
\left(  \frac{e^{\nu \, z^2 \,
{\cal D}_S} -1 - {\nu \, z^2 \,
{\cal D}_S} 
}{{\cal D}_S} \right)  {\tilde {\cal E}}_2 (S)  \;,
\end{equation}
where the operator on the right hand side is defined by the power series.

Next,
let us add to $ \Xi_1 (z, S, \Upsilon)$ the term
proportional to $I_1(S) = \partial \omega(S) / \partial S$ that appears in ${\widehat \Omega}$ given in  \eqref{Sig}, so that now we consider
\begin{equation}
{\widehat \Xi}_1 (z, S, \Upsilon) =  \sum_{n = 1}^{\infty}  \frac{\nu^n}{n!} 
\;  z^{2n} \; I_n(S) \;.
\end{equation}
As we review in appendix \ref{sec-rkb},
the $I_n$, with $n\geq 3$, can be expressed in terms of $1$st Rankin-Cohen brackets for modular forms, 
while $I_2$ can be expressed in terms of the quasi-modular form $I_1$ by making use of $1$st Rankin-Cohen brackets for quasi-modular
forms of depth $1$ \cite{martin},
\begin{equation}
I_2 = -  \frac{1}{4 g} [ I_1, g]_1 \;,
\end{equation}
where $g(S) = \vartheta_2^8 (S)$ has weight $4$. 
 We thus have the following proposition. \\

\noindent
{\bf Proposition:} Let $g(S) = \vartheta_2^8 (S)$. Then,
${\widehat \Xi}_1 (z, S, \Upsilon)$ can be expressed as
\begin{eqnarray}
{\widehat \Xi}_1 (z, S, \Upsilon) 
= \nu  \, z^2
\sum_{n=0}^{\infty} \frac{(-1)^n}{(n+1)!} \, \left(
\frac{ \nu \, z^2}{4} \right)^n
\, \left( \frac{1}{g}  [\cdot, g]_1 \right)^n \, I_1(S) \;,
 \label{Sig1RC}
\end{eqnarray}
where
\begin{equation}
 \left( \frac{1}{g}  [\cdot, g]_1 \right)^n \, I_1 \equiv 
  \frac{1}{g} \, [\frac{1}{g}  [  \dots [ \frac{1}{g} [I_1, g]_1, \dots, g]_1, g]_1 \;,
\end{equation}
with the understanding that this equals $I_1$ when $n=0$. 
In the last step, the bracket $[\cdot, \cdot]_1$ denotes
the $1$st Rankin-Cohen bracket for quasi-modular
forms of depth $1$.

\begin{proof}
This follows immediately, using the results of appendix \ref{sec-rkb}.

\end{proof}

The expression \eqref{Sig1RC}  exhibits a formal similarity with the differential \eqref{adk}
of the exponential map $\exp : \mathfrak{g}
\rightarrow G$ from the Lie algebra   $\mathfrak{g}$ into the linear group $G$.
\\

Now consider 
the $n$th Serre-Rankin-Cohen bracket \eqref{rcmodcov} for $\Gamma_0(2)$,
\begin{equation}
\frac{1}{g} \, SRC_n (I_2,g) (S)  =  (-1)^n \, \begin{pmatrix}
n + 3\\
n
\end{pmatrix} \, 
I_{n+2} (S) \;,
\end{equation}
where we used \eqref{covtheta8}. Thus, we can write ${\widehat \Xi}_1 (z, S, \Upsilon) $
as
\begin{eqnarray}
\label{Xieholz}
{\widehat \Xi}_1 (z, S, \Upsilon) &=& \nu \, z^2 \, I_1(S) + 
 \sum_{n = 0}^{\infty}  \frac{1}{(n+2)!} \, \nu^{n+2} 
\;  z^{2(n+2)} \; I_{n+2} (S) \\
&=& \nu \, z^2 \, I_1(S) + 6 \,  \frac{\nu^2 z^4 }{g} \,  \sum_{n = 0}^{\infty}  \frac{(-1)^n
\; n!}{  (n+2)! \;
(n+3)!} \, \left( \nu  z^{2} \right)^n \,  SRC_n (I_2,g) (S) \;.
\nonumber
\end{eqnarray}
We relate this to the formal deformation Eholzer product \eqref{ehol} by means of the 
generalized hypergeometric  function 
\begin{equation}
{}_2 F_2 (a, b, c, d; x) =  \sum_{n =0}^{\infty} \frac{ (a)_n \, (b)_n}{(c)_n \, (d)_n} \, \frac{x^n}{n!} \;,
\end{equation}
where $(a)_n = a (a+1) \dots (a + n-1)$. Using
\begin{equation}
{}_2 F_2 (1, 1, 3, 4; x) = 12  \sum_{n =0}^{\infty} \frac{ n!}{(n+2)! \, (n+3)!} \, x^n \;,
\end{equation}
we obtain the formal expression
\begin{eqnarray}
{\widehat \Xi}_1 (z, S, \Upsilon) 
&=& \nu \, z^2 \, I_1(S) +  \frac{\nu^2 z^4 }{ 4 \pi i  g} \,  \ointctrclockwise \frac{d \hbar}{\hbar} \, {}_2 F_2 (1, 1, 3,4; - \frac{\nu z^2}{\hbar}) \, I_2  \#  g \;,
\end{eqnarray}
where $I_2 \# g$ is given as in \eqref{ehol}, with $\hbar \in \mathbb{C}$ a deformation parameter, and the integration contour encloses the origin.
It would be interesting to further study the relation of ${\widehat \Xi}_1 (z, S, \Upsilon)$ with formal
deformation.

\subsubsection{Deformed Calogero model}

We return to \eqref{hatOmecons} and deduce a Hamilton-Jacobi equation 
from it, as follows.

Taking the square of the second equation in \eqref{hatOmecons}, and suitably combining
it with the first equation in \eqref{hatOmecons}, we obtain
\begin{eqnarray}
\Bigg[ 8  \, \Upsilon^2 \, \alpha \,
  \frac{\partial \widehat \Omega}{\partial S} + \left(  
   \frac{\partial  \widehat \Omega}{\partial z}  - \frac{2 \Upsilon \, \gamma }{ z} 
  \right)^2 \Bigg]^\prime = \Delta^2  \,
  \left[ 8  \, \Upsilon^2 \, \alpha \,  \, \frac{\partial \widehat \Omega}{\partial S} +  
  \left(  \frac{\partial  \widehat \Omega}{\partial z}  - \frac{2 \Upsilon \, \gamma }{ z}  \right)^2 \right] \;.
  \end{eqnarray}
 This implies that the combination 
 $\left[ 8 \, \Upsilon^2 \,  \alpha \, \partial \widehat \Omega/\partial S +
  (\partial \widehat \Omega/\partial z  - 2 \Upsilon \, \gamma / z )^2  \right]$ transforms as a modular form
  of weight $2$ (with a trivial multiplier system) under $\Gamma_0(2)_S$ transformations.  
 We denote  this combination by $V(S,z)/z^2$, where $V(S,z)$ denotes a modular
invariant function, 
\begin{equation}
8  \, \Upsilon^2 \, \alpha \,  \, \frac{\partial \widehat \Omega}{\partial S} = - \left[
  \left(  \frac{\partial  \widehat \Omega}{\partial z}  - \frac{2 \Upsilon \, \gamma }{ z}  \right)^2 - 
  \frac{V(S,z)}{z^2} \right] \;.
  \label{calpot}
\end{equation}
Then, inserting the expression \eqref{Sig} for $\widehat \Omega$ into 
\eqref{calpot} gives
\begin{eqnarray}
 \nu 
\, \left( \frac{\partial \Xi}{\partial S} + \frac{1}{2 \gamma}   \, I_1 (S) \, z \, \frac{\partial  \Xi}{\partial z} \right)
&=& -  \frac{1}{8 \gamma} \left(  \frac{\partial  \Xi}{\partial z} \right)^2 
+ \frac{1}{2z}  \frac{\partial  \Xi}{\partial z}  
-  \nu^2  
 \, I_2 (S) \, z^2\nonumber\\
&& - \left( \frac{4  \, (\gamma \,  \Upsilon)^2- V(S,z) }{8 \gamma \, \Upsilon^2 \, z^2} \right) \;,
  \label{diffhatOm22}
\end{eqnarray}
where $\nu = \alpha \Upsilon/\gamma$, as before. 
By comparing this equation with \eqref{PDEXi}, we infer
\begin{equation}
V(S,z) = 4  \, (\gamma \,  \Upsilon)^2 + 2 \, \Upsilon^2 \, z^2
\left(     \left(
\frac{\partial \Xi}{\partial z}  \right)^2 + 2 \nu^2 \, I_2 (S)\, z^3 \, 
\frac{\partial \Xi}{\partial z}
\right) \;.
\label{defpot}
\end{equation}
\\

\noindent
{\bf Proposition:}
The partial differential equation \eqref{calpot} is a Hamilton-Jacobi equation with Hamilton's 
principal function ${\cal S} (t, z)$ given by
\begin{equation}
{\cal S} (t, z) = 
 { \widehat \Omega} (t, z)  - 2 \Upsilon \, \gamma \, \ln z 
\label{SHJ}
\end{equation}
with $t =
S/ (4 \Upsilon^2 \, \alpha)$. While ${\cal S}$ is not invariant under $\Gamma_0(2)$ transformations, 
\begin{equation}
{\cal S} (t, z) \rightarrow {\cal S} (t, z) - \frac{z^2}{2 \Delta}  \, \frac{\partial \Delta}{\partial t} \;,
\label{transfprinc}
\end{equation}
the combination ${\cal S} - {\cal S}_1$, with ${\cal S}_1 (t, z) =  \frac12 \,  \frac{z^2}{t + \bar t}$, is invariant.

\begin{proof}
Setting $t =
S/ (4 \Upsilon^2 \, \alpha)$ 
we write \eqref{calpot} as 
\begin{eqnarray}
 \frac{\partial ( \widehat \Omega - 2 \Upsilon \, \gamma \, \ln z)
}{\partial t} = - \frac12 \left[
  \left(  \frac{\partial ( \widehat \Omega - 2 \Upsilon \, \gamma \, \ln z)
}{\partial z} \right)^2 -  \frac{V(t,z)}{z^2}
\right] \;.
  \end{eqnarray}
Using \eqref{SHJ}, we obtain
\begin{equation}
- \frac{\partial {\cal S}}{\partial t} = {\cal H} (\frac{\partial {\cal S}}{\partial z}, z, t) 
\end{equation}
with
\begin{equation}
{\cal H} (p,z, t) = \frac12 \left(  p^2 - \frac{V(t,z)}{z^2}
  \right)\;.
\label{hamical}
\end{equation}
Next, using \eqref{omSGam02} and \eqref{Sig}, we infer the transformation behaviour \eqref{transfprinc}.
We can compensate for the term proportional to $\partial \Delta/\partial t$ on the 
right hand side of \eqref{transfprinc} by considering the $\Gamma_0(2)$ invariant
combination ${\cal S} -
{\cal S}_1$, where
\begin{equation}
{\cal S}_1 =  \frac12 \,  \frac{z^2}{t + \bar t} \;,
\end{equation}
which satisfies
\begin{equation}
- \frac{\partial {\cal S}_1}{\partial t} =  \frac12 \left( \frac{ \partial {\cal S}_1}{\partial z} \right)^2 \;.
\end{equation}
When $t$ and $z$ are real, 
${\cal S}_1$ describes Hamilton's principal function for a free particle.

\end{proof}

Using \eqref{Sig},  we note that the combination ${\cal S} - {\cal S}_1$, when expressed in terms of $S$ and $z$,
takes the form
\begin{equation}
{\cal S} (S,z) - {\cal S}_1 (S,z) = 
 \Upsilon \left[ \omega (S) - 2 \gamma \ln z  + \nu \, z^2 \, {\hat I}_1 (S, \bar S)
+ 
 \Xi (z, S, \Upsilon) \right] \;,
\end{equation}
with ${\hat I}_1$ given in \eqref{I1hat}.

When truncating $V(S,z)$ in \eqref{defpot}
 to the constant term, $V(S,z) = 4  \, (\gamma \,  \Upsilon)^2$, the Hamiltonian \eqref{hamical} 
 becomes ${\cal H} (p,z) = \tfrac12 (p^2 - g^2/z^2)$, with $g^2 = 4 (\Upsilon \, \gamma)^2 > 0$.
For real $z$, ${\cal H} (p,z)$  is the conformal
mechanics Hamiltonian of  \cite{deAlfaro:1976vlx} (with negative coupling constant, though), 
also related to the two-particle rational Calogero model
\cite{Avan:2012dd}. When $V(S,z) = 4  \, (\gamma \,  \Upsilon)^2$,
the PDE \eqref{diffhatOm22}, subject to $\Xi \vert_{z=0} = 0$ and $(\partial_{z^2} \Xi) \vert_{z=0} = 0$,
is solved by \eqref{xi1} in the approximation that we drop the term quadratic in $\partial \Xi/\partial z$.

On the other hand, the full $V(S,z)$ in \eqref{defpot} results in a time-dependent Hamiltonian 
 \eqref{hamical}, which can be viewed as a time-dependent deformation of the rational Calogero Hamiltonian 
 by an infinite set of terms involving powers of $I_n(S)$, starting with terms quadratic in $I_n(S)$.

\subsection{Charge bilinears in the STU model}

In the STU model, 
BPS black holes may carry electric/magnetic charges $(q_I, p^I)$, with $I=0,1,2,3$. Under $\Gamma_0(2)_S$ they transform as follows \cite{Cardoso:2008fr},
\begin{equation}
    \label{eq:S-charge-duality}
    \begin{array}{rcl}
      p^0 &\!\to\!& d \, p^0 + c\, p^1 \;,\\
      p^1 &\!\to\!& a \, p^1 + b \, p^0 \;,\\
      p^2 &\!\to\!& d\, p^2 - c \,q_3  \;,\\
      p^3 &\!\to\!& d\, p^3 - c \,q_2  \;,
    \end{array}
    \qquad
    \begin{array}{rcl}
      q_0 &\!\to\!& a\,  q_0 -b\,q_1 \;, \\
      q_1 &\!\to\!& d \,q_1 -c\, q_0 \;,\\
      q_2 &\!\to\!& a \, q_2 -b\, p^3 \;,\\
      q_3 &\!\to\!& a\, q_3 - b\, p^2  \;. 
    \end{array}
\end{equation}
These charges can be assembled into three charge bilinears,
\begin{eqnarray}
n &=& - q_0 p^1 + q_2 q_3 \;, \nonumber\\
m &=& p^0 q_1 + p^2 p^3 \;, \nonumber\\
l &=& q_0 p^0 - q_1 p^1 + q_2 p^2 + q_3 p^3 \;.
\label{bilin}
\end{eqnarray}
These bilinears transform as a triplet under $\Gamma_0(2)_S$ \cite{Cardoso:2008fr},
\begin{eqnarray}
\label{chargebiltrafo}
\begin{pmatrix}
n\\
m\\
l\\
\end{pmatrix} \rightarrow 
\begin{pmatrix}
a^2 & \; b^2 & \;  - ab\\
c^2 & \; d^2 & \;  - cd \\
- 2 ac &\;  - 2 bd &\;  ad + bc 
\end{pmatrix} \begin{pmatrix}
n\\
m\\
l\\
\end{pmatrix} \;,
\end{eqnarray}
and the $\Gamma_0(2)$ invariant norm of this vector is $ 4 nm - l^2$.

\subsection{Large single-center BPS black holes \label{sec:largeBH}}

Now we turn to the computation of the quantum entropy function \eqref{quantumentro} for large single-center BPS black holes in the STU model.  \\

\noindent
{\bf Definition:}
A large single-center BPS black hole is a dyonic spherically symmetric BPS black hole carrying electric/magnetic charges $(q_I, p^I)$ such that 
the charge bilinears $m, \, n$ and the charge 
combination $4 n m - l^2$ are positive, which ensures that the black hole
has a non-vanishing horizon area, proportional to $\sqrt{4 n m - l^2}$, at
the two-derivative level. \\

At the two-derivative level, the horizon area $A_{\rm BH}$ equals $A_{\rm BH}= 4 \pi (S + \bar S) \, m$
\cite{Behrndt:1996jn},
where $S$ denotes the value at the horizon.
Hence $S + {\bar S} >  0$ 
for a large BPS black hole, which implies that
$p^0$ and $p^1$ cannot be simultaneously zero.\footnote{Imposing the magnetic attractor equation
\eqref{Xphi} gives
${\rm Re} \, S = (p^1 \, \phi^0 - p^0 \, \phi^1) / |\phi^0 + i p^0|^2 $.}

The semi-classical macroscopic entropy 
${\cal S}_{\rm BH}$ of a BPS black hole equals ${\cal S}_{\rm BH} = \pi H(\phi_*, p, q)$,
with  $H(\phi_*, p, q)$ given by \eqref{Hsymp}. The quantum entropy function
computes corrections to the semi-classical entropy.

To compute the quantum entropy function, we will work
in a regime where $T$ and $U$ are large, so as to be able to use \eqref{Sig}.
We will expand
$H(\phi, p, q)$ given in \eqref{Hexp} around large values ${\rm Re} \, T_0, {\rm Re} \, U_0$ defined below in \eqref{reTU0}.
These values, which depend on $Y^0$ and on $S$, are invariant under $\Gamma_0(2)_S$
transformations.  When evaluated at the horizon of the BPS black hole, $T_0$ and $U_0$ become
entirely expressed in terms of the charges carried by the black hole, and this 
implies that we will have to choose the charges
of the black hole in such a way as to ensure that the horizon values of ${\rm Re} \, T_0$ and ${\rm Re} \, U_0$ are large.

We will now evaluate  $H(\phi, p, q)$  on the attractor values $\phi^2_*, \phi^3_*$ that satisfy
\begin{equation}
F_a - {\bar F}_{a} = i q_a \;\;\;,\;\;\; a=2, 3 \;.
\label{atrrphi23}
\end{equation}
We will denote the resulting expression by
$H(\tau_1, \tau_2, p, q)$, which we subsequently expand 
around large values ${\rm Re} \, T_0, {\rm Re} \, U_0$.
In doing so, we will keep all the charges, including $p^0$.

\subsubsection{$H(\tau_1, \tau_2, p, q)$ \label{sec:spH}}

We set $Y^2 = \tfrac12 (\phi^2 + i p^2)$ and $Y^3 = \tfrac12 (\phi^3 + i p^3)$, and we 
solve 
 the attractor equations \eqref{atrrphi23}.
We obtain
\begin{eqnarray}
\phi^2_* &=& \frac{2}{S + \bar S} \left( - q_3 - \frac{i}{2} (S - \bar S) \, p^2 - 2i \, \Delta_U
 \right) \;, 
\nonumber\\
\phi^3_* &=& \frac{2}{S + \bar S} \left( - q_2 - \frac{i}{2} (S - \bar S) \, p^3 - 2i \, \Delta_T
\right) \;,
\label{att23}
\end{eqnarray}
where
\begin{eqnarray}
\Delta_T &=& \frac{1}{Y^0} \, \frac{\partial \Omega}{\partial T} - {\rm c.c} \;, \nonumber\\
\Delta_U &=& \frac{1}{Y^0} \, \frac{\partial \Omega}{\partial U} - {\rm c.c} \;,
\end{eqnarray}
and we note that
$\Delta_T$ and $\Delta_U$ depend on $Y^0, S$ (and on their complex conjugates) as well as on 
$\phi^2_*, \phi^3_*$.
This yields 
\begin{equation}
T =  -i \frac{Y^2_*}{Y^0} =  -i \frac{( \phi^2_* + i p^2)}{2 \, Y^0}
= T_0 + t \;\;\;,\;\;\; U = -i \frac{Y^3_*}{Y^0} = -i \frac{( \phi^3_* + i p^3)}{2 \, Y^0}
=  U_0 + u \;,
\label{saddleTU}
\end{equation}
where
\begin{eqnarray}
T_0 &=& - \frac{i}{Y^0 (S + \bar S)} \left( - q_3 + i {\bar S} \, p^2 \right) \;, \nonumber\\
U_0 &=& - \frac{i}{Y^0 (S + \bar S)} \left( - q_2 + i {\bar S} \, p^3 \right) \;, \nonumber\\
t &=& - \frac{2}{Y^0 (S + \bar S)} \, \Delta_U \;, \nonumber\\
u &=&  - \frac{2}{Y^0 (S + \bar S)} \, \Delta_T \;.
\label{valuestu}
\end{eqnarray}
We infer
\begin{eqnarray}
T_0 + {\bar T}_0 &=& \frac{p^1 p^2 + p^0 q_3}{|Y^0|^2 (S + \bar S)} \;, \nonumber\\
U_0 + {\bar U}_0 &=& \frac{p^1 p^3 + p^0 q_2}{|Y^0|^2 (S + \bar S)} \;. 
\label{reTU0}
\end{eqnarray}
Note that $T_0$ and $U_0$ are inert under $\Gamma_0(2)_S$ transformations, as can be checked by using the
transformation rules \eqref{eq:S-charge-duality} and \eqref{SY0transf};
$t$ and $u$, on the other hand, are not inert under $\Gamma_0(2)_S$ transformations.

Next, we evaluate $H(\phi, p, q)$ given in \eqref{Hexp} at the values $T, U$ in \eqref{saddleTU}.
To this end, we use the parametrization \cite{LopesCardoso:2006ugz}
\begin{equation}
Y^0 = \frac{p^1 + i {\bar S} p^0}{S + \bar S} \;\;\;,\;\;\; Y^1 = \frac{i S p^1 - |S|^2 p^0}{S + \bar S} \;.
\label{Y0Y1S}
\end{equation}
We will also set 
\begin{equation}
\tau = i S = \tau_1 + i \tau_2 \;\;\;,\;\,\; {\bar \tau} = -i {\bar S} = \tau_1 - i \tau_2 
\label{tauS}
\end{equation}
in the following.
Then, $H(\phi, p, q)$ becomes
\begin{eqnarray}
 H(\tau_1, \tau_2, p, q) = \frac{n + l \tau_1 + m \tau_1^2 + m \tau_2^2}{\tau_2} +
 4 \left(\Omega + \bar \Omega \right) 
 + 4 \frac{\Delta_T \Delta_U}{\tau_2} \;,
 \label{valH23}
 \end{eqnarray}
where we made use of the charge bilinears \eqref{bilin}. This yields
\begin{eqnarray}
 H(\tau_1, \tau_2, p, q) &=& \frac{n + l \tau_1 + m \tau_1^2 + m \tau_2^2}{\tau_2} +
 4 \left(\Omega + \bar \Omega \right) \nonumber\\
 && + \frac{4}{\tau_2 \, (Y^0)^2 } 
 \frac{\partial \Omega}{\partial T} \,  \frac{\partial \Omega}{\partial U}
 + \frac{4}{\tau_2 \, ( {\bar Y}^0)^2 } 
 \frac{\partial {\bar \Omega}}{\partial {\bar T}} \,  \frac{\partial {\bar \Omega}}{\partial {\bar U}} \nonumber\\
 && - \frac{4}{\tau_2 \,  |Y^0|^2 } \left(
 \frac{\partial \Omega}{\partial  T} \,  \frac{\partial {\bar \Omega}}{\partial {\bar U}} + 
 \frac{\partial {\bar \Omega}}{\partial  {\bar T}} \,  \frac{\partial \Omega}{\partial  U} 
  \right) \;.
  \label{Hpqinter}
  \end{eqnarray}
  The terms on the right hand side are evaluated at $\phi^2_*, \phi^3_*$.
The first term in this expression is invariant under $\Gamma_0(2)_S$. When 
extremized with respect to $\tau_1$ and $\tau_2$ (c.f. \eqref{saddt1t2}), it
yields the entropy of  a large BPS black hole at the two-derivative level.
 The terms proportional to $1/|Y^0|^2$ are also invariant under $\Gamma_0(2)_S$, since both
 $\partial \Omega/ \partial T$ and  $\partial \Omega/ \partial U$ are invariant \cite{Cardoso:2008fr}.

Now we expand $H(\tau_1, \tau_2, p, q)$ around large values of ${\rm Re} \, T_0,  {\rm Re} \, U_0$. First, 
we expand 
\begin{eqnarray}
\omega (T) &=& \omega (T_0)  - \frac{2}{Y^0 (S + \bar S)} \, \Delta_U \, \frac{\partial \omega}{\partial T}\vert_{T_0}
+ \dots \;, \nonumber\\
\omega (U) &=& \omega (U_0)  - \frac{2}{Y^0 (S + \bar S)} \, \Delta_T \, \frac{\partial \omega}{\partial U}\vert_{U_0}
+ \dots \;, 
\label{approxomTU}
\end{eqnarray}
and drop 
terms
that involve higher order
derivatives of $\omega$ with respect to  $T$ and $U$ in view of \eqref{wTU}.
Inserting this in the combination
\begin{eqnarray}
\Omega +  \frac{2}{(S + \bar S) \, (Y^0)^2 } 
 \frac{\partial \Omega}{\partial T} \,  \frac{\partial \Omega}{\partial U} \;
\end{eqnarray}
that appears in \eqref{Hpqinter}, we obtain, using \eqref{Sig},
\begin{eqnarray}
\Omega +  \frac{2}{(S + \bar S) \, (Y^0)^2 } 
 \frac{\partial \Omega}{\partial T} \,  \frac{\partial \Omega}{\partial U} 
 &=& \Upsilon \left( \omega (S) + \omega (T_0) + \omega (U_0) \right) \\
&& + \frac{\Upsilon^2}{(Y^0)^2} \, \frac{\alpha}{\gamma} \,  \hat{I}_1 (S, \bar S)  +\Upsilon \,  \Xi(Y^0, S, \Upsilon)
\nonumber\\
  && + \frac{2 \Upsilon \bar \Upsilon}{(S + \bar S)\,  |Y^0|^2 } \left(
 \frac{\partial \omega}{\partial  T} \,  \frac{\partial {\bar \omega}}{\partial {\bar U}} + 
 \frac{\partial {\bar \omega}}{\partial  {\bar T}} \,  \frac{\partial \omega}{\partial  U} 
  \right)\vert_{T_0, U_0}
 \;,  \nonumber
 \end{eqnarray}
where 
$\hat{I}_1 (S, \bar S) $ is given in \eqref{I1hat}.

Inserting the above into \eqref{Hpqinter} gives
\begin{eqnarray}
 H(\tau_1, \tau_2, p, q) &=& \frac{n + l \tau_1 + m \tau_1^2 + m \tau_2^2}{\tau_2} +
 4 \Big(\Upsilon \left(  \omega (T_0) + \omega (U_0) \right) + {\rm c.c.} \Big) \nonumber\\
 && + 4  \Big( \Upsilon \,  \omega (\tau) + {\rm  c.c.} \Big) \nonumber\\
  && + 4 \left( \frac{ \Upsilon^2}{(Y^0)^2} \, \frac{\alpha}{\gamma} \, \hat{I}_1 (\tau, \bar \tau) + 
  \Upsilon \,  \Xi(Y^0, \tau, \Upsilon) + {\rm c.c.} 
 \right) 
 \nonumber\\ 
  && + \frac{8 \, \alpha \,  \Upsilon \bar \Upsilon}{\tau_2 \,  |Y^0|^2 } 
 \;,
  \label{saddle23H}
  \end{eqnarray}
with $Y^0$ expressed in terms of $\tau = iS$ and charges $(p^0,p^1)$ as in \eqref{Y0Y1S}, and where in the last line we 
 replaced 
$ \left(  \frac{\partial \omega}{\partial  T} \, \frac{\partial {\bar \omega}}{\partial {\bar U}} + 
 \frac{\partial {\bar \omega}}{\partial  {\bar T}} \,  \frac{\partial \omega}{\partial  U} 
  \right)\vert_{T_0, U_0}  $ by $2 \alpha$ in view of \eqref{alpha}.

Summarizing, \eqref{saddle23H} gives the value 
of $H(\phi, p, q)$ evaluated at $\phi^2_*$ and $\phi^3_*$, in the approximation
where the real part of $T_0, U_0$  is taken to be large, so that terms involving higher derivatives of $\omega$ with
respect to $T$ and $U$ can be dropped.

In this approximation, all the terms in \eqref{saddle23H}
are invariant under $\Gamma_0(2)_S$ transformations, except for the term in the second line, whose
duality invariance can be repaired by adding a term proportional to $\ln Y^0$ and its
complex conjugate to $H(\tau_1, \tau_2, p, q)$,
\begin{eqnarray}
 \label{HlnY0}
 H(\tau_1, \tau_2, p, q) + 8 \gamma \Big( \Upsilon \,  \ln Y^0 + {\rm c.c.} \Big)
 &=& \frac{n + l \tau_1 + m \tau_1^2 + m \tau_2^2}{\tau_2} 
 \\
 && +
 4 \Big(\Upsilon \left(  \omega (T_0) + \omega (U_0) \right) + {\rm c.c.} \Big) \nonumber\\
 && + 4  \Big( \Upsilon \, ( \omega (\tau) + 2 \gamma \, \ln Y^0 )
   + {\rm  c.c.} \Big) \nonumber\\
  && + 4 \left( \frac{ \Upsilon^2}{(Y^0)^2} \, \frac{\alpha}{\gamma}\, \hat{I}_1 (\tau,  \bar \tau) + 
  \Upsilon \,  \Xi(Y^0, \tau, \Upsilon) + {\rm c.c.} 
\right) 
 \nonumber\\ 
  && + \frac{8 \, \alpha\, \Upsilon \bar \Upsilon}{\tau_2 \,  |Y^0|^2 } 
 \;. \nonumber
   \end{eqnarray}
 
We recall that for BPS black holes the horizon value of $\Upsilon$ is real and given by
$\Upsilon = -64$, and that 
\begin{equation}
4 \pi \, \Upsilon \, \gamma =1 \;.
\label{upsgam1}
\end{equation}
The combination on the left hand side of 
\eqref{HlnY0} is thus $H(\tau_1, \tau_2, p, q) + (4/\pi)  \,  \ln |Y^0| $. 
This combination will play a role in the quantum entropy function below, c.f. \eqref{psiapprox}.

\subsubsection{Evaluating $W(q,p)$ 
beyond saddle point approximation 
 \label{sec:beyspa}}

In \eqref{Wapproxsymp} we displayed the value of the quantum entropy function $ W(q,p)$ in a saddle
point approximation. Now, 
we proceed with the evaluation of 
 $W(q,p)$ beyond the saddle point
approximation. In doing so, we will impose approximations 
that we will clearly delineate in what follows.

We decompose
\begin{equation}
Y^0 = \ft12 \left( \phi^0 + i p^0 \right) \;\;\;,\;\;\; Y^1 = \ft12 \left( \phi^1 + i p^1 \right) \;.
\label{Y0p}
\end{equation}
Using  \eqref{Y0Y1S} and \eqref{tauS}
we infer
\begin{equation}
\phi^0 = \frac{p^1 - \tau_1 \, p^0}{\tau_2} \;\;\;,\;\;\; 
\phi^1 = - p^0 \, \tau_2 + \frac{\tau_1}{\tau_2} \left(p^1 - \tau_1 \, p^0 \right) \;,
\label{p0p}
\end{equation}
to obtain
\begin{equation}
d\phi^0 \wedge d \phi^1 = \frac{\left( (\phi^0)^2 + (p^0)^2 \right) }{\tau_2} \, d \tau_1 \wedge d \tau_2
= \frac{4 |Y^0|^2}{\tau_2}  \, d \tau_1 \wedge d \tau_2 \;.
\end{equation}
Then, the quantum entropy function reads (using the approximate measure factor \eqref{approxmeas} with $\chi =0$ ) 
\begin{equation}
{W}(q,p) = 4 \int \frac{d \tau_1 \, d \tau_2}{\tau_2} \, d \phi^a \, e^{\pi [ 4 \, {\rm Im} F(\phi + i p) -  q \cdot \phi] } \, |Y^0|^{4} \, 
e^{{- \cal K}^{(0)}}
\;\;\;,\;\;\; a = 2, 3 \;.
\label{quantumtau}
\end{equation}
Next, we integrate out  $\phi^2$ and $\phi^3$,  by expanding 
the exponent 
$4 \, {\rm Im} F(\phi + i p) -  q \cdot \phi $
around
the attractor values $\phi^2_*$ and $\phi^3_*$ computed in \eqref{att23},
and retaining only quadratic fluctuations in $\phi^2$ and $\phi^3$. 
The associated quadratic form takes the form given in 
\eqref{Hexpsadd}, with the indices $I, J$ restricted to $I, J =2,3$.
We approximate the resulting fluctuation determinant by replacing $F$ by $F^{(0)}$,
in which case it takes the value $\tau_2^2$. Thus, in this approximation,
the Gaussian integration\footnote{In doing so, we view 
$(\phi^2, \phi^3)$ as local coordinates on $\mathbb{C}^2$, and we choose
an appropriate path of integration
 to obtain a well defined Gaussian integral.}
 over fluctuations in $\phi^2, \phi^3$
 yields a factor
\begin{equation}
1/\tau_2 \;,
\end{equation}
and we obtain the following approximate expression for the
quantum entropy function, 
\begin{equation}
{ W}(q,p) = 4 \int \frac{d \tau_1 \, d \tau_2 }{\tau_2^2} \, 
\, e^{\pi \, {H}(\tau_1, \tau_2, p, q)  } \,  |Y^0|^{4} \,  e^{{- \cal K}^{(0)}}
\;,
\label{quantum-approx-STU}
\end{equation}
with ${H}(\tau_1, \tau_2, p, q)  $ given by 
\eqref{Hpqinter}. Here, $e^{{- \cal K}^{(0)}}$ is evaluated at the
values $T, U$ given in \eqref{saddleTU}.

To proceed, we perform  three more approximations. First, we approximate
$e^{{- \cal K}^{(0)}}$ in \eqref{quantum-approx-STU} by replacing the values $T, U$
by the values
$T_0, U_0$ given in \eqref{reTU0},
\begin{equation}
 e^{- {\cal K}^{(0)} 
(\tau_1, \tau_2, T_0, U_0)} = \frac{(p^1 p^2 + p^0 q_3)(p^1 p^3 + p^0 q_2)}{2  |Y^0|^2 \tau_2}\;.
\end{equation} 
Using
\begin{equation}
(p^1 p^2 + p^0 q_3)(p^1 p^3 + p^0 q_2) = (p^1)^2 \, m + p^0 p^1 \, l + (p^0)^2 \, n \;,
\end{equation}
where $m,l,n$ denote the charge bilinears introduced in \eqref{bilin},
we obtain
\begin{equation}
 |Y^0|^4 \, e^{- {\cal K}^{(0)} 
(\tau_1, \tau_2, T_0, U_0)} = |p^1 - \tau p^0|^2 \, \frac{
[(p^1)^2 \, m + p^0 p^1 \, l + (p^0)^2 \, n] }{8 \tau_2^3} \;,
\end{equation} 
and hence,
\begin{equation}
{W}(q,p) = \ft12 \int \frac{d \tau_1 \, d \tau_2 }{\tau_2^5} \, 
\, e^{\pi \, {H}(\tau_1, \tau_2, p, q)  } \,
 |p^1 - \tau  p^0|^2 \, \bigg( (p^1)^2 \, m + p^0 p^1 \, l + (p^0)^2 \, n \bigg)
\;.
\label{quantum-approx-STU2}
\end{equation}
Using 
\begin{equation}
 (p^1)^2 \, m =  |p^1 - \tau p^0|^2 \, m + 2 \tau_1 \, p^0 p^1 \, m - (p^0)^2 (\tau_1^2 + \tau_2^2)
 \, m \;,
 \end{equation}
we obtain
\begin{equation}
(p^1)^2 \, m + p^0 p^1 \, l + (p^0)^2 \, n = |p^1 -\tau p^0|^2 \, m +  p^0 p^1 
\left( l + 2 \tau_1\, m \right) + (p^0)^2 \left(n - m (\tau_1^2 + \tau_2^2) \right) \;.
\end{equation}
Hence we write \eqref{quantum-approx-STU2} as
\begin{eqnarray}
{W}(q,p) &=& \ft12 \int \frac{d \tau_1  \, d \tau_2 }{\tau_2^5} \, 
\, e^{\pi \, {H}(\tau_1, \tau_2, p, q)  } \,
 |p^1 - \tau p^0|^2 \nonumber\\
&& 
  \Big[
  |p^1 -\tau p^0|^2 \, m
 +  
  p^0 p^1 
\left( l + 2 \tau_1\, m \right) + (p^0)^2 \left(n - m (\tau_1^2 + \tau_2^2) \right) 
  \Big]
\;.
\label{quantum-approx-STU22}
\end{eqnarray}
Now we note 
that the two combinations
\begin{equation}
 l + 2 \tau_1\, m   \;\;\;,\;\;\; n - m (\tau_1^2 + \tau_2^2) 
 \label{attrtau2d}
 \end{equation}
 vanish at the extremum of 
 the combination $ (n + l \tau_1 + m \tau_1^2 + m \tau_2^2)/\tau_2 $
 that appears in the exponential of  ${H}(\tau_1, \tau_2, p, q)  $ in
 \eqref{quantum-approx-STU22}. This yields 
the
 attractor value for $\tau$ at the two-derivative level,
 \begin{equation}
 \tau_1^* = - \frac{l}{2m} \;\;\;,\;\;\;  \tau_2^* = \sqrt{\frac{4 n m - l^2}{4 m^2}} > 0 \;.
 \label{saddt1t2}
 \end{equation}
 Therefore, imposing the vanishing of \eqref{attrtau2d}
 in  \eqref{quantum-approx-STU22} leads to the approximate result
\begin{eqnarray}
{W}(q,p) &=& \ft12 \int \frac{d \tau_1  \, d \tau_2 }{\tau_2^5} \, 
\, e^{\pi \, {H}(\tau_1, \tau_2, p, q)  } 
 |p^1 - \tau p^0|^4 \, m
 \;.
\label{quantum-approx-STU2222}
\end{eqnarray}
The third approximation consists in replacing ${H}(\tau_1, \tau_2, p, q) $ in
\eqref{Hpqinter} by \eqref{saddle23H}. This is achieved by
taking ${\rm Re} \, T_0, {\rm Re} \, U_0$ to be large. Since $T_0, U_0$ depend on $\tau$, c.f.
\eqref{valuestu}, we replace $\tau$ by the saddle point values \eqref{saddt1t2}
in the expression for $T_0, U_0$. Thus, from now on, $T_0, U_0$ will
refer to the attractor values of $T, U$ at the two-derivative level, which are
entirely expressed in terms of the charges of the BPS black hole.
Then, requiring the real part of $T_0, U_0$ to be large, translates into
a condition on the values of the charges.

Summarising, under these approximations, \eqref{quantum-approx-STU2222}
becomes 
\begin{eqnarray}
\label{quantum-approx-STU222}
{W}(q,p) &=& \ft12 
\,
e^{ 
4 \pi  \left[
\Upsilon \left(  \omega (T_0) + \omega (U_0)  \right)+ c.c. \right] } \,
\int \frac{d \tau_1 \, d \tau_2 }{\tau_2^5} \, 
\, e^{\frac{\pi}{\tau_2}  (n + l \tau_1 + m \tau_1^2 + m \tau_2^2)  + \pi \Lambda }  \,
 |p^1 - \tau p^0|^4 \, m
 \;, \nonumber\\
\end{eqnarray}
where we introduced 
\begin{eqnarray}
\label{lambdafull}
\Lambda =
     4  \left(  \Upsilon \,  \omega (\tau) + 
     \frac{ \Upsilon^2}{(Y^0)^2} \,  \frac{\alpha}{\gamma} \, \hat{I}_1 ( \tau,   \bar \tau) 
   +
   \Upsilon \,  \Xi(Y^0, \tau , \Upsilon) + {\rm c.c.} 
  \right) 
    + \frac{8 \alpha \, \Upsilon \bar \Upsilon}{\tau_2 \,  |Y^0|^2 } 
   \:,
   \end{eqnarray}
and where we recall that $Y^0$ is expressed in terms of $p^0, p^1$ and $\tau$ through
\eqref{Y0Y1S}.

In what follows, we will use the approximate result 
\eqref{quantum-approx-STU222} as a starting point for various
considerations. One should keep in mind that there are subleading corrections to 
\eqref{quantum-approx-STU222} that will not be considered in this paper.

The integrand in \eqref{quantum-approx-STU222} is not invariant under
$\Gamma_0(2)$-transformations of $(\tau_1, \tau_2)$. To obtain an integrand
that is invariant under $\Gamma_0(2)$-transformations, we perform a rewriting
 of \eqref{quantum-approx-STU222}
following an approach given in \cite{Gomes:2015xcf,Murthy:2015zzy}. 
This requires a certain assumption, as follows. 
We recall that the measure \eqref{approxmeas} that we used in \eqref{quantumtau} is an approximate measure. Following \cite{Gomes:2015xcf,Murthy:2015zzy}, let us
assume that there are subleading corrections to the measure such that $m$ in \eqref{quantum-approx-STU222} gets replaced by the combination
$\left( m + \tfrac12 \frac{d \Lambda}{d \tau_2} + \tfrac{1}{2 \pi} \frac{d \ln |p^1 - \tau p^0|^4}{d \tau_2}
\right)   $, in which case \eqref{quantum-approx-STU222} becomes
\begin{eqnarray}
{W}(q,p) &=& \ft12
\,
e^{ 
4 \pi  \left[
\Upsilon \left(  \omega (T_0) + \omega (U_0)  \right)+ c.c. \right] } \,
 \int \frac{d \tau_1 \,  d \tau_2 }{\tau_2^5} \, 
\, e^{\frac{\pi}{\tau_2}  (n + l \tau_1 + m \tau_1^2 + m \tau_2^2)  + \pi \Lambda }  \,
 |p^1 - \tau p^0|^4 \nonumber\\
 && \qquad \qquad  \qquad \qquad \qquad \qquad 
  \left( m +\tfrac12 \frac{d \Lambda}{d \tau_2} + \frac{1}{2 \pi} \frac{d \ln |p^1 - \tau p^0|^4}{d \tau_2}
 \right)  \;. 
\label{quantum-approx-STULam}
\end{eqnarray}
Now, consider the integrand of \eqref{quantum-approx-STULam} and compute
\begin{eqnarray}
\label{TDmass}
&& \frac{d}{d \tau_2} \left( \frac{1}{\tau_2^5} \, 
\, e^{\frac{\pi}{\tau_2}  (n + l \tau_1 + m \tau_1^2 + m \tau_2^2)  +  \pi \Lambda } \; |p^1 - \tau p^0|^4
\right) \\
&=& \frac{\pi}{\tau_2^5}
\left( 2 m + \frac{d \Lambda}{d \tau_2}
+ \frac{1}{\pi} \frac{d \ln |p^1 - \tau p^0|^4}{d \tau_2}
 \right) e^{\frac{\pi}{\tau_2}  (n + l \tau_1 + m \tau_1^2 + m \tau_2^2)  + 
\pi \Lambda } \, |p^1 - \tau p^0|^4  \nonumber\\
&&- \frac{1}{\tau_2^6} \left(5 + \frac{\pi}{\tau_2}  (n + l \tau_1 + m \tau_1^2 + m \tau_2^2) 
\right) e^{\frac{\pi}{\tau_2}  (n + l \tau_1 + m \tau_1^2 + m \tau_2^2)  + \pi \Lambda  } \, 
 |p^1 - \tau p^0|^4  \;.\nonumber
\end{eqnarray}
Then, dropping an overall constant,
we obtain for \eqref{quantum-approx-STULam},
\begin{eqnarray}
\label{quantum-approx-STU3}
{W}(q,p) &=& 
\,
e^{ 
4 \pi  \left[
\Upsilon \left(  \omega (T_0) + \omega (U_0)  \right)+ c.c. \right] } \, \\
&& \int \frac{d \tau_1 \,  d \tau_2 }{\tau_2^6} \, 
\left(5 + \frac{\pi}{\tau_2}  (n + l \tau_1 + m \tau_1^2 + m \tau_2^2) 
\right)
\, e^{\frac{\pi}{\tau_2}  (n + l \tau_1 + m \tau_1^2 + m \tau_2^2)  + \pi \Lambda  }  
|p^1 - \tau  p^0|^4 \;,  \nonumber
\end{eqnarray}
where we discarded the total derivative term in \eqref{TDmass}.
We write \eqref{quantum-approx-STU3} as
\begin{equation}
{W}(q,p) =   \int \frac{d \tau_1 \,  d \tau_2 }{\tau_2^2} \,
e^{\Psi (\tau_1, \tau_2, p, q)} \;,
\label{micro-approx}
\end{equation}
where
\begin{equation}
e^{\Psi (\tau_1, \tau_2, p, q)} = \left(5 + \frac{\pi}{\tau_2}  (n + l \tau_1 + m \tau_1^2 + m \tau_2^2) \right)
\, e^{ \pi H (\tau_1, \tau_2, p, q) + 4 \ln |Y^0 (\tau_1, \tau_2, p^0, p^1)|}
\;,
\label{psiapprox}
\end{equation}
with the combination $\pi H (\tau_1, \tau_2, p, q) + 4 \ln |Y^0(\tau_1, \tau_2, p^0, p^1)|$ given in \eqref{HlnY0}
(recall that $4 \pi \Upsilon \gamma = 1$).
Note that $\Psi$ is invariant under $\Gamma_0(2)_S$ transformations.

The purpose of the various manipulations described above was to 
bring $W(q,p)$ into the form \eqref{micro-approx}, which is invariant under $\Gamma_0(2)_S$ transformations.
This was achieved by subtracting a total derivative term from \eqref{quantum-approx-STULam}.
The form of $W(q,p)$ given in \eqref{micro-approx}
will provide the link
with the microstate counting proposal that we will discuss in the next section.
Thenceforth, we will refer to \eqref{micro-approx} as the quantum entropy function.

To fully specify $W(q,p)$, we also need to choose an integration contour $\cal C$
that passes through $(\tau_1^*, \tau_2^*)$ given in
\eqref{saddt1t2}. 
This requires extending $(\tau_1, \tau_2) \in \mathbb{R}^2$ to 
$(\tau_1, \tau_2) \in \mathbb{C}^2$. The contour $\cal C$ will be discussed in 
 subsection \ref{sec:contour}, using a construction given in \cite{Sen:2007qy}.
We thus have the following proposition.\\

\noindent
{\bf Proposition:}
For large black holes, and for large attractor values $T_0, U_0$, the quantum entropy function ${W}(q,p)$
of the STU model is approximately given by
\begin{equation}
{W}(q,p) = \int_{{\cal C}} 
\frac{d \tau_1 d \tau_2}{\tau_2^2} \, e^{\Psi (\tau_1, \tau_2, p, q)} \;.
\label{quantumint}
\end{equation}
\\

Finally, we note the following behaviour of the quantum entropy function \eqref{quantumint}:\\

\noindent
{\bf Proposition:} The quantum entropy function \eqref{quantumint} 
is encoded in the function
\begin{equation}
\hat{F} (Y, \Upsilon) = F( Y, \Upsilon) + 4i \ \Upsilon \gamma \ln Y^0 \;,
\end{equation}
where  $F( Y, \Upsilon)$ denotes the Wilsonian function 
\begin{eqnarray}
\label{omegaIn}
F(Y,  \Upsilon) &=& F^{(0)} (Y) + 2 i \Omega (Y, \Upsilon) \\
&=&   - \frac{Y^1 Y^2 Y^3}{Y^0} + 
2 i \Upsilon \Big[ \omega(S) + \omega(T) + \omega (U) 
+ \frac{\Upsilon}{(Y^0)^2} \, 
\frac{\alpha}{\gamma} \, I_1(S)  + 
\Xi (Y^0, S, \Upsilon) 
\Big] \;. \nonumber
\end{eqnarray}
The function $\hat{F} $ is the one that was obtained recently in  \cite{CdWM}
using the duality symmetries of the model, in the limit of large $T, U$, see above
\eqref{om1}.

\begin{proof}
It is straightforward to verify that for large values of $T_0, U_0$, the 
quantum entropy function \eqref{quantumint} takes the form 
\begin{eqnarray}
{W} (q,p) = \int_{\cal C} \frac{d\tau_1 \, d\tau_2}{\tau_2^2}
\left(5 + \frac{\pi}{\tau_2}  (n + l \tau_1 + m \tau_1^2 + m \tau_2^2) \right) \,
 e^{\pi \left[ 4 {\rm Im} \hat{F}(\phi + i p) - q \cdot \phi \right]} \vert_{\phi^2_*, \phi^3_*}
  \;,
  \nonumber\\
\end{eqnarray}
where $\phi^2_*, \phi^3_*$, given in \eqref{att23}, are evaluated
at large attractor values of $T_0, U_0$ (c.f. the discussion below \eqref{quantum-approx-STU2222}),
and where $Y^0$ is given by \eqref{Y0p} and
\eqref{p0p}.

\end{proof}

\subsection{Choice of contour \label{sec:contour} }

The quantum entropy function \eqref{quantumint} has a form that is suggestive 
of a Siegel modular form. Namely,
following \cite{Banerjee:2008ky}, we perform 
the change of variables 
\begin{equation}
\label{rst12}
\rho \equiv \tau_1+i\tau_2 \;,
\quad
\sigma \equiv -\tau_1+i\tau_2 \;,
\end{equation}
in which case
\begin{equation}
\label{tt12rs}
\tau_1 = \frac12 (\rho - \sigma) \;\;\;,\;\;\; \tau_2 = -  \frac{i}{2} (\rho + \sigma) \;,
\end{equation}
and 
\begin{equation}
\frac{\pi}{\tau_2}  \left(n + l \tau_1 + m \, \tau_1^2 + m \, \tau_2^2 \right) = -  \frac{2 \pi i}{\rho + \sigma}
\left( -n - \tfrac12 l (\rho - \sigma) + m \,  \rho \sigma \right) \;.
\end{equation}
Next, introducing $k=2$, and
 using  $\Upsilon = - 64$ as well as $\Upsilon \alpha/\gamma = \pi/4$,
we write ${W} (q,p)$ in \eqref{quantumint} as 
\begin{eqnarray}
\label{hatWmicroprep}
&& W(q,p) \,
e^{   \ln |\vartheta_2 (T_0)|^8  \ln |\vartheta_2 (U_0)|^8 } =
 \int_{\cal C}
\frac{d\rho\, d\sigma}{(\rho+\sigma)^2}
\left( k + 3   -  2 \pi i
\left[\frac{\rho\sigma}{\rho+\sigma} \, m-\frac{1}{\rho+\sigma} \, n- \tfrac12 \frac{\rho-\sigma}{\rho+\sigma} \,  l \right] 
 \right) \;\nonumber\\
&& \exp{\left\lbrace- 2\pi i\left[\frac{\rho\sigma}{\rho+\sigma} \, m-\frac{1}{\rho+\sigma} \, n- \tfrac12 \frac{\rho-\sigma}{\rho+\sigma} \,  l \right] -(k+2)\ln(\rho+\sigma) + \pi \, {\tilde \Lambda} 
   \right\rbrace} \;, \nonumber\\
   \end{eqnarray}
up to an overall constant, where 
\begin{eqnarray}
\label{tillamD}
\pi  \, {\tilde \Lambda} &=& -  \tfrac12 \ln \vartheta_2^8 (\rho) - \tfrac12 \ln \vartheta_2^8 (\sigma) 
 + 2 \ln (p^1 - \rho \, p^0) + 2 \ln (p^1 + \sigma \, p^0) + D(\rho, \sigma) \;, \nonumber\\
  D(\rho, \sigma)  &=&   - 256 \pi  \left( \frac{ \pi}{4 \, (Y^0)^2} \,  
   \hat{I}_1 ( \rho,  \sigma) 
+ \frac{ \pi}{4 \, ({\tilde Y}^0)^2} \,  
 \hat{I}_1 ( \sigma,  \rho) 
   +  \Xi(Y^0, \rho)
   +   \Xi({\tilde Y}^0, \sigma)
 \right) \nonumber\\
&& + 
\frac{  i  \pi  }{(\rho + \sigma)\,  Y^0( \rho, \sigma) \, {\tilde Y}^0( \rho, \sigma)} 
  \;. 
  \end{eqnarray}
 Here, $\Xi(Y^0,  \rho)$ is given by \eqref{Inexp} and \eqref{coefficientsalphanmn}, and
 \begin{eqnarray}
Y^0( \rho, \sigma) &=& i  \, \frac{p^1 + \sigma \, p^0}{\rho + \sigma} \;, \nonumber\\
{\tilde Y}^0(\rho, \sigma) &=& i \,  \frac{p^1 - \rho \,  p^0}{\rho + \sigma} = Y^0( \rho, \sigma) - i p^0  \;, \nonumber\\
\hat{I}_1 (\rho, \sigma) &=& i \left( \frac{\partial \omega}{\partial \rho} - \frac{2 \gamma}{\rho + \sigma} \right)
\;.
\label{Y^0tildeY0I1}
\end{eqnarray}
The integrand in \eqref{hatWmicroprep} exhibits a dependence on the parameter $k=2$ (both in the exponent and in the measure)
that is characteristic of
a microstate counting formula based on a Siegel modular form $\Phi_2$ of weight $k=2$ \cite{David:2006yn},
see \eqref{Phi2int} and \eqref{sieg-asym} below. 
Thus, we expect the microstate counting formula for large BPS black holes to involve $\Phi_2$.
Therefore, we choose a contour ${\cal C}$ in \eqref{quantumint} that
captures the information about the locus of the second-order
zero of the Siegel modular form which yields the leading microstate degeneracy.
This contour was constructed in  \cite{Sen:2007qy}. It is given by 
the image of a surface ${\cal B}'$ in the complex $(\tau_1, \tau_2)$-plane,
as follows.

Consider Siegel's upper half plane ${\cal H}_2$ with local coordinates
$\tilde \rho = \tilde \rho_1 + i \tilde  \rho_2, \; \tilde  \sigma = \tilde \sigma_1 + i \tilde \sigma_2, \; 
\tilde v= \tilde v_1 + i \tilde v_2$, 
where
$\tilde \rho_2 > 0, \; \tilde \sigma_2 > 0, \;
 \tilde \rho_2 \, \tilde \sigma_2 - {\tilde v}_2^2 > 0$, c.f. 
\eqref{Omrvs}.
The surface  ${\cal B}'  \subset{\cal H}_2$ is obtained by intersecting 
$\tilde \rho_2 = \eta_1 (\lambda), \,  \tilde \sigma_2 = \eta_2 (\lambda), \,  \tilde v_2 = \eta_3 (\lambda), $ with 
${\tilde \rho} {\tilde \sigma} - {\tilde v}^2 + {\tilde v} =0$.  The latter describes the locus of the second-order
zero of the Siegel modular form that is relevant for the microstate proposal. The intersection yields a family of curves 
  given by \cite{Sen:2007qy}, 
\begin{eqnarray}
\tilde \rho_2 &=& \eta_1 (\lambda) \;\;\;,\;\;  \tilde \sigma_2 = \eta_2 (\lambda) \;\;\;,\;\;\;
\tilde v_2 = \eta_3 (\lambda) \;,
\nonumber\\
\tilde \rho_1 &=& - \left( \frac{\eta_1(\lambda)  \, \tilde \sigma_1 -  \eta_3(\lambda) \, ( 2 \tilde v_1 -1) }{\eta_2(\lambda)} \right) \;,
\\
&& \frac{\eta_1(\lambda)}{\eta_2(\lambda)} \, (\tilde \sigma_1)^2 + \left( \tilde v_1 - \frac12 \right)^2 - 2 \, \frac{\eta_3(\lambda) }{\eta_2(\lambda)} \, {\tilde \sigma}_1
\left( \tilde v_1 - \frac12 \right) = \frac14 - \left( \eta_1(\lambda) \eta_2 (\lambda) - \eta_3^2 (\lambda) \right) \;. \nonumber
\end{eqnarray}
The last equation describes an ellipse in the $(\tilde \sigma_1, \tilde v_1)$-plane so long as $ 0 \leq \eta_1(\lambda) \eta_2 (\lambda) - \eta_3^2 (\lambda) \leq 1/4$,
\begin{eqnarray}
\label{ellipseeta}
\frac{\left( \eta_1(\lambda) \eta_2 (\lambda) - \eta_3^2 (\lambda) \right)}{\eta_2^2(\lambda)} 
  \, (\tilde \sigma_1)^2 + \left( \tilde v_1 - \frac12 - \, \frac{\eta_3(\lambda) }{\eta_2(\lambda)} \, {\tilde \sigma}_1 \right)^2
= \frac14 - \left( \eta_1(\lambda) \eta_2 (\lambda) - \eta_3^2 (\lambda) \right) \;. \nonumber\\
\end{eqnarray}
 The surface ${\cal B}' $ is obtained by picking
\begin{eqnarray}
\eta_1(\lambda) = \lambda \, \frac{n}{\sqrt{4 m n -l^2}} \;\;\;,\;\;\; \eta_2(\lambda) = \lambda \, \frac{m}{\sqrt{4 m n -l^2}} \;\;\;,\;\;\;
 \eta_3(\lambda) =  \lambda \, \frac{l}{2 \, \sqrt{4 m n -l^2}} \;,
 \end{eqnarray}
 and restricting $\lambda$ to lie in the range $0 < \epsilon \leq \lambda \leq 1$, so that 
 $0 <  \eta_1(\lambda) \eta_2 (\lambda) - \eta_3^2 (\lambda) = \lambda^2/4 \leq 1/4$.
 The requirement $\epsilon >0$ ensures the Siegel upper half plane conditions $\tilde \rho_2 > 0, \; \tilde \sigma_2 > 0, \;
 \tilde \rho_2 \, \tilde \sigma_2 - {\tilde v}_2^2 > 0$.
 At $\lambda = 1$, the surface ${\cal B}' $ shrinks to a point, 
 \begin{eqnarray}
 \tilde \sigma_1 &=& 0 \;\;\;,\;\;\; \tilde v_1 = \frac12 \;\;\;,\;\;\; \tilde \rho_1 = 0 \;, \nonumber\\
  \tilde \sigma_2 &=& \frac{m}{\sqrt{4 m n -l^2}}  \;\;\;,\;\;\; \tilde v_2 = \frac{l}{2 \, \sqrt{4 m n -l^2}} \;\;\;\;,\;\;\; \tilde \rho_2 =  \frac{n}{\sqrt{4 m n -l^2}}  \;,
 \label{l1pt}
   \end{eqnarray}
  while at
 $\lambda = \epsilon >0$ the ellipse has maximal radii.
 
 Next, introducing variables $(\rho, \sigma, v)$ as \cite{Jatkar:2005bh}
 \begin{eqnarray}
 \rho &=& \frac{\tilde \rho \, \tilde \sigma - {\tilde v}^2}{\tilde \sigma} \;\;\;,\;\;\; 
 \sigma = \frac{\tilde \rho \, \tilde \sigma - ({\tilde v} -1)^2}{\tilde \sigma} \;\;\;,\;\;\; v = \frac{{\tilde \rho} {\tilde \sigma} - {\tilde v}^2 + {\tilde v}}{\tilde \sigma} \;,
 \end{eqnarray}
 it follows that on the locus  ${\tilde \rho} {\tilde \sigma} - {\tilde v}^2 + {\tilde v} =0$,
 \begin{equation}
\rho = 
- \frac{\tilde v}{\tilde \sigma}
 \;\;\;,\;\;\; 
 \sigma = \frac{\tilde v -1}{\tilde \sigma}
\;\;\;,\;\;\; v =  0 \;.
\label{rsv0}
\end{equation}
 Then, using \eqref{tt12rs}, we express
 $(\tau_1, \tau_2)$ in terms of $(\tilde \sigma, \tilde v)$ on the locus $v=0$,
 \begin{equation}
\tau_1 = \frac12 (\rho - \sigma) = - \frac{(\tilde v -\tfrac12)}{\tilde \sigma} \;\;\;,\;\;\; \tau_2 = -  \frac{i}{2} (\rho + \sigma) = \frac{i}{2 \, \tilde \sigma} \;.
\label{loctt12}
\end{equation}
We parametrize the ellipse \eqref{ellipseeta} in terms of an angle $0 \leq \theta < 2 \pi$,
 \begin{eqnarray}
 {\tilde \sigma}_1 = \frac{\sqrt{1 - \lambda^2} \, \cos \theta}{2 \tau_2^*} \;\;\;,\;\;\; {\tilde v}_1 - \frac12 - \frac{l}{2m} \, {\tilde \sigma}_1 = 
 \frac{\sqrt{1 - \lambda^2} \, \sin \theta}{2 } \;,
 \end{eqnarray}
 where $\tau_2^*$ denotes the attractor value given in \eqref{saddt1t2}.
Then, using \eqref{loctt12}, we obtain
 \begin{eqnarray}
 \label{arcpar}
 \tau_1 &=& \frac{1}{\lambda^2 + (1 - \lambda^2) \cos^2 \theta} \left[ \tau_1^* \, \lambda^2 +   \tau_1^* \, (1 - \lambda^2) \cos^2 \theta 
- \tau_2^* (1-\lambda^2) \sin \theta \cos \theta  \right] \nonumber\\
&& + i \,  \frac{ \lambda \, \sqrt{1 - \lambda^2}}{\lambda^2 + (1 - \lambda^2) \cos^2 \theta} \,  \tau_2^* \sin \theta  \;,
 \nonumber\\
 \tau_2 &=& \frac{\tau_2^*}{\lambda^2 + (1 - \lambda^2) \cos^2 \theta} \left[ \lambda + i \, \sqrt{1 - \lambda^2} \, \cos \theta \right] \;.
 \end{eqnarray}
Since at $\lambda = 1$ $(\tau_1, \tau_2)$ equals  $(\tau_1^*, \tau_2^*)$,
  the image of the surface ${\cal B}'$ in the complex $(\tau_1, \tau_2)$-plane is a surface that passes through the attractor point
 \eqref{saddt1t2}.

 Next, for $(\rho, \sigma)$ given in \eqref{rsv0}, we verify that ${\rm Im} \, \rho > 0, \; {\rm Im } \, \sigma > 0$. The condition 
  ${\rm Im} \, \rho > 0$ gives
  \begin{equation}
 - \frac12 <
  \tilde v_1 - \frac12 - \, \frac{\eta_3(\lambda) }{\eta_2(\lambda)} \, {\tilde \sigma}_1   \;,
  \label{imcond1}
  \end{equation}
  while the condition  ${\rm Im} \,  \sigma > 0$ results in 
 \begin{equation}
  \tilde v_1 - \frac12 - \, \frac{\eta_3(\lambda) }{\eta_2(\lambda)} \, {\tilde \sigma}_1 < \frac12\;.
 \label {imcond2}
 \end{equation}
 Inspection of \eqref{ellipseeta} shows that both conditions are satisfied, for if $| \tilde v_1 - \frac12 - \, \frac{\eta_3(\lambda) }{\eta_2(\lambda)} \, {\tilde \sigma}_1| \geq 1/2$,  \eqref{ellipseeta} does not have a solution for $0 <  \eta_1(\lambda) \eta_2 (\lambda) - \eta_3^2 (\lambda) \leq 1/4$.
 Then,  defining $q = \exp [ 2 \pi i \, \rho]$ and $\tilde q =  \exp [ 2 \pi i \, \sigma]$, we obtain $|q| < 1, \, |\tilde q| < 1$, which is a necessary
 condition for defining the modular forms $\vartheta_2^8 (\rho), \, \vartheta_2^8 (\sigma)$, c.f. \eqref{phi2vvth}. This gives another justification
 for  having to require
 $\epsilon > 0$.  
 
 Let us return to the integrand in \eqref{hatWmicroprep} and analyse the behaviour of the real part of the exponent.
 The three terms proportional to the charge bilinears, when evaluated on \eqref{rsv0}, yield 
 the contribution $ \pi \lambda [ m (\tau_1^*)^2 +  m (\tau_2^*)^2 + n + l
 \tau_1^* ]/\tau_2^*$  \cite{Sen:2007qy}, which is finite for $0 < \epsilon \leq \lambda \leq 1$. 
 Now, consider the term
\begin{equation}
2 \ln | Y^0  \, {\tilde Y}^0 | = 2 \ln \left\vert \frac{p^1}{\tau_2} - p^0 \, \frac{\tau_1}{\tau_2} + i p^0 \right\vert + 
2 \ln  \left\vert \frac{p^1}{\tau_2} - p^0 \, \frac{\tau_1}{\tau_2} - i p^0 \right\vert
\end{equation}
contained in $\tilde \Lambda$.
In the parametrization \eqref{arcpar}, this expression ceases to be defined when $\lambda^2 + (1- \lambda^2) \cos^2 \theta = 0$,
which occurs when $\cos \theta =0$ and $\lambda \rightarrow 0$. In this limit, $\tau_1/ \tau_2$ remains finite, while $|\tau_2| \rightarrow \infty$,
and hence $ 2 \ln | Y^0  \, {\tilde Y}^0 | \rightarrow - \infty$, which yields an exponentially damped contribution.
Next, let us consider the contribution $D(\rho, \sigma)$ in \eqref{tillamD}, which was determined in the context of the Wilsonian effective action as 
power series in inverse powers of $(Y^0)^2$. This requires taking $|Y^0|$ to be large. To ensure that $|Y^0|$ remains large on the
integration contour ${\cal C}$, we impose a charge dependent lower bound on $\epsilon$, as follows. 
Computing ${\rm Re} \, Y^0( \rho, \sigma) = {\rm Re}  \, {\tilde Y}^0( \rho, \sigma) $ on ${\cal C}$, we obtain
\begin{equation}
{\rm Re} \, Y^0( \rho, \sigma) = {\tilde \sigma}_2 \, p^1 + {\tilde v}_2 \, p^0 \;.
\label{reYlam}
\end{equation}
 At $\lambda = \epsilon$, we infer from \eqref{reYlam}
that ${\rm Re} \, Y^0= \epsilon \left( m \, p^1 + \tfrac12 \, l \, p^0  \right)/\sqrt{4 mn - l^2}$. By taking $ {\cal Q} \equiv |m \, p^1 + \tfrac12 \, l \, p^0|
/ \sqrt{4 mn - l^2} \gg 1$
and choosing $\epsilon = 1/\sqrt{\cal Q}$ we can ensure that $| {\rm Re} \, Y^0| \gg 1$. 
Demanding $ {\cal Q}  \gg 1$ imposes a condition on the charges carried by the BPS black hole.
Presumably, were we to have at our disposal the exact
expression for the function $H$,  there would be no need for such a charge dependent lower bound on $\epsilon$.
Noting that $|\rho + \sigma| = 2 |\tau_2| = 2 \tau_2^*/\sqrt{\lambda^2 + (1- \lambda^2) \cos^2 \theta }
\geq 2 \tau_2^* \gg 1$, we see that the first two terms and the last term in the expression for $D(\rho, \sigma)$ are well-behaved on $\cal C$.
$\Xi(Y^0,  \rho)$, on the other hand, is given by \eqref{Inexp}. As shown in appendix \ref{sec:jacfor}, the series converges in an open
neighbourhood of $(q, z) = (0,0)$, where $q = \exp [ 2 \pi i \, \rho]$ and $z = 1/Y^0$.  To ensure that we work in this neighbourhood,
we impose a further condition on the charges, as follows.
In the parametrization  \eqref{arcpar}, we obtain
\begin{equation}
|q| = e^{ - 2 \pi ( {\rm Im} \, \tau_1 + {\rm Re} \,  \tau_2 ) }=e^{ -2 \pi \tau_2^*  \lambda \left( \sqrt{1-\lambda^2} \, \sin \theta + 1 \right)/
(\lambda^2 + (1 - \lambda^2) \cos^2 \theta)} \;.
\end{equation}
To ensure that $|q| \ll 1$ when $\lambda = \epsilon$,
 we demand $\tau_2^* \epsilon \gg 1$, which results in 
$(\tau_2^*)^2 \gg {\cal Q} \gg  1$, which constitutes a further condition on the charges carried by the BPS black hole. In this way, 
we conclude that the real part of the exponent of the integrand in \eqref{hatWmicroprep} is well-behaved in the range $\epsilon = 1/\sqrt{\cal Q} \leq 
\lambda \leq 1$.

Under $\Gamma_0(2)$-transformations, the critical point $(\tau_1^*, \tau_2^*)$
gets mapped to a new critical point $(\tilde{\tau}_1^*, \tilde{\tau}_2^*)$, given 
in terms of transformed charge bilinears ${\tilde m}, \tilde{n}, \tilde{l}$.
The transformed contour $\tilde{\cal C}$ will pass through
$(\tilde{\tau}_1^*, \tilde{\tau}_2^*)$.

\section{Microstate proposal \label{sec:microstatesec}}

The quantum entropy function computes the macroscopic entropy
of a BPS black hole. Ideally, we would like to reproduce it by state counting.
Microstate counting formulae for BPS black holes in $N=4,8$ superstring
theories \cite{Dijkgraaf:1996it,Shih:2005qf,Jatkar:2005bh,David:2006ru} suggest that the state counting will be based on modular objects.
In the $N=4$ context, these are Siegel modular forms, whose Fourier expansion
yields integer coefficients that count microstates of $\tfrac14$ BPS black holes.
Encoding microstate degeneracies of BPS black holes in terms of modular
forms is a powerful principle that we will also use in
the context of the $N=2$ STU model.  This gives a microstate proposal that can be tested
by a state counting process as and when it is divised.

As already mentioned in the previous subsection, the approximate result for the quantum entropy function ${W} (q,p)$ 
 in \eqref{hatWmicroprep} exhibits a dependence on the parameter $k=2$ 
that is characteristic of
a microstate counting formula based on a Siegel modular form $\Phi_2$ of weight $k=2$ \cite{David:2006yn}.
Thus, we expect the microstate counting formula for large BPS black holes to involve $\Phi_2$.
On the other hand,
the integrand in \eqref{hatWmicroprep} also depends on $Y^0$, on ${\tilde Y}^0$ as well as on $\ln (p^1 - \rho \, p^0)$
and $\ln (p^1 + \sigma\, p^0)$, which means 
that the counting formula cannot solely be given in terms of Siegel modular 
form $\Phi_2$.  The microstate counting
formula will, in particular, have to depend on the charges $(p^0, p^1)$, which transform as a doublet under
$\Gamma_0(2)_S$, c.f. \eqref{eq:S-charge-duality}.

Let us first focus on the Siegel modular form $\Phi_2$.
For a given weight $k$ with respect to a subgroup of the full modular group 
${\rm Sp} (4, \mathbb{Z})$, there may exist one or more Siegel modular forms.
We propose that the Siegel modular form $\Phi_2 (\rho, \sigma, v)$ relevant for the microstate counting formula of large BPS 
black holes in the STU model is the Siegel modular form of weight $k=2$ 
briefly discussed in \cite{David:2006ru} in
the context of $N=4$ BPS black holes,
which in the limit $v \rightarrow 0$
behaves as
\begin{equation}
\Phi_2 (\rho, \sigma, v) = 4 \pi^2 \, 2^{-16} \, v^2 \, \vartheta_2^8 (\rho) \, \vartheta_2^8 (\sigma)  + {\cal O}(v^4)
\;.
\label{phi2vvth}
\end{equation}
We note that  $\Phi_2$ is related to the Siegel modular forms $\Phi_6$ arising in the heterotic $\mathbb{Z}_2$ CHL orbifold model
and $\Phi_{10}$ arising in the heterotic string theory on $T^6$ by $\Phi_2 = \Phi_6^2 / \Phi_{10}$, up to a normalization constant.
This Siegel modular form can be constructed as follows \cite{Jatkar:2005bh}.\\

\noindent
{\bf Proposition:} There exists a Siegel modular form $\Phi_2 (\rho, \sigma, v)$ of weight $k=2$, 
symmetric in $\rho$ and $\sigma$, with the property 
\begin{equation}
\Phi_2 (\rho, \sigma, v) = 4 \pi v^2 \,  \, 2^{-16} \, \vartheta_2^8 (\rho) \, \vartheta_2^8 (\sigma) + {\cal O} (v^4)
\label{Phi2the8the8}
\end{equation}
as $v \rightarrow 0$, that can be
constructed by applying a Hecke lift to the Jacobi form 
\begin{eqnarray}
\phi_{2,1} (\rho, z) = \frac{\vartheta_1^2 (\rho, z)}{ \eta^6 (\rho)} \, 2^{-8} \,  \vartheta_2^8 (\rho) 
\end{eqnarray}
of weight $k=2$ and index $m=1$.

\begin{proof}
We refer to appendix \ref{sec:Heckelift} for the proof, which is based on the construction given in \cite{Jatkar:2005bh}. Our proof of the property \eqref{Phi2the8the8}
relies on the relation \eqref{absrel}, which we prove by considering Hecke eigenforms.

Note that under $\Gamma_0(2)$ transformations, ${\vartheta_1^2 (\rho, z) / \eta^6 (\rho)}$ transforms as a modular form
of weight $-2$, while  $\vartheta_2^8 (\rho)$ transforms with weight $4$, so that the total weight is $k=2$.
Also note that none of these two factors has a non-trivial
multiplier system.

\end{proof}

As mentioned above, the microstate counting formula for large BPS black holes cannot be solely based on the Siegel
modular form $\Phi_2$. It should be encoded in various modular objects, one of
them being $\Phi_2$. In the following, we make a proposal for reproducing
${W}(q,p) \, 
e^{   \ln |\vartheta_2 (T_0)|^8 + \ln |\vartheta_2 (U_0)|^8 }$
in terms of modular objects.\\

\noindent
{\bf Proposition:} 
The approximate expression for the quantum entropy function given in \eqref{hatWmicroprep} is 
captured by the following integral in Siegel's upper half plane
${\cal H}_2$,
\begin{eqnarray}
&&\int_{\mathcal{C}'} \frac{d \rho d \sigma dv}{(2v - \rho - \sigma)^5} \;
\exp{\left\lbrace-2\pi i \left[\frac{v^2-\rho\sigma}{2v-\rho-\sigma} \, m+\frac{1}{2v-\rho-\sigma} \, n
+   \tfrac12 \frac{\rho-\sigma}{2v-\rho-\sigma} \, l \right]\right\rbrace}\nonumber\\
&& \hskip 3cm \frac{G(\rho, Y^0 (\sigma, \rho) ) \, G(\sigma, {\tilde Y}^0 (\rho, \sigma) )   \, 
\exp{[
\frac{i \pi  }{(\rho + \sigma)\,  Y^0( \rho, \sigma) \, {\tilde Y}^0( \rho, \sigma)} ] }
}{\Phi_2(\rho, \sigma, v)} \;,
\label{microprop}
\end{eqnarray}
where 
\begin{eqnarray}
\label{FF}
G(\rho, Y^0) &=& \exp\left[ \left[ \tfrac12  \ln \vartheta_2^8 (\rho) + 2  \ln (p^1 - \rho \, p^0) \right]
- \frac{64 \pi^2 }{(Y^0)^2} \, \hat{I}_1 ( \rho, \sigma) - 256 \pi \, \Xi (Y^0, \rho)
\right]\;, \nonumber\\
G(\sigma, {\tilde Y}^0) &=&
\exp\left[ \left[ \tfrac12  \ln \vartheta_2^8 (\sigma) + 2  \ln (p^1 + \sigma\, p^0) \right]
- \frac{64 \pi^2 }{( {\tilde Y}^0)^2} \, \hat{I}_1 ( \sigma,   \rho) - 256 \pi \, \Xi ({\tilde Y}^0, \sigma)
\right]\;. \nonumber\\
\end{eqnarray}
Here, $Y^0$, ${\tilde Y}^0$ and $\hat{I}_1$ are given by \eqref{Y^0tildeY0I1},
and $\Xi$
is a solution to the non-linear PDE \eqref{PDEXi}. The contour
$\mathcal{C}'$ denotes a contour that encircles $v=0$, and that in the 
$(\tau_1, \tau_2)$-plane 
is identified with the contour ${\cal C}$  discussed in subsection \ref{sec:contour}.
Note that $G$ is given in terms of three distinct building blocks.

\begin{proof}

Let us begin by considering the integral 
\begin{eqnarray}
\label{Phi2int}
 \int_{\mathcal{C}'} \frac{d \rho d \sigma dv}{(2v - \rho - \sigma)^5} \;
\exp{\left\lbrace-2\pi i \left[\frac{v^2-\rho\sigma}{2v-\rho-\sigma} \, m+\frac{1}{2v-\rho-\sigma} \, n+ 
 \tfrac12 \frac{\rho-\sigma}{2v-\rho-\sigma} 
\, l\right]\right\rbrace}
\frac{1}{\Phi_2(\rho, \sigma, v)} \;.
\nonumber\\
\end{eqnarray}
As shown in \cite{Jatkar:2005bh}, the exponent in \eqref{Phi2int} as well as
\begin{equation}
\frac{d \rho d \sigma dv}{(2v - \rho - \sigma)^5 \,\Phi_2(\rho, \sigma, v) } \;
\label{phiintgr}
\end{equation}
are invariant under symplectic transformations acting on Siegel's upper
half plane that belong to the subgroup
$H \subset {\rm Sp} (4, \mathbb{Z})$, which 
consists of elements $h \in H$ given by
\begin{equation}
h = g_1(a, b, c, d) \, g_2  \, g_1 (a, -b, -c, d) \,  (g_2)^{-1} = 
\begin{pmatrix}
A & B\\
C & D
\end{pmatrix} = 
\begin{pmatrix}
a &0 & b  & 0\\
0 & a & 0  & -b \\
c & 0 & d & 0 \\
0 & -c & 0 & d
\end{pmatrix} \;\;,\;\; \begin{pmatrix}
a & b\\
c & d
\end{pmatrix} \in \Gamma_0(2) \;,
\label{gggg22}
\end{equation}
with $g_1$ and $g_2$ given in \eqref{g1} and \eqref{g2}, respectively.
Indeed, using \eqref{eq:SP2}, one finds that under $H$-transformations,
\begin{eqnarray}
\rho &\rightarrow& \frac{ \rho + a c (- \rho \sigma + v^2) + b c (\rho - \sigma) + bd }{\Delta} \;, 
\nonumber\\
\sigma & \rightarrow & \frac{ \sigma + b c (\sigma - \rho) - bd + ac (\rho \sigma - v^2) }{\Delta} \;, \nonumber\\
v & \rightarrow & \frac{v}{\Delta} \;, \nonumber\\
2 v - \rho - \sigma  & \rightarrow & \frac{2 v - \rho - \sigma }{\Delta } \;, \nonumber\\
\Phi_2 (\rho, \sigma, v) & \rightarrow & \Delta^2 \, \Phi_2 (\rho, \sigma, v) \;, \nonumber\\
\Delta &=& c^2 ( - \rho \sigma + v^2) + cd (\rho -\sigma) + d^2 \:.
\label{Hvn0}
\end{eqnarray}
The charge bilinears $(m,n,l)$ transform as in \eqref{chargebiltrafo}. Using these transformation rules,
one establishes that the exponent in \eqref{Phi2int} as well as \eqref{phiintgr}
are invariant under $H$-transformations.

The integral \eqref{Phi2int} depends on a contour
$\mathcal{C}'$, which transforms as follows under $H$. Under $H$, $v$ gets mapped
to $v/\Delta$, see \eqref{Hvn0}. Thus, a small contour around $v =0$ gets mapped to a small contour
around $v=0$. On the other hand, $\rho$ and $\sigma$ get mapped to $\rho'$ and $\sigma'$ according to
\eqref{Hvn0}. As mentioned in subsection \ref{sec:contour},
this means that in the limit $v=0$, 
the contour  ${\cal C}$ passing through the attractor point $(\tau^{*}_1, \tau^{*}_2)$ 
gets transformed into a new contour that passes 
through the transformed attractor point.

In the limit $v=0$, the transformation rules \eqref{Hvn0}
result in
(using $a d - bc =1$)
\begin{equation}
\Delta = (d + c \rho) (d - c \sigma) 
\end{equation}
and
\begin{eqnarray}
\rho &\rightarrow& \frac{ a \rho + b}{c \rho + d} \;, \nonumber\\
\sigma & \rightarrow & \frac{ a \sigma - b}{- c \sigma + d } \;, \nonumber\\
\rho + \sigma  & \rightarrow & \frac{ \rho + \sigma }{\Delta } \;.
\label{Hv0}
\end{eqnarray}
Supplementing this by (c.f. \eqref{eq:S-charge-duality})
\begin{eqnarray}
p^1 &\rightarrow & a \, p^1 + b \, p^0 \;, \nonumber\\
p^0 &\rightarrow & d \, p^0  + c \, p^1 \;,
\end{eqnarray}
we infer
\begin{eqnarray}
p^1 - p^0 \rho &\rightarrow & \frac{p^1 - p^0 \rho}{c \rho + d} \;, \nonumber\\
p^1 + p^0 \sigma &\rightarrow & \frac{p^1 +  p^0 \sigma}{- c \sigma + d} \;, \nonumber\\
Y^0 (\rho, \sigma)  & \rightarrow & (c \rho + d) \, Y^0 (\rho, \sigma) \:, \nonumber\\
{\tilde Y}^0 (\rho, \sigma)  & \rightarrow & (- c \sigma + d) \, {\tilde Y}^0 (\rho, \sigma) \:.
\end{eqnarray}
It follows that each of the three building blocks that make up
$G(\rho, Y^0 (\sigma, \rho) ) $ and 
$G(\sigma, {\tilde Y}^0 (\rho, \sigma) )$ in 
\eqref{FF} is invariant under $H$ in the limit $v=0$,
and hence the whole integrand  in 
\eqref{microprop} is invariant under $H$ in this limit. 
 
Why is the subgroup  $H \subset Sp(4, \mathbb{Z})$ of relevance? For once, 
it
implements the $\Gamma_0(2)_S$-symmetry
of the STU model, c.f. \eqref{gggg22}. 
This is similar to the role played by $H$ in the CHL $\mathbb{Z}_2$-orbifold
counting formula discussed in \cite{Jatkar:2005bh}.
In addition, 
in the limit $v=0$, $H$-transformations result in
a simultaneous transformation of $\rho, \sigma, Y^0, {\tilde Y}^0$
that is consistent with the attractor point identification $\rho = i S, \, \sigma = i {\bar S}, 
\, Y^0 = (p^1 + i {\bar S} p^0)/( S + \bar S), \,
{\tilde Y}^0 = (p^1- i {S} p^0 ) / (S + \bar S)$.

Performing the contour integration over $v$ in \eqref{Phi2int} using \eqref{Phi2the8the8}
results in (with $k=2$)
\begin{eqnarray}
\label{sieg-asym}
&&  \int_{\cal C}
\frac{d\rho \,  d\sigma}{(\rho+\sigma)^2}
\left( k + 3   -  2 \pi i
\left[\frac{\rho\sigma}{\rho+\sigma} \, m-\frac{1}{\rho+\sigma} \, n- \tfrac12 \frac{\rho-\sigma}{\rho+\sigma} \,  l \right] 
 \right) \;\\
&& \exp{\left\lbrace- 2\pi i\left[\frac{\rho\sigma}{\rho+\sigma} \, m-\frac{1}{\rho+\sigma} \, n- \tfrac12 \frac{\rho-\sigma}{\rho+\sigma} \,  l \right]-  
 \ln \vartheta^8_2(\rho)-  \ln \vartheta^8_2(\sigma)
-(k+2)\ln(\rho+\sigma) 
   \right\rbrace} \;. \nonumber
   \end{eqnarray}
Similarly, performing the contour integration over $v$ in  \eqref{microprop} yields the quantum
entropy function integral
${W}(q,p) \, 
e^{   \ln |\vartheta_2 (T_0)|^8 + \ln |\vartheta_2 (U_0)|^8 }$
 given in
\eqref{hatWmicroprep}.

\end{proof}

Thus, our proposal for a 
microstate counting formula for large BPS black holes which reproduces
the corresponding approximate quantum entropy function \eqref{hatWmicroprep} is
\begin{eqnarray}
\label{apppp0p1}
d^{(p^0, p^1)} (m,n,l) &=& 
e^{  -  \ln |\vartheta_2 (T_0)|^8 - \ln |\vartheta_2 (U_0)|^8 } \,
\int_{\mathcal{C}'} \frac{d \rho d \sigma dv}{(2v - \rho - \sigma)^5} \; \nonumber\\
&&
\exp{\left\lbrace-2\pi i \left[\frac{v^2-\rho\sigma}{2v-\rho-\sigma} \, m+\frac{1}{2v-\rho-\sigma} \, n
+ \tfrac12  \frac{\rho-\sigma}{2v-\rho-\sigma} \, l \right]\right\rbrace}\nonumber\\
&& 
\frac{G(\rho, Y^0 (\sigma, \rho) ) \, G(\sigma, {\tilde Y}^0 (\rho, \sigma) )   \, 
\exp{[
\frac{i \pi  }{(\rho + \sigma)\,  Y^0( \rho, \sigma) \, {\tilde Y}^0( \rho, \sigma)} ] }
}{\Phi_2(\rho, \sigma, v)} \;.
\end{eqnarray}
Being approximate, this formula, which is based on modular objects, gives
a non-integer value of $d^{(p^0, p^1)} (m,n,l)$.
Note that we have attached the label $(p^0, p^1)$ to the degeneracy 
$d^{(p^0, p^1)} (m,n,l)$, to indicate that it also depends on the 
$\Gamma_0(2)_S$ doublet $(p^0, p^1)$.


\section{Conclusions \label{sec:concl}}

We conclude with a brief summary and a few observations.

We computed the quantum entropy function \eqref{quantumtau} for 
large BPS black holes in the $N=2$ model
of Sen and Vafa, by integrating out the moduli $\phi^2$ and $\phi^3$ in
Gaussian approximation, to arrive at the intermediate result 
\eqref{quantum-approx-STU222},
which we then converted into the integral
\eqref{quantumint} by adding a total derivative term \eqref{TDmass}.
In doing so, we resorted to various approximations.
The integrand in \eqref{quantumint}  is invariant under
$\Gamma_0(2)$ transformations of $(\tau_1, \tau_2)$. 
However, it also depends on $T_0, U_0$, and is not
invariant under $\Gamma_0(2)$ transformations of $T_0, U_0$. This 
is due to the fact that when evaluating \eqref{quantumtau}, we expanded around
background values $T_0$ and $U_0$ which we took to be large.
There will then be subleading corrections in $T_0, U_0$ that will
restore the invariance under $\Gamma_0(2)$ transformations of $T_0, U_0$.

Proceeding in the manner described above, namely integrating out the moduli 
$\{\phi^a\}_{a=2,3}$ and retaining the dependence on $\phi^0, \phi^1$, one
obtains an integral which resembles, in part, the expression obtained
by integrating the inverse of a Siegel modular form $\Phi_2 (\rho, \sigma, v)$ of weight $2$
along a closed contour surrounding $v=0$, as in \eqref{Phi2int}. 
However, the result of the
quantum
entropy function calculation also depends on the $\Gamma_0(2)_S$ doublet $(p^0, p^1)$ through 
the dependence on $Y^0$, and hence, a microstate counting formula cannot
solely be based on the inverse of a Siegel modular forms, since the evaluation
of the latter gives a result that depends only on the three charge bilinears $m,n,l$,
but not on the individual charges $(p^0, p^1)$. Thus, the proposal for a microstate 
counting formula has to depend on an additional modular object $G(\rho, Y^0(\sigma,
\rho))$, c.f.  \eqref{FF}.

We can generalize the above discussion to  a certain class of
$N=2$ models with $n_V$ vector multiplets, as follows.
We take these models to have a holomorphic
Wilsonian function $F(Y, \Upsilon) = F^{(0)} (Y) + 2i \Omega (Y, \Upsilon)$
with 
a heterotic type prepotential of the form
\begin{equation}
F^{(0)} (Y) = - \frac{Y^1 Y^a \eta_{ab} Y^b}{Y^0} + \dots \;\;\;,\;\;\; a = 2, \dots, n_V \:.
\label{F0hetty}
\end{equation}
The ellipsis in \eqref{F0hetty} stands for one-loop corrections that involve $Y^0, Y^a$, but not $Y^1$,
since $Y^1/Y^0$ serves as the loop-counting parameter in heterotic string theory.
Moreover, introducing $\tau = Y^1/Y^0$, we 
assume that  the $\tau$-dependence of the first gravitational 
coupling function $\omega^{(1)}$ in $\Omega$ is encoded in a modular
form of a certain weight under (a subgroup of) $SL(2, \mathbb{Z})$. 
Suppressing the dependence on the other moduli, we set  $4 \pi \Upsilon \omega^{(1)} (\tau) = g(\tau)$.
This modular form will
then be related to the seed of an associated Siegel modular form, as we will discuss momentarily.

Setting $Y^I = \tfrac12 (\phi^I + i p^I)$, and 
using the approximate measure factor \eqref{approxmeas}, the quantum entropy function for large BPS black holes
in these models becomes
\begin{equation}
{W}(q,p) =  \int \frac{d \tau_1 \, d \tau_2}{\tau_2} \, d \phi^a \, e^{\pi [ 4 \, {\rm Im} F(\phi + i p) -  q \cdot \phi] } \, |Y^0|^{4 - \chi/12} \, 
e^{{- \cal K}^{(0)}} \;,
\end{equation}
with $e^{{- \cal K}^{(0)}}$ given by \eqref{KF0}, and with 
 $\chi = 2 (n_V - n_H + 1 )$
 determined in terms of the number of vector and
 hyper multiplets of the $N=2$ model ($n_V$ and $n_H$, respectively). Following the steps in subsection \ref{sec:beyspa} and integrating out 
the $(n_V-1)$ moduli
$\phi^a $ in Gaussian approximation gives the approximate result
\begin{equation}
{W}(q,p) =  \int \frac{d \tau_1 \, d \tau_2}{\tau_2^{(n_V + 1)/2}} \, e^{\pi \, H(\tau_1, \tau_2, p, q) } \, |Y^0|^{4 - \chi/12} \, 
e^{{- \cal K}^{(0)}} \;,
\end{equation}
with
\begin{equation}
|Y^0|^{4 - \chi/12} \,  e^{- {\cal K}^{(0)}} = \frac{|p^1 - \tau p^0|^{2 - \chi/12}}{  \tau_2^{3 - \chi/12}}
\left((p^1)^2 \, m +  p^0 p^1 \, l + (p^0)^2 \, n \right) \;,
\end{equation} 
up to an overall constant. 
Here,  $H(\tau_1, \tau_2, p, q) $ takes a form similar to \eqref{saddle23H},
\begin{eqnarray}
 \pi H(\tau_1, \tau_2, p, q) = \pi  \frac{n + l \tau_1 + m \tau_1^2 + m \tau_2^2}{\tau_2} +
 \Big( g  (\tau) + {\rm  c.c.} \Big)  + h(Y^0, {\bar Y}^0, \tau_1, \tau_2) 
  \;,
  \end{eqnarray}
where $h$ encodes the dependence on $Y^0 (\tau_1, \tau_2)$ due to 
presence of the higher gravitational
coupling functions, and where we have suppressed the dependence on the other moduli.

Then, the analogue of the intermediate result \eqref{quantum-approx-STU2222}
is
\begin{eqnarray}
{W}(q,p) &=& \ft12 \int \frac{d \tau_1  \, d \tau_2 }{\tau_2^{(n_V + 7)/2 - \chi/12}} \, 
\, e^{\pi \, {H}(\tau_1, \tau_2, p, q)  } 
 |p^1 - \tau p^0|^{4 - \chi/12} \, m
 \;.
\label{quantum-approx-het}
\end{eqnarray}
By adding an appropriate total derivative term to \eqref{quantum-approx-het}, the latter
can be brought into a form analogous to \eqref{quantum-approx-STU3}, 
\begin{eqnarray}
\label{Whetapp}
{W}(q,p) &=& 
 \int \frac{d \tau_1 \,  d \tau_2 }{\tau_2^{(n_V + 9)/2 - \chi/12}} \,
\left( \frac{n_V + 7}{2} - \frac{\chi}{12}
+ \frac{\pi}{\tau_2}  (n + l \tau_1 + m \tau_1^2 + m \tau_2^2) 
\right) e^{\pi \, H(\tau_1, \tau_2, p, q) }
\nonumber\\
&& \qquad  |p^1 - \tau p^0|^{4 - \chi/12} \;.
\label{quantum-approx-genhet}
\end{eqnarray}
Introducing 
\begin{equation}
k =  \frac{n_V + 1}{2} - \frac{\chi}{12} = \frac{2(n_V + 1) + n_H }{6} > 0 \;,
\end{equation}
we write \eqref{quantum-approx-genhet} as 
\begin{eqnarray}
{W}(q,p) &=& 
 \int_{\cal C} \frac{d \tau_1 \,  d \tau_2 }{\tau_2^{2 }} \,
\left( k + 3
+ \frac{\pi}{\tau_2}  (n + l \tau_1 + m \tau_1^2 + m \tau_2^2) 
\right) \nonumber\\
&& \qquad e^{\pi \, H(\tau_1, \tau_2, p, q) + (2 - \chi/24) \ln |p^1 - \tau p^0|^2  - (k+2) \ln \tau_2 }\;,
\label{Wk2}
\end{eqnarray}
with an appropriately chosen contour ${\cal C}$.
 Note the
dependence of the measure on $k+3$, which 
suggests a microstate counting formula based on  a Siegel modular form $\Phi(\rho, \sigma,v)$ of weight
$k$, with the property that as $v \rightarrow 0$,
\begin{equation}
\Phi_k (\rho, \sigma, v) \sim v^2  \, f_{k + 2} (\rho)  \, f_{k + 2} (\sigma) \;,
\label{siegkv0}
\end{equation}
where
$f_{k + 2} $ is a modular form of weight $k+2$ under (a subgroup of) $SL(2, \mathbb{Z})$.  
Requiring $k$ to be integer valued imposes a restriction on the allowed values of $(n_V, n_H)$.
We assume that this Siegel modular form can be constructed by applying a Hecke lift to a Jacobi form $\phi_{k,1} (\tau,z)$
of weight $k$ and index $1$ (c.f. \eqref{seedjac}),
\begin{equation}
\phi_{k,1} (\tau,z) = \frac{\vartheta_1^2(\tau,z)}{\eta^6(\tau)} \, f_{k + 2} (\tau) \;.
\label{seedhet}
\end{equation}
$f_{k + 2} (\tau) $ should be related to the modular form $g(\tau)$ that appears
in the first gravitational coupling function $\omega^{(1)}$, as mentioned below \eqref{F0hetty}, but need not coincide with it.

Evaluating
the analogue of \eqref{Phi2int}, 
\begin{eqnarray}
 \int_{\mathcal{C}'} \frac{d \rho d \sigma dv}{(2v - \rho - \sigma)^{k+3}} \;
\exp{\left\lbrace-2\pi i \left[\frac{v^2-\rho\sigma}{2v-\rho-\sigma} \, m+\frac{1}{2v-\rho-\sigma} \, n+ 
\tfrac12 \frac{\rho-\sigma}{2v-\rho-\sigma} \, l\right]\right\rbrace}
\frac{1}{\Phi_k(\rho, \sigma, v)} 
\nonumber\\
\end{eqnarray}
by performing a contour integration over $v$ around $v=0$ gives \cite{David:2006yn}
\begin{eqnarray}
&& 
 \int_{\cal C}
\frac{d\rho \,  d\sigma}{(\rho+\sigma)^{k+4}}
\left( k + 3   -  2 \pi i
\left[\frac{\rho\sigma}{\rho+\sigma} \, m-\frac{1}{\rho+\sigma} \, n- \tfrac12 \frac{\rho-\sigma}{\rho+\sigma} \,  l \right] 
 \right) \;\\
&& \exp{\left\lbrace- 2\pi i\left[\frac{\rho\sigma}{\rho+\sigma} \, m-\frac{1}{\rho+\sigma} \, n
- \tfrac12 \frac{\rho-\sigma}{\rho+\sigma} \,  l \right]
-   \ln f_{k+2}(\rho)-  \ln f_{k+2} (\sigma)
   \right\rbrace} \;. \nonumber
   \end{eqnarray}
Using \eqref{rst12} 
 this reproduces part of
\eqref{Wk2}. This then needs to be further supplemented by modular objects that depend on $(p^0, p^1)$
through $Y^0 (\rho, \sigma)$, as in \eqref{apppp0p1}.

An example of a model in this class is the FHSV model, which has $n_V = 11, \chi =0$, $k=6$ and modular subgroup $\Gamma_0(2)$. 
For this model, $g(\tau) = - \tfrac12 \ln \eta^{24} (2 \tau)$, which we write as 
\begin{equation}
g (\tau) = -   \ln ( \vartheta_2^{8} (\tau) {\cal E}_4 (\tau)) + \tfrac12 \ln {\cal E}_4 (\tau)  \;,
\end{equation}
up to a constant (here we used the relation given below \eqref{theta2eisen}).
This suggests to use as seed \eqref{seedhet} for the Hecke lift the cusp form
$f_{k+2} (\tau) =  \vartheta_2^{8} (\tau)  {\cal E}_4 (\tau)$, which has weight $k+2 =8$, 
has
trivial multiplier system and equals $[ \eta (\tau) \eta (2 \tau) ]^8$, up to a normalization constant
(see below \eqref{seedjac}). Since the vector space of cusp forms of weight $8$
has dimension one, $f_{k+2} (\tau)$ is a Hecke eigenform, and hence one deduces the property \eqref{siegkv0}
in a manner similar to the one given below \eqref{F2ff}. 
Using $g(\tau) = - \ln f_{k+2} (\tau)  + \tfrac12 \ln {\cal E}_4 (\tau)$, we note that
 the transformation of the second term on the right hand side of $g(\tau)$ under
 $\Gamma_0(2)$-transformations is precisely compensated by the term
 $2 \ln (p^1 - \tau p^0)$ in \eqref{Wk2}, and hence, the combination $(g(\tau) + {\rm c.c.}) + 2 \ln | p^1 - \tau p^0|^2 - 8 \ln \tau_2$
 in the exponent of \eqref{Wk2}  is invariant under  $\Gamma_0(2)$-transformations. We note that our proposal differs from the one made in \cite{David:2007tq}.

\begin{figure}
\centering
\includegraphics[width=0.4 \linewidth] {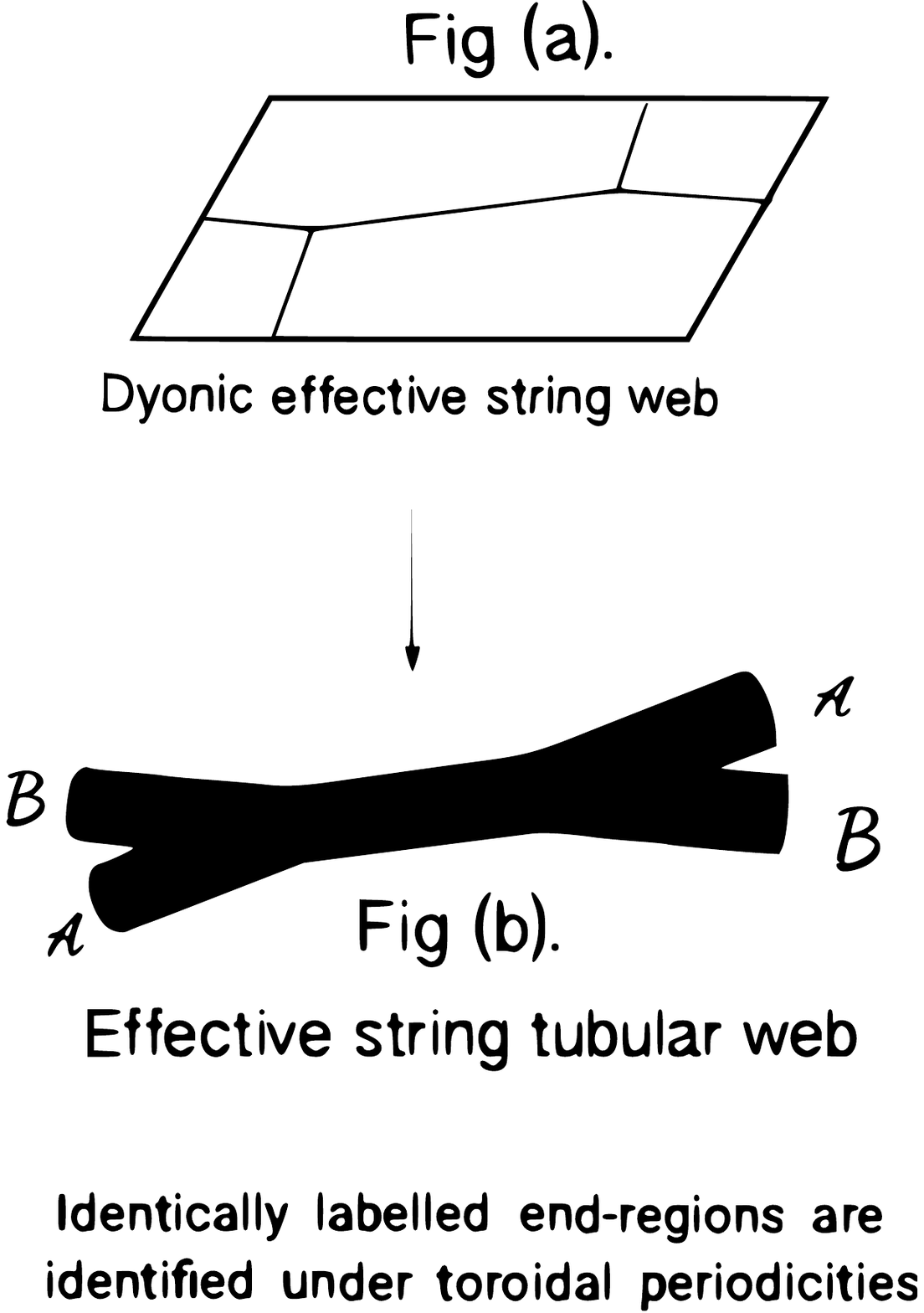}
\caption{String web picture}
\label{swp}
\end{figure}

We conclude by 
suggesting a string web picture
of our counting proposal.
We have proposed an approximate degeneracy formula for  $\frac{1}{2}$ BPS states that gravitate to form a large black hole at 
strong 't Hooft coupling  in terms of a contour integral of the inverse of a Siegel modular form and of
additional subleading contributions expressed as a series in
$1/(Y^0)^2$.
A physical picture of this mathematical structure is suggested by the $N=4$ picture discussed in
\cite{Gaiotto:2005hc,Dabholkar:2006bj,Dabholkar:2007vk}, whereby a dyonic BPS black hole is viewed as a string web wrapping a two-torus
in type IIB string theory. In this picture, a dyonic black hole in type IIB is described in terms of a web of
effective strings. Viewing the six-dimensional
compact manifold as a fibration over a two-torus, these effective strings wrap either one of 
the two cycles of the two-torus. Electric states correspond to wrappings on
 one of these cycles, while magnetic states correspond to wrappings on the other cycle.
 This effective description is valid at points in moduli space where
 the volume of the fibre is small compared to the volume of the two-torus.
 The electric-magnetic duality group $\Gamma_0(2)$ is viewed as a subgroup of the large diffeomorphism group of the two-torus.
 A computation of the 
 Euclidean partition function of this string network requires time to be compactified, so that the arms of the torus are tube-like and the resultant partition function of the string web 
 is a genus two partition function of an effective string theory  that encodes BPS states of the original type II theory. 
  In the present $N=2$ case, topological string theory is a candidate for this effective string theory, since
  amplitude calculations in topological string theory
   result in corrections to the free energy of the type II BPS partition function. 
   In this picture, the world sheet path integral will encode contributions from worldsheet instantons corresponding to higher genera string webs. 
  These contributions should correspond to the series in $1/(Y^0)^2$
   which appears in the proposed degeneracy formula, and which is absent in the $N=4$ case as a result of the larger supersymmetry in the system.


\subsection*{Acknowledgements}
We would like to thank Justin David, Bernard de Wit, Abhiram Kidambi,
Albrecht Klemm,
Swapna Mahapatra and Ashoke Sen for valuable discussions. We thank 
the Max Planck Institute for
Gravitational Physics (AEI, Golm), and the Kavli Institute for
Theoretical Physics (KITP) for hospitality during various stages of
this work. This research was supported in part by the National Science
Foundation under Grant No. NSF PHY17-48958. This work was partially
supported by FCT/Portugal through UID/MAT/04459/2013, through
FCT-DFRH-Bolsa SFRH/BPD/101955/2014
 (S. Nampuri) and through a LisMath PhD fellowship (D. Polini).


\appendix
\section{Modular forms for $SL(2, \mathbb{Z})$ \label{sec:modsl} }

We review basic properties of modular forms for $SL(2, \mathbb{Z})$, following \cite{DiaShur}.

Let  ${\cal H} = \{ \tau \in \mathbb{C}: {\rm Im} \,  \tau > 0 \}$ denote the complex upper half plane, and 
let ${\cal H}^* = {\cal H} \cup \{ \infty \}$. 
\\

\noindent
{\bf Definition:}
Let $k \in \mathbb{Z}$.
A meromorphic function $f :  {\cal  H} \rightarrow \mathbb{C}$ is weakly 
modular of weight $k$ if, $\forall \tau \in {\cal H}$ ,
\begin{equation}
f( \gamma(\tau)) = ( c \tau + d)^k \, f(\tau) \;,
\label{fgamat}
\end{equation}
where 
\begin{equation}
\gamma (\tau) = \frac{ a \tau + b}{c \tau + d} \;,
\label{gamt}
\end{equation}
with
\begin{equation}
\begin{pmatrix}
a & b \\
c & d
\end{pmatrix} \in SL (2, \mathbb{Z}) \;,
\end{equation} 
where 
\begin{eqnarray}
SL (2, \mathbb{Z}) = \left\{ 
\begin{pmatrix}
a & b \\
c & d
\end{pmatrix}: a, b, c, d \in \mathbb{Z}, \, ad - bc =1 
\right\} \;.
\end{eqnarray}
\\

\noindent
{\it Remark:}  $SL (2, \mathbb{Z})$ is generated by two elements, 
\begin{equation}
\begin{pmatrix}
1 & 1 \\
0 & 1
\end{pmatrix} : \tau \mapsto \tau + 1
 \;\;\;,\;\;\;  \begin{pmatrix}
0 & - 1 \\
1 & 0
\end{pmatrix} : \tau \mapsto - \frac{1}{\tau} 
\;.
\end{equation} 
\\

\noindent
{\it Remark:} 
Consider an open subset ${\cal F} \subset
{\cal H}$ such that no two distinct points of ${\cal F}$ are equivalent under the action
of $SL(2, \mathbb{Z})$, and every $\tau \in {\cal H}$ is equivalent
to a point in the closure ${\cal D} \equiv \overline{\cal F}$.  Then, ${\cal D}$ is
called  a fundamental domain for $SL(2, \mathbb{Z})$,
\begin{equation}
{\cal D} = \{ \tau \in {\cal H} \vert - \frac12 \leq {\rm Re} \, \tau \leq   \frac12 \;\; {\rm and} \;\; |\tau| \geq 1 \} \;.
\end{equation}
\\

\noindent
{\it Remark:} $f$ is weakly 
modular of weight $k$ if $f(\tau + 1) = f(\tau), \; f(-1/\tau) = \tau^k \, f(\tau)$.
\\

\noindent
{\bf Definition:} Modular forms are weakly modular functions that are also holomorphic on ${\cal H}$
and at $ \infty$, i.e. on ${\cal H}^*$.
\\

\noindent
{\it Remark:}
To show that a weakly modular function $f$
is holomorphic
at $\infty$, it suffices to show that $f(\tau)$ is bounded at ${\rm Im} \,  \tau \rightarrow \infty$, i.e. there exists $C \in \mathbb{R}$ such that $|f(\tau)| \leq C $ $\forall \tau$ with ${\rm Im} \, \tau \gg 1$. 
\\

Let $D = \{ q \in \mathbb{C}: |q| < 1 \}$ denote the open complex unit disc, and $D'$ the punctured disc $D' = D - \{0\}$. 
Consider the map 
\begin{equation}
\tau \mapsto e^{2 \pi i \tau} = q \;,  
\end{equation}
which takes ${\cal H}^*$ to $D$.
A modular form $f$  is $\mathbb{Z}$-periodic, since
$f(\tau +1) = f(\tau)$. Then, $f$ has the Fourier expansion
\begin{equation}
f(\tau) = \sum_{n =0}^{\infty} a_n \, q^n  \;\;\;,\;\;\;   q = e^{2 \pi i \tau}  \;.
\end{equation}
\\

\noindent
{\bf Definition:} The set of modular forms of weight $k  \in \mathbb{Z}$ is denoted by ${\cal M}_k (SL(2, \mathbb{Z}))$.
\\

\noindent
{\it Remark:} 
${\cal M}_k (SL(2, \mathbb{Z}))$ forms a finite-dimensional vector space over $\mathbb{C}$, and the direct sum
\begin{equation}
{\cal M} (SL(2, \mathbb{Z})) = \bigoplus_{k \in \mathbb{Z}} {\cal M}_k (SL(2, \mathbb{Z}))
\end{equation}
forms a graded ring.\\

\noindent
{\bf Definition:} A cusp form of weight $k  \in \mathbb{Z}$ is a modular form of weight $k$
whose Fourier expansion has a coefficient $a_0 = 0$, i.e.
\begin{equation}
f(\tau) = \sum_{n =1}^{\infty} a_n \, q^n  \;\;\; , \;\;\; q = e^{2 \pi i \tau}  \;.
\end{equation}
The limit point $\infty$ represents the cusp of $SL(2, \mathbb{Z})$. The modular
images of $\infty$ are the rational numbers $\gamma (\infty) = a/c \in \mathbb{Q}$. 
\\

\noindent
{\bf Definition:} The set of cusp forms of weight $k  \in \mathbb{Z}$ is denoted by ${\cal S}_k (SL(2, \mathbb{Z}))$.
\\

\noindent
{\bf Example:} 
The discriminant function $\Delta (\tau)
= \eta^{24}(\tau)$  is a cusp form with $\Delta \in {\cal S}_{12} (SL(2, \mathbb{Z}))$. It has
a simple zero at $q=0$.\\

\noindent
{\bf Example:}  Let $k \in \mathbb{N}$ with $k \geq 2$. The 
Eisenstein series, defined by
\begin{equation}
G_{2k} (\tau) = \sum_{(m,n) \in \mathbb{Z}^2 - \{(0,0)\}} \frac{1}{(m + n \, \tau)^{2k}} \;\;\;,\;\;\; \tau \in {\cal H} \;,
\label{ESLnon}
\end{equation}
is a modular form of weight $2k$.
It has  the following Fourier expansion,
\begin{equation}
G_{2k} (\tau) = 2  \zeta (2k) + \frac{2(2 \pi i )^{2k}}{(2k-1)!} \, \sum_{n=1}^{\infty} \sigma_{2k-1}(n) \, q^n \;\;\;,\;\;\; q= e^{2 \pi i \tau} \;,
\label{G2ksl}
\end{equation}
where the sum $ \sigma_{p}(n) = \sum_{d | n} d^p$ is over positive divisors of $n$, and
$\zeta (z)$ is Riemann's zeta function.

Setting $k=1$ in \eqref{G2ksl} yields $G_2$,
a quasi-modular form of weight $2$ and depth $s=1$, i.e. a holomorphic function $G_2 : {\cal H}^*
 \rightarrow \mathbb{C}$ that, compared
to \eqref{fgamat},
transforms with an additional shift proportional to $c/(c \tau + d)$
under \eqref{gamt},
\begin{equation}
G_2 \left(\frac{a \tau + b}{c \tau + d} \right) = (c \tau + d)^2 \, G_2 (\tau) - 2 \pi i \, c \, (c \tau + d) \;.
\label{G2quasi}
\end{equation}
More generally, a quasi-modular function of weight $k$ and depth $s$  is defined as follows  \cite{martin}:
\\

\noindent
{\bf Definition:} Let $k \in \mathbb{Z}$ and $s \in \mathbb{N}$. A holomorphic function 
  $f : {\cal H}^* \rightarrow \mathbb{C}$ is a quasi-modular form of weight $k$ and depth $s$
if there exist holomorphic functions $Q_1(f), \dots, Q_s(f)$ on ${\cal H}$ such that
\begin{equation}
(c \tau + d)^{-k} \, f\left(\frac{a \tau + b}{c \tau + d} \right) = f(\tau) + \sum_{i=1}^{s} Q_i(f)(\tau) \, \left( \frac{c}{c \tau + d} \right)^i
\end{equation}
for all $a,b,c,d \in \mathbb{Z}$ with $a d - b c =1$, and such that $Q_s (f)$ is not identically zero.\\

The normalized Eisenstein
series are $E_{2k} = G_{2k}/(  2  \zeta (2k) )$. For $k \geq 2$, 
the normalized Eisenstein functions can also be defined by \cite{BGHZ}
\begin{equation}
E_{2k} (\tau) = \tfrac12 \,  \sum_{(m,n) \in \mathbb{Z}^2 - \{(0,0)\}, \; \gcd (m,n) = 1} \frac{1}{(m + n \, \tau)^{2k}} 
= \sum_{n > 0, m \in \mathbb{Z}, \; \gcd (m,n) = 1} \frac{1}{(m + n \, \tau)^{2k}} \;.
\label{ESL}
\end{equation}
\\

\noindent
{\bf Definition:} 
A weakly holomorphic modular form of weight $k$ \cite{Dabholkar:2012nd}
is a weakly modular function $f: {\cal H} 
\rightarrow \mathbb{C}$ that is holomorphic on ${\cal H}$, with a pole at $q =0$.
Its Fourier expansion is given by 
\begin{equation}
f(\tau)  = \sum_{n = - N }^{\infty} a_n \, q^n \;\;\; , \;\;\; q = e^{2 \pi i \tau}  \;,
\end{equation}
with $ N \in \mathbb{N}$. Thus, 
$f(\tau)$ grow as $q^{-N}$ at $ \infty$.\\

\noindent
{\bf Example:} $Z (\tau) = 1/\Delta(\tau)$ is a weakly holomorphic modular form of weight $-12$ with a simple
pole at $q =0$. Hence
\begin{equation}
Z (\tau) = \sum_{n=-1}^{\infty} a_n \, q^n  \;\;\; , \;\;\; q = e^{2 \pi i \tau}  \;.
\end{equation}
For large $n$, the Fourier coefficients $a_n$ grow as 
\begin{equation}
a_n \sim e^{4 \pi \sqrt{n}} \;.
\label{small-asympt}
\end{equation}
Thus, they exhibit exponential growth, as required for the microstate
degeneracy of small BPS black holes in $N=4$ string theories \cite{Dabholkar:2004yr}.\\

Depending on the nature of the modular form, the growth property of its Fourier coefficients
can be markedly different \cite{Dabholkar:2012nd, BrDah}: \\

\noindent
{\bf Growth conditions:}

\begin{enumerate}

\item $f \in {\cal S}_k (SL(2, \mathbb{Z}))$: $a_n = {\cal O} (n^{k/2})$ as $n \rightarrow \infty$;

\item $f \in {\cal M}_k (SL(2, \mathbb{Z})), \; f \notin {\cal S}_k (SL(2, \mathbb{Z}))$: $a_n = {\cal O} (n^{k-1})$ as $n \rightarrow \infty$;

\item $f$ weakly holomorphic modular form of weight $k$: $a_n = {\cal O} (e^{C  \sqrt{n}})$ as $n \rightarrow \infty$ for some $C> 0$.

\end{enumerate}

\section{Congruence subgroups of $SL(2, \mathbb{Z})$ \label{sec:G02}}

We review basic properties of modular forms for congruence subgroups of 
$SL(2, \mathbb{Z})$, following \cite{DiaShur}. \\

\noindent
{\bf Definition:} Let $N$ be a positive integer. The principal congruence subgroup of 
$SL(2, \mathbb{Z})$ of level $N$ is
\begin{eqnarray}
\Gamma (N) = \left\{ 
\begin{pmatrix}
a & b\\
c & d\\
\end{pmatrix} \in SL(2, \mathbb{Z}): 
\begin{pmatrix}
a & b\\
c & d\\
\end{pmatrix} \equiv 
\begin{pmatrix}
1 & 0\\
0 & 1\\
\end{pmatrix} \mod N
\right\} \;.
\end{eqnarray}
In particular, $\Gamma (1) = SL(2, \mathbb{Z})$.\\

\noindent
{\bf Definition:} A subgroup $\Gamma$ of $SL(2, \mathbb{Z})$ is a congruence subgroup
if $\Gamma(N) \subset \Gamma$ for some $N \in \mathbb{Z}^+$.
The least such $N$ is called the level of $\Gamma$. 
\\

\noindent
{\bf Example:}
Let $N$ be a positive integer. Let
\begin{eqnarray}
\Gamma_0 (N) = \left\{ 
\begin{pmatrix}
a & b\\
c & d\\
\end{pmatrix} \in SL(2, \mathbb{Z}): 
\begin{pmatrix}
a & b\\
c & d\\
\end{pmatrix} \equiv 
\begin{pmatrix}
* & *\\
0 & *\\
\end{pmatrix} \mod N
\right\} \;,
\end{eqnarray}
\begin{eqnarray}
\Gamma_1 (N) = \left\{ 
\begin{pmatrix}
a & b\\
c & d\\
\end{pmatrix} \in SL(2, \mathbb{Z}): 
\begin{pmatrix}
a & b\\
c & d\\
\end{pmatrix} \equiv 
\begin{pmatrix}
1 & *\\
0 & 1\\
\end{pmatrix} \mod N
\right\} \;,
\end{eqnarray}
where $*$ can take any value in $\mathbb{Z}$. Hence, $\Gamma(N) \subset \Gamma_1 (N) 
\subset \Gamma_0 (N) \subset SL(2, \mathbb{Z})$.

In particular, $\Gamma (2) \subset 
\Gamma_1 (2) =  \Gamma_0 (2)$, since $a, d = 1 \mod 2$. This is the case of interest 
 for this paper, $\Gamma =  \Gamma_0 (2)$.\\

\noindent
{\it Remark:}
$\Gamma_0(2)$ is generated by two elements, 
\begin{equation}
 \begin{pmatrix}
 1 & 0\\
 2 & 1 
 \end{pmatrix} = S^2 \left (S \, T^{-2} \, S \right)  \;\;\;,\;\;\; T = 
  \begin{pmatrix}
 1 & 1\\
 0 & 1 
 \end{pmatrix} \;,
  \end{equation}
where
\begin{equation}
S =  
 \begin{pmatrix}
 0 & -1\\
 1 & 0 
 \end{pmatrix} \;.
 \end{equation}
\\

\noindent
{\it Remark:} Each congruence subgroup $\Gamma$ of  $SL(2, \mathbb{Z})$
contains a translation matrix of the form 
\begin{eqnarray}
\begin{pmatrix}
1 & h\\
0 & 1\\
\end{pmatrix} : \tau \mapsto \tau + h 
\label{hZper}
\end{eqnarray}
for some minimal $h \in \mathbb{Z}^+$. Hence, every function $f: \mathcal{H} \rightarrow
\mathbb{C}$ that is weakly modular with respect to $\Gamma$ is $h \, \mathbb{Z}$-periodic,
and has a corresponding function $g : D' \rightarrow \mathbb{C}$ with $f(\tau) = g(q_h)$, 
where $q_h = e^{2 \pi i \tau/h}$. Then, $f$ is defined to be holomorphic at $\infty$ if $g$ 
extends holomorphically to $q=0$, in which case
\begin{equation}
f(\tau) = \sum_{n =0}^{\infty} a_n \, q_h^n \;\;,\;\; q_h = e^{2 \pi i \tau/h} \;.
\label{gqh}
\end{equation}
\\

\noindent
{To define modular forms for a congruence subgroup  $\Gamma$, one adjoins not only $ \infty$ to $\mathcal{H} $,
but also the rational numbers $\mathbb{Q}$. Then, one identifies all elements in 
$\{ \infty\} \cup 
\mathbb{Q}$ that are $\Gamma$-equivalent. When $\Gamma = SL(2, \mathbb{Z})$, all
rational numbers are $\Gamma$-equivalent to $\infty$, i.e. $\gamma (- d/c) = \infty$, where
$c, d \in \mathbb{Z}$ with $c \neq 0$. When  $\Gamma \subset SL(2, \mathbb{Z})$ is a proper
subgroup, fewer points are $\Gamma$-equivalent. 
\\

\noindent
{\bf Definition:} 
A $\Gamma$-equivalence class of points in $\{\infty\} \cup \mathbb{Q}$ is called a cusp of $\Gamma$.\\

\noindent
{\bf Example:} 
For $\Gamma = \Gamma_0 (2)$ 
there are two cusps,
$0$ and $\infty$ \cite{BrDah}.
The orbit of $\infty$
consists of the set of  rational numbers
of the form $a/c$ with $a \neq 0$, $c = 0 \mod 2$ and $\gcd (a, c)=1$. The orbit of $0$
consists of the set of  rational numbers
of the form $b/d$ with $\gcd (b,d)=1$.
Thus, every element of 
$\{ \infty\} \cup \mathbb{Q}$ is in exactly one of the two orbits.
Since there is no element in $\Gamma_0 (2)$ that maps $0$ to $\infty$, there are two cusps. \\

\begin{figure}
\centering
\includegraphics[width=0.6 \linewidth] {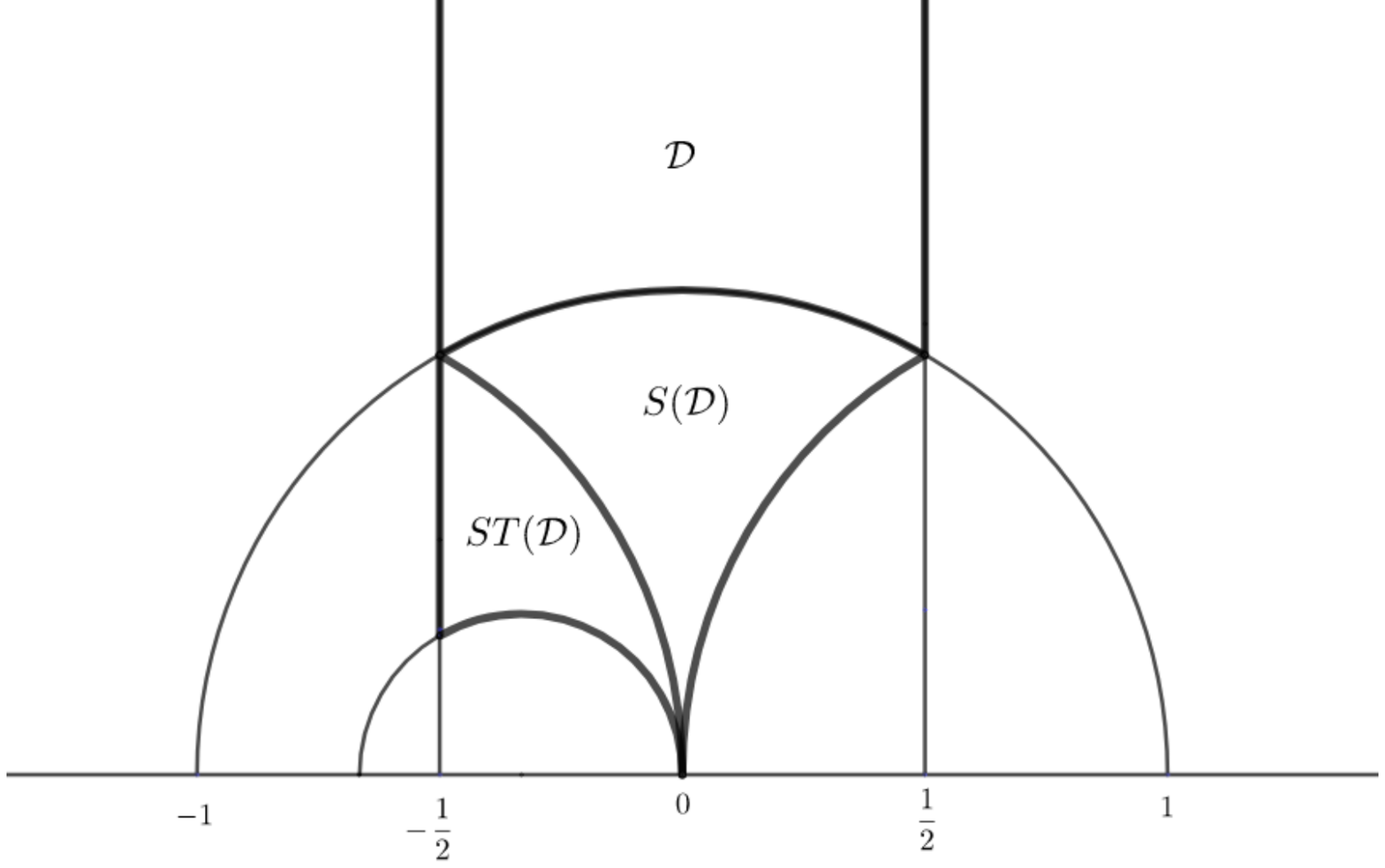}
\vskip - 10mm 
\caption{A fundamental domain for $\Gamma_0(2)$}
\label{fomg2}
\end{figure}

\noindent
{\bf Proposition (fundamental domain):} Let $\Gamma$ be a congruence subgroup of 
$SL(2, \mathbb{Z})$, and let $R$ be a set of coset representatives for the quotient $\Gamma \backslash SL(2, \mathbb{Z})$. Then, the set ${\cal D}_{\Gamma} = \cup_{\tilde{\gamma} \in R} \; \tilde{\gamma}\,
{\cal D}$ is a fundamental domain for $\Gamma$ \cite{BrDah}. Here, ${\cal D}$ denotes a fundamental domain
for $SL(2, \mathbb{Z})$.\\

\noindent
{\bf Example:} Let $\Gamma = \Gamma_0(2)$.  A system of coset representatives
for $\Gamma \backslash SL(2, \mathbb{Z})$ is \cite{BrDah}
\begin{eqnarray}
R = \left\{  \begin{pmatrix}
 1 & 0\\
 0 & 1 
 \end{pmatrix}, 
  \begin{pmatrix}
 0 & -1\\
 1 & 0
 \end{pmatrix}, 
  \begin{pmatrix}
 0 & -1\\
 1 & 1 
 \end{pmatrix}
\right\} =  \left\{ \mathbb{I}, S, S T \right\} \;,
\end{eqnarray}
i.e. if $\alpha \in SL(2, \mathbb{Z})$, then $\alpha = \gamma \, {\tilde \gamma}$ with $\gamma \in
\Gamma$ and $\tilde{\gamma} \in R$. A fundamental domain for $\Gamma_0(2)$ is thus given by
${\cal D}_{\Gamma_0(2)} = {\cal D} \cup
 S ({\cal D})  \cup S T ({\cal D})$ \cite{BrDah,Angelantonj:2013eja}, see Figure \ref{fomg2}.
Note that the two cusps, $0$ and $\infty$, are in the closure of ${\cal D}_{\Gamma_0(2)}$
in the Riemann sphere, but not in ${\cal H}$ \cite{BrDah}.\\

\noindent
{\bf Definition:} Let $\Gamma$ be a congruence subgroup of $SL(2, \mathbb{Z})$, and let $k \in 
\mathbb{Z}$. A modular form of weight $k$ for the subgroup $\Gamma$ is a holomorphic function
$f : \mathcal{H} \rightarrow \mathbb{C}$ that is weakly modular of weight $k$ for $\Gamma$ and 
holomorphic at all cusps of $\Gamma$. 
\\

\noindent
{\bf Definition:}  Let $\Gamma$ be a congruence subgroup of $SL(2, \mathbb{Z})$, and let $k \in 
\mathbb{Z}$.
Writing any element $s \in \{ \infty \}
 \cup \mathbb{Q}$ as $s  = \gamma (\infty)$ for some $\gamma \in SL(2, \mathbb{Z})$, holomorphy of $f$ at $s$ is defined in terms
 of holomorphy of $f [\gamma]_k$
 at $\infty$ for all $\gamma \in SL(2, \mathbb{Z})$. Here, $f [\gamma]_k : {\cal H} \rightarrow \mathbb{C}$ is defined by
 $ ( f [\gamma]_k ) (\tau) = (c \tau + d)^{-k} \, f( \gamma(\tau))$. If $a_0 = 0$ in the Fourier expansion of $f [\gamma]_k$ 
 for all $\gamma \in SL(2, \mathbb{Z})$, then $f$ is called a cusp form of weight $k$
with respect to $\Gamma$.
  \\

\noindent
{\it Remark:} For $\Gamma$, the finite-dimensional vector space over $\mathbb{C}$ of modular forms of weight $k$ is denoted by ${\cal M}_k (\Gamma)$,
and the vector subspace of cusp forms of weight $k$ is denoted by ${\cal S}_k (\Gamma)$. The direct
sums
\begin{equation}
{\cal M} (\Gamma) = \bigoplus_{k \in \mathbb{Z}} {\cal M}_k (\Gamma) \;\;\;,\;\;\; 
{\cal S} (\Gamma) = \bigoplus_{k \in \mathbb{Z}} {\cal S}_k (\Gamma) 
\end{equation}
form a graded ring.\\

\noindent
{\bf Example:} An example of a modular form of weight $2$ for  $\Gamma_0(N)$ 
with $N \geq 2$
is
$N G_2 (N \tau) - G_2 (\tau)$, where $G_2 (\tau)$ denotes the Eisenstein
series of weight $2$ for $SL(2, \mathbb{Z})$ introduced below \eqref{G2ksl}. \\

\noindent
{\bf Proposition:} Let $\Gamma = \Gamma_0(2)$. Let $k \in \mathbb{N},  \;k \geq 2$.
Then, 
$\dim  {\cal M}_k(\Gamma) = \left[\frac{k}{4} \right] +1$ 
and $\dim {\cal S}_k(\Gamma) = \left[\frac{k}{4} \right] -1$  for $k$ even, while
$\dim {\cal M}_k(\Gamma) = \dim {\cal S}_k(\Gamma) =0$ for $k$ odd \cite{DiaShur}.
Here, $[n]$ denotes the integer part of $n$.\\

\noindent
{\bf Example:}  Let $\Gamma = \Gamma_0(2)$. Then, 
$\dim {\cal M}_2(\Gamma) = 1$, 
$\dim {\cal M}_4(\Gamma) = 2$ 
and $\dim {\cal S}_4(\Gamma) = 0$. The vector space ${\cal M}_2(\Gamma)$
is generated by \cite{ACH,hahn}
\begin{equation}
{\tilde {\cal E}}_2 (\tau) = \frac12 \left(3 \, {\cal E}_2 (\tau) - E_2 (\tau) \right) = 2 E_2(2 \tau) - E_2 (\tau) \;.
\label{calE2mod}
\end{equation}
Here, $E_2 (\tau) = G_2 (\tau)/( 2 \zeta(2))$ denotes the normalized quasi-modular form of weight $2$ for $SL(2, \mathbb{Z})$, while
${\cal E}_2 (\tau)$ denotes one of the normalized Eisenstein series on $\Gamma_0(2)$, see \eqref{cE2q}.
\\

\noindent
{\bf Definition:} Let $k \geq 2$. For $\Gamma_0(2)$, define an Eisenstein series by
\cite{Schoene}
\begin{eqnarray}
G_{2k, N=2} (\tau) &=& 
\sum_{(m,n) \in \mathbb{Z}^2, (m,n) = (1,0) \mod 2} \frac{1 }{\left( m + n  \, \tau \right)^{2k} }\;, \nonumber\\
&=&
\sum_{(m,n) \in \mathbb{Z}^2 - \{(0,0)\}} \frac{\chi_2 (m) }{\left( m + 2 \, n  \, \tau \right)^{2k} }\;,
\end{eqnarray}
where 
\begin{eqnarray}
\chi_N (m) = \left\{
\begin{matrix}
1\;, \;\; {\rm if} \;\; \gcd (m, N) = 1 \\
0 \;, \;\; {\rm if} \;\; \gcd (m, N) > 1  \;.
\end{matrix}
\right.
\end{eqnarray}
$\chi_N (m) $ is called the principal character $\mod N$. \\
Define the normalized
Eisenstein series ${\cal E}_{2k} (\tau)$ by $G_{2k, N=2} = 2 \zeta_2 (2k) \, {\cal E}_{2k} $,
where
\begin{equation}
\zeta_N (2k) = \sum_{r=1}^{\infty} \chi_N (r) \, r^{-2k} \;.
\end{equation}
\\

\noindent
{\it Remark:}
For $N=1$, $\zeta_1 (2k) =  \zeta(2k)$, while for $N=2$, $\zeta_2 (2k) = (1-2^{-2k} ) \, \zeta(2k)$.
\\

\noindent
{\bf Proposition:} Let $k \geq 2$. Then, $G_{2k, N=2} (\tau)$ are modular forms of weight $2k$
for $\Gamma_0(2)$ \cite{Schoene}.
\\

\noindent
{\bf Proposition:} Let $k \geq 2$.
The normalized Eisenstein functions ${\cal E}_{2k} (\tau)$ are given by \cite{toth}
\begin{equation}
{\cal E}_{2k} (\tau)
=  \sum_{n > 0, m \in \mathbb{Z}, \; \gcd (m, 2 n ) = 1} \frac{1}{(m + 2 \, n \, 
 \tau)^{2k}} \;.
\label{calEG}
\end{equation}
\begin{proof}
Write \cite{toth}
\begin{eqnarray}
G_{2k, N=2} (\tau) &=& \sum_{r = 1}^{\infty}
\sum_{(m,n) \in \mathbb{Z}^2 - \{(0,0)\}, \, \gcd (m,n) = r} \; \frac{\chi_2 (r) \;
\chi_2 (m/r)
}{r^{2k} \, 
\left( m/r  + 2 \, (n/r)  \, \tau \right)^{2k} } \nonumber\\
&=& \zeta_2 (2k) \, \sum_{(c,d) \in \mathbb{Z}^2 - \{(0,0)\}, \, \gcd (c,d) = 1} 
\frac{
\chi_2 (d)
}{
\left( d  + 2 \, c \, \tau \right)^{2k} } \nonumber\\
&=& \zeta_2(2k) \, \sum_{(c,d) \in \mathbb{Z}^2 - \{(0,0)\}, \, \gcd (2 c,d) = 1} 
\frac{1}{
\left( d  + 2 \, c \, \tau \right)^{2k} } \;.
\end{eqnarray}
\end{proof}

\noindent
{\bf Proposition:} Let $k \geq 2$. Then, ${\cal E}_{2k}$ has the following expansion
around the cusp $\infty$ \cite{ACH},
\begin{equation}
{\cal E}_{2k} (\tau) = 1 + \frac{4k}{(1- 2^{2k}) B_{2k}}\, \sum_{n=1}^{\infty} \frac{(-1)^n \, n^{2k-1} \, q^n}{1 - q^n} \;\;\;,\;\;\;
q = e^{2 \pi i \tau} \;,
\label{eisenstein_gamma02}
\end{equation}
with $B_k$ the $k$-th Bernoulli number. ${\cal E}_{2k} (\tau) $ can also be written as 
\begin{equation}
{\cal E}_{2k} (\tau) = 1 + \frac{4k}{(1- 2^{2k}) B_{2k}}\, \sum_{n=1}^{\infty} {\tilde \sigma}_{2k-1} (n) \, q^n 
 \;\;\;,\;\;\;
q = e^{2 \pi i \tau} \;,
\label{eisenstein_gamma02sig}
\end{equation}
where the sum
\begin{equation}
{\tilde \sigma}_{2k-1} (n) = \sum_{d|n} (-1)^d \, d^{2k-1} 
\end{equation}
is over positive divisors of $n$.

\begin{proof}
Using absolute convergence, we write
\begin{eqnarray}
G_{2k, N=2} (\tau) &=& 
\sum_{(m,n) \in \mathbb{Z}^2 - \{(0,0)\}} \frac{1 }{\left( m + 2 \, n  \, \tau \right)^{2k} }
- 
\sum_{(m,n) \in \mathbb{Z}^2 - \{(0,0)\}} \frac{1 }{\left( 2 \, m + 2 \, n  \, \tau \right)^{2k} } 
\nonumber\\
&=& G_{2k} (2 \tau) - 2^ {- 2k} \, G_{2k} (\tau)  \;,
\end{eqnarray}
where $G_{2k}$ denote Eisenstein functions for $SL(2, \mathbb{Z})$. Using \eqref{G2ksl} in the form
\begin{equation}
G_{2k} (\tau) = 2  \zeta (2k) + \frac{2 ( -1)^{k}(2 \pi )^{2k}}{(2k-1)!} \, \sum_{m=1}^{\infty} \sum_{d=1}^{\infty}  d^{2k-1}\, q^{dm} \;\;\;,\;\;\; q= e^{2 \pi i \tau} \;,
\end{equation}
we obtain, using absolute convergence,
\begin{eqnarray}
G_{2k, N=2} (\tau) &=& 2  \zeta_2 (2k) + 2^{-2k} \frac{2 ( -1)^{k}(2 \pi )^{2k}}{(2k-1)!} \,
\left[ 2  \sum_{m=1}^{\infty} \sum_{d=1}^{\infty} (2d)^{2k-1} \, q^{2 d m } -  \sum_{m=1}^{\infty} \sum_{d=1}^{\infty} d^{2k-1} \, q^{d m} 
\right] \nonumber\\
&=& 2 \zeta_2 (2k) + 2^{-2k} \frac{2 ( -1)^{k}(2 \pi )^{2k}}{(2k-1)!} \,  \sum_{m=1}^{\infty} \sum_{d=1}^{\infty} (-1)^d \,  d^{2k-1} \, q^{d m} 
\nonumber\\
&=& 2 \zeta_2 (2k) + 2^{-2k} \frac{2 ( -1)^{k}(2 \pi )^{2k}}{(2k-1)!} \,   \sum_{d=1}^{\infty} (-1)^d \,  d^{2k-1} \, \frac{q^d}{1- q^d}
\nonumber\\
&=& 2 \zeta_2 (2k) \left[1 +  \frac{ 4k}{(1 - 2^{2k}) \, B_{2k} } \,   \sum_{d=1}^{\infty} (-1)^d \,  d^{2k-1} \, \frac{q^d}{1- q^d} \right] \;.
\end{eqnarray}
Expanding in a geometric series, 
\begin{eqnarray}
G_{2k, N=2} (\tau) &=& 
 2 \zeta_2 (2k) \left[1 +  \frac{ 4k}{(1 - 2^{2k}) \, B_{2k} } \,   \sum_{d=1}^{\infty} \sum_{m=1}^{\infty}
  (-1)^d \,  d^{2k-1} \, q^{d m} \right] \;,
\end{eqnarray}
and interchanging the order of summations yields \eqref{eisenstein_gamma02sig}.

\end{proof}

\noindent
{\bf Definition:} 
Setting $k=1$ in \eqref{eisenstein_gamma02} defines the second Eisenstein series ${\cal E}_2$ \cite{ACH,hahn},
\begin{equation}
{\cal E}_{2} (\tau) = 1 - 8 \, \sum_{n=1}^{\infty} \frac{(-1)^n \, n
 \, q^n}{1 - q^n} \;\;\;,\;\;\;
q = e^{2 \pi i \tau} \;.
 \label{cE2q}
\end{equation}
\\
\noindent
{\bf Proposition:}
${\cal E}_{2} $ can be expressed as \cite{hahn}
\begin{equation}
{\cal E}_{2} = q \frac{d}{dq} \ln \vartheta_2^8 \;,
\end{equation}
where the theta function  $\vartheta_2$ has the product representation
\begin{equation}
\vartheta_2 (\tau) = 2 \, q^{1/8}  \, \prod_{n=1}^{\infty}  (1-q^n) (1 + q^n )^2 \;\;\;,\;\;\;
q = e^{2 \pi i \tau} \;,
\label{thet2}
\end{equation}
valid in the open complex unit disc
$D = \{ q \in \mathbb{C}: |q| < 1 \}$. Thus, ${\cal E}_{2} $
 is a quasi-modular form for $\Gamma_0(2)$, 
\begin{equation}
{\cal E}_{2} (\tau) = \frac{1}{\pi \gamma} \frac{\partial \omega (S)}{\partial S} = 
\frac{1}{\pi \gamma} \, I_1(S) \;\;\;,\;\;\; S = -i \tau \;.
\end{equation}
\begin{proof}
Using \eqref{thet2} and \eqref{cE2q} as well as absolute convergence, we compute
\begin{eqnarray}
q \frac{d}{dq} \ln \vartheta_2^8 - {\cal E}_2 = 16 \left( \sum_{n=1}^{\infty} n \, 
\frac{q^n}{1+ q^n}  - \sum_{n=1}^{\infty} (2n-1) \, 
\frac{q^{2n-1}}{1- q^{2n-1}}  \right) \;.
\label{diffE2E2}
\end{eqnarray}
Using the relation
\begin{equation}
\frac{x}{1 + x} = \frac{x}{1-x} - \frac{2  x^2}{1-x^2} \;,
\end{equation}
it follows that the right hand side of \eqref{diffE2E2} vanishes. Since $I_1$ is
quasi-modular, so is ${\cal E}_{2} $.

\end{proof}

\noindent
{\it Remark:}
The vector space ${\cal M}_4(\Gamma)$, which has $\dim {\cal M}_4(\Gamma) = 2$, is generated by \cite{ACH}
${\tilde {\cal E}}_2^2 $ and ${\cal E}_{4}$.  Therefore, $\vartheta_2^8 (\tau)$, which has weight $4$,
is given by a linear combination of ${\tilde {\cal E}}_2^2 $ and ${\cal E}_{4}$, namely \cite{ACH} 
\begin{equation}
2^{-8} \, \vartheta_2^8 (\tau) = \frac{1}{64} \left( {\tilde {\cal E}}_2^2 (\tau) - {\cal E}_{4} (\tau) \right)
= \frac{1}{240} \left( E_4 (\tau) - E_4 (2 \tau) \right) \;.
\label{theta2eisen}
\end{equation}
Note that $\vartheta_2^8 (\tau) $ is not a cusp form of $\Gamma_0(2)$, and that ${\cal E}_{4} (\tau) 
= 16 \, \eta^{12}(\tau)/\vartheta_2^4 (\tau)$. 
\\

\noindent
{\it Remark:} $\tilde{\cal E}_2 (\tau)$ has the following $q$-expansion, 
\begin{equation}
\tilde{\cal E}_{2} (\tau) = 1 + 24 \, \sum_{n=1}^{\infty} \left( \sum_{d | n ,  \;\;\; d  \; {\rm odd} } d \right) 
q^n \;\;\;,\;\;\;
q = e^{2 \pi i \tau} \;.
 \label{cE2qtil}
\end{equation}
\\

\noindent
{\bf Proposition:} 
The series ${\cal E}_2, \tilde{\cal E}_2, {\cal E}_4$ satisfy the differential equations
\cite{ACH} 
\begin{eqnarray}
q \frac{ d {\cal E}_2}{d q} &=& \frac14 \left( ({\cal E}_2)^2 - {\cal E}_4 \right) \;, \nonumber\\
q \frac{ d \tilde{\cal E}_2}{d q} &=& \frac12 \left( \tilde{\cal E}_2 \, {\cal E}_2   - {\cal E}_4 \right) \;, \nonumber\\
q \frac{ d {\cal E}_4}{d q} &=&  {\cal E}_2 \, {\cal E}_4 - \tilde{\cal E}_2 \, {\cal E}_4  \;.
\label{dereisen}
\end{eqnarray}
\\

\noindent
{\bf Definition:} 
Given a modular form $f(S)$ of $\Gamma_0(2)$ of weight $2k$, the operator \cite{ACH}
\begin{equation}
f \rightarrow {\cal D}_S f = \partial_S f + \frac{k}{\gamma} \, \frac{\partial \omega(S)}{\partial S} \, f
\label{calDf2k}
\end{equation}
maps ${\cal M}_{2k} ({\Gamma_0(2)})$ to ${\cal M}_{2k+2} ({\Gamma_0(2)})$. This operator can be written as 
\begin{equation}
- \frac{1}{2 \pi} \, f \rightarrow \left( q \frac{ d }{d q} - \frac{k}{2} \, {\cal E}_2 \right) f \;.
\label{oper_covd}
\end{equation}
\\

Using the above relations, 
one readily verifies that \cite{hahn}
\begin{equation}
{\cal D}_S \vartheta_2^8 (S) =0 \;,
\label{covtheta8}
\end{equation}
as well as 
\begin{eqnarray}
I_2(S) &=&\tfrac12 \,  \pi^2 \, \gamma \, {\cal E}_4 \;, \nonumber\\
I_3(S) &=& \pi^3 \, \gamma \, \tilde{\cal E}_2 \, {\cal E}_4 \;, \nonumber\\
I_4(S)  &=& \pi^4 \, \gamma \left( {\cal E}_4^2 + 2 (\tilde{\cal E}_2)^2 \, {\cal E}_4 \right) \;.
\label{IcalE}
\end{eqnarray}
Acting with the operator \eqref{oper_covd} on $\tilde{\cal E}_2$ gives
\begin{equation}
- \frac{1}{2 \pi} {\cal D}_S \tilde{\cal E}_2 = - \tfrac12 \, {\cal E}_4 \;,
\end{equation}
and hence we infer
\begin{equation}
I_2 = \tfrac12 \,  \pi \, \gamma \, {\cal D}_S \tilde{\cal E}_2 \;.
\label{I2E2}
\end{equation}
\\

\noindent
{\bf Proposition:} 
For $k\geq 2$, the $I_k(S)$ are given by
\begin{equation}
I_k (S) =\gamma \,  \sum_{m + 2n = k, \, m \geq 0, n \geq 1} a_{m,n} \, (\tilde{\cal E}_2)^m \, ({\cal E}_4)^n \;,
\label{Ikeps2eps4}
\end{equation}
where $a_{m,n} $ are real, positive constants.

\begin{proof}

The proof is by induction.  The claim holds for $k=2$. Assume that it holds for a $k_0$ with $k_0 >2$.
Then, consider operating with \eqref{oper_covd} on a summand $(\tilde{\cal E}_2)^m \, ({\cal E}_4)^n$ of weight $2k_0 = 2m + 4n$. Using the relations \eqref{dereisen}, one infers
\begin{equation}
 \left( q \frac{ d }{d q} - \frac{k}{2} \, {\cal E}_2 \right) (\tilde{\cal E}_2)^m \, ({\cal E}_4)^n =
 - \tfrac12 \, m \, (\tilde{\cal E}_2)^{m-1} \, ({\cal E}_4)^{n+1} - n (\tilde{\cal E}_2)^{m+1} \, 
 ({\cal E}_4)^n \;.
 \label{qdqeis}
 \end{equation}
 Note that the terms on the right hand side have weight $2k_0 +2$, and that they have the same structure as 
 in  \eqref{Ikeps2eps4}.
    Thus, when operating
 with \eqref{oper_covd} on $I_{k_0}$, 
 the resulting sum is of the form \eqref{Ikeps2eps4} with $m,n$ satisfying the relations $m \geq 0, \, n \geq 1$ and $m + 2n = k_0 + 1$.
  Multiplying \eqref{qdqeis} with $(- 2 \pi)$, so as to obtain ${\cal D}_S$ on the left hand side, we infer that the coefficients $a_{m,n}$ of $I_{k_0 + 1}$
  are real and positive. 

\end{proof}

Next, we discuss the growth properties of the Fourier coefficients
of modular forms for $\Gamma_0(2)$. \\

\noindent
{\bf Proposition:}  Let $f \in {\cal S}_k (\Gamma_0(2))$,  with $k \in \mathbb{N}$, 
\begin{equation} 
f(\tau) = \sum_{n=1}^{\infty} a_n \, q^n \;\;\;,\;\;\; q = e^{2 \pi i \tau} \;.
\label{four_cusp}
\end{equation} 
Then $|a_n| \leq C \, n^{k/2}$ \;  $\forall \, n \in \mathbb{N} $
for some $C \in \mathbb{R}$ with $C > 0$
 \cite{DiaShur}.
\\

\noindent
{\bf Proposition:}  Let $\Gamma = \Gamma_0(2)$. 
Consider the 
Eisenstein series \eqref{eisenstein_gamma02sig} with $k \geq 2$, 
\begin{equation} 
{\cal E}_{2k} (\tau) = \sum_{n=0}^{\infty} a_n \, q^n \;\;\;,\;\;\; q = e^{2 \pi i \tau} \;.
\label{fqeis}
\end{equation} 
Then  \cite{BGHZ}
\begin{equation}
|a_n| \leq C \, n^{2k-1} \;\;\; \forall \, n \in \mathbb{N}
\label{growthbm}
\end{equation}
for some $C \in \mathbb{R}$ with $C > 0$. 

\begin{proof}

For $k \geq 2$, 
using \cite{BGHZ}
\begin{equation}
| {\tilde \sigma}_{2k-1} (n) | \leq \sum_{d|n} \, d^{2k-1} \leq n^{2k-1} \, \zeta(2k-1) \:,
\end{equation}
we infer
\begin{equation}
|{\cal E}_{2k} (\tau)| \leq 1 + C \sum_{n=1}^{\infty}  n^{2k-1} \, |q|^n \;,
\label{eisengam02}
\end{equation}
with $C = |4k \, \zeta(2k-1)/(1 - 2^{2k} ) B_{2k}|$. Hence $|a_n| \leq C \, n^{2k-1}$ for $n \in \mathbb{N}$.

\end{proof}

\noindent
{\bf Proposition:}  Consider the modular form $\tilde{\cal E}_2$ of weight
$2$ given in \eqref{cE2qtil},
\begin{equation} 
\tilde{\cal E}_2(\tau) = \sum_{n=0}^{\infty} a_n \, q^n \;\;\;,\;\;\; q = e^{2 \pi i \tau} \;.
\label{qexpE2}
\end{equation} 
Then  
\begin{equation}
|a_n| \leq C \, n \;\;\; \forall \, n \in \mathbb{N}
\end{equation}
for some $C \in \mathbb{R}$ with $C > 0$. 

\begin{proof}

We consider $\tilde{\cal E}^2_2$ and use \eqref{theta2eisen},
\begin{equation}
 {\tilde {\cal E}}_2^2 (\tau) = {\cal E}_{4} (\tau) 
+ \frac{8}{30} \left( E_4 (\tau) - E_4 (2 \tau) \right) \;.
\end{equation}
The right hand side is a linear combination of 
modular forms of weight $4$. Each of them  has
$q$-expansion coefficients $a_n$ that exhibit the property $|a_n| \leq C \, n^ 3$, c.f. \eqref{growthbm}.
Hence, if we denote the $q$-expansion of $\tilde{\cal E}_2^2$ by $\tilde{\cal E}_2^2 = \sum_{N \geq  0} c_N q^N$, the coefficients $c_N$ will 
also satisfy $|c_N| \leq D\, N^ 3$, 
for some constant  
$D > 0$.  Now consider \eqref{qexpE2}, and let us assume
that its coefficients $a_n$ satisfy the bound  $|a_n| \leq A \, n^ p$, with $p \in \mathbb{N}$, $A>0$. Then
$|\tilde{\cal E}_2|^2  \leq  \left( 1 + A \, \sum_{m\geq 1} m^p |q|^m \right)  \left( 1 + A \,  \sum_{n\geq 1} n^{p} |q|^n \right) $.
Using absolute convergence, we obtain, for $N >0$, $|c_N| \leq  \left(2A \, N^{p} +A^2 \, 
 \sum_{m + n = N, \; m, n \geq 1} \; m^p \, n^p \right) $. We now place an upper bound on the sum, for a given $N$, as follows. Extremizing $m^p \, n^p = (N-n)^p  \, n^{p}$ with respect to $n$
gives $n_* = N/2$, in which case $m^p_* \, n_*^p = N^{2p}/2^{2p} $.
Using that the number of partitions of $N$ into two positive integers $m$ and $n$ with $m + n =N$ is
$N-1$, we obtain the bound $\sum_{m + n = N, \; m, n \geq 1}
 \; m^p \, n^p \leq m^p_* \, n_*^p \, 
(N-1)  \leq B \, N^{2p+1}$, for some constant  $B > 0$. 
Hence, $|c_N| \leq D \, N^{2p+1}$,  for some constant  
$D > 0$. On the other hand, we had already concluded that 
$|c_N| \leq D\, N^ 3$, so that $p=1$.

\end{proof}

\noindent
{\bf Proposition:} Let $\Gamma = \Gamma_0(2)$. Consider the 
modular form $I_k$ of weight $2k$
given in \eqref{Ikeps2eps4}, with $k \geq 2$. We express its $q$-expansion
as
\begin{equation} 
I_k(\tau) = \sum_{n=0}^{\infty} a_n \, q^n \;\;\;,\;\;\; q = e^{2 \pi i \tau} \;.
\end{equation}
Then
\begin{equation}
|a_n| \leq C \, n^{2k-1} \;\;\; \forall \, n \in \mathbb{N}
\label{growthIn}
\end{equation}
for some $C \in \mathbb{R}$ with $C > 0$.

\begin{proof}

The claim holds for $k=2$, since $I_2 \propto {\cal E}_4$.
Let us consider the case $k > 2$.

We begin 
by considering a summand ${\cal E}_4^2$ of weight $8$ in \eqref{Ikeps2eps4}. Then, using
\eqref{growthbm}, we obtain
$|{\cal E}_4^2|  \leq  \left( 1 + a \, \sum_{m\geq 1} m^3 |q|^m \right)  \left( 1 + a \,  \sum_{n\geq 1} n^{3} |q|^n \right) $ for some constant $a > 0$.
Denoting the $q$-expansion of ${\cal E}_4^2$ by ${\cal E}_4^2 = \sum_{N \geq  0}
c_N q^N$, and following the same steps as in the proof given above, we obtain the
bound
 $|c_N | \leq D \, N^{8-1}$,  for some constant  
$D > 0$.

Next, consider a summand ${\cal E}_4^p$ of weight $4p$ in \eqref{Ikeps2eps4}. We proceed by induction.  
Denoting the $q$-expansion of ${\cal E}_4^p$ by ${\cal E}_4^p = \sum_{n \geq  0}
b(n) q^n$, and assuming $|b(n)| \leq b \, n^{4p-1}$,  for some constant  
$b > 0$, we proceed to show that the expansion coefficients of ${\cal E}_4^{p+1}$ exhibit the growth property
$|c_N| \leq D \, N^{4(p+1)-1}$. 
Namely, proceeding as before, we obtain $|{\cal E}_4 \,  {\cal E}_4^p|  \leq  \left( 1 + a \, \sum_{m\geq 1} m^3 |q|^m \right)  \left( 1 + b \,  \sum_{n\geq 1} n^{4p-1} |q|^n \right) $ for some constants $a, b > 0$. Extremizing $(N-n)^3 \, n^{4p-1}$ with respect to $n$ gives $n_* = N (4p-1)/(4p+2)$,
in which case $(N-n)_*^3 \, n_*^{4p-1} = N^{4p + 2} \,  3^3 (4p-1)^{4p-1}/(4 p+2)^{4p+2}$. Hence,
we obtain the bound $\sum_{m + n = N, \; m, n \geq 1} \; m^3 \, n^{4p-1} \leq m^3_* \, n_*^{4p-1} \, 
(N-1)  \leq B \, N^{4p +3}$ for some constant $B>0$. Thus, the expansion coefficients of ${\cal E}_4^{p+1}$ exhibit the growth property
$|c_N| \leq D \, N^{4(p+1)-1}$,
 as we
wanted to show. Note that this  growth goes as $N^{2k-1}$ with $2k = 4 (p+1)$.

Proceeding in a similar manner, one finds that 
the expansion coefficients of ${\tilde {\cal E}}_2 \, {\cal E}_4^{p}$ exhibit the growth property
$|c_N| \leq D \, N^{4p+1}$.  By induction, one then shows that  the expansion coefficients of ${\tilde {\cal E}}_2^l \, {\cal E}_4^{p}$ exhibit the growth property
$|c_N| \leq D \, N^{4p+2l -1}$. 
This  growth goes as $N^{2k-1}$ with $2k = 4p + 2l$.

Hence we conclude that expansion coefficients  of $I_k$ in \eqref{Ikeps2eps4}, with $k \geq 2$,
exhibit the growth property  \eqref{growthIn}.

\end{proof}

\section{Rankin-Cohen brackets \label{sec-rkb}}

\noindent
{\bf Definition:}  Let $n \in \mathbb{N}_0$.
The $n$th Rankin-Cohen bracket  is a bilinear, differential operator
that acts on modular forms $f, g$ of $SL(2, \mathbb{Z})$, of weight $k \in \mathbb{N}$ and 
$l \in \mathbb{N}$, respectively, by \cite{zagierrc} 
\begin{equation}
[ f, g]_n (\tau) = \frac{1}{(2 \pi i)^n} \, \sum_{r =0}^n (-)^r \, \begin{pmatrix}
k + n -1\\
n-r
\end{pmatrix} \, \begin{pmatrix}
l+ n -1\\
r
\end{pmatrix} \, 
f^{(r)}  (\tau) \, g^{(n-r)}  (\tau) \;,
\label{rcmod}
\end{equation}
where $f^{(r)} (\tau)$ denotes the $r$th derivative of $f$ with respect to $\tau$, and similarly for $g^{(n-r)} (\tau)$. 

In the following, we drop the normalization
factor $\frac{1}{(2 \pi i)^n} $. 
\\

\noindent
{\bf Proposition:} 
The $n$th Rankin-Cohen 
bracket $[ f, g]_n$ is a modular form for $SL(2, \mathbb{Z})$,
of weight $k + l + 2n$ \cite{zagierrc}.  \\

\noindent
{\bf Example:} 
Let $f(S)$ be a modular form for $SL(2, \mathbb{Z})$ of weight $k$, and  let $g(S) = \vartheta_2^8 (S)$, which
has weight $l=4$.
Then,
\begin{equation}
\frac{1}{g} \, [ f, g]_1 = -4 {\cal D}_S f \;,
\end{equation}
where (c.f. \eqref{calDf2k})
\begin{equation}
{\cal D}_S f = \left( \partial_S - \frac{k}{4} \, \partial_S \ln g \right) f \:.
\end{equation}

In particular, consider the case $f(S) = I_2 (S)$, which we recall is proportional to the Eisenstein series ${\cal E}_4$ of weight $4$, c.f.
\eqref{IcalE}. Then,
\begin{eqnarray}
\frac{1}{g} \, [ I_2, g]_1 &=& -4 I_3 \;, \nonumber\\
\frac{1}{g} \, [ I_2, g]_2 &=& 10 \left( I_4- \frac{4}{\gamma} I_2^2 \right) \;, \nonumber\\
\frac{1}{g} \, [ I_2, g]_3 &=& -20 \left( I_5 - \frac{16}{\gamma} I_2 \, I_3 \right) \:.
\end{eqnarray}

For $n >2$ we obtain 
\begin{equation}
I_n = - \frac{1}{4g} \, [I_{n-1}, g]_1 = \left( \frac{-1}{4} \right)^2 \,  \frac{1}{g} \, [\frac{1}{g} [I_{n-2}, g]_1, g]_1
= \left( \frac{-1}{4} \right)^{n-2} \,  \frac{1}{g} \, [\frac{1}{g}  [  \dots [ \frac{1}{g} [I_2, g]_1, \dots, g]_1, g]_1 \;.
\label{InRC}
\end{equation}
\\

The above definition can be extended to include quasi-modular forms as well
\cite{martin}. Here we focus on quasi-modular forms of weight $k$ and depth $s=1$ of $SL(2, \mathbb{Z})$.
An example thereof is provided by \eqref{G2quasi}. \\

\noindent
{\bf Definition:} 
Let $f$ be a quasi-modular form of weight $k$ and depth $s=1$,
 and let $g$ be a modular form of weight $l$, respectively, for $SL(2, \mathbb{Z})$. Then, their $n$th Rankin-Cohen
bracket  is given by
\begin{equation}
[ f, g]_n (\tau) = \frac{1}{(2 \pi i)^n} \, \sum_{r =0}^n (-)^r \, \begin{pmatrix}
k -s + n -1\\
n-r
\end{pmatrix} \, \begin{pmatrix}
l+ n -1\\
r
\end{pmatrix} \, 
f^{(r)} (\tau) \, g^{(n-r)} (\tau) \;,
\end{equation}
where $f^{(r)} (\tau)$ denotes the $r$th derivative of $f$ with respect to $\tau$, and similarly for $g^{(n-r)} (\tau)$. 

In the following, we drop the normalization
factor $\frac{1}{(2 \pi i)^n} $.
\\

\noindent
{\bf Proposition:} 
The $n$th Rankin-Cohen
bracket $[ f, g]_n$ is a quasi-modular form for $SL(2, \mathbb{Z})$,
of weight $k + l + 2n$ and depth $\leq s = 1$ \cite{martin}.
\\

\noindent
{\bf Example:} 
As an application, set
\begin{equation}
f(S) = I_1 (S) = \frac{\partial \omega}{\partial S}  \;\;\;,\;\;\; g(S) = \vartheta_2^8 (S) \;,
\end{equation}
with weights $k=2$ and $l =4$, respectively. The depth of $f$ is $s=1$. Then
\begin{equation}
\frac{1}{g} \, [ f, g]_1 =  -4 \left(  \frac{\partial^2 \omega}{\partial S^2} +
\frac{1}{2 \gamma}  \left( \frac{\partial \omega}{\partial S} \right)^2 \right) = -4 I_2(S)
\;,
\label{I1I2rc}
\end{equation}
which yields a modular form of weight $4$ and depth $s=0$.
\\

\noindent
{\it Remark:}
The Rankin-Cohen bracket \eqref{rcmod} endows
the graded vector space $M_*$
of modular forms of $SL(2, \mathbb{Z})$, $M_* \equiv  \-
\oplus_{k \geq 0} \, {\cal M}_k (SL(2, \mathbb{Z}))$,
with 
an infinite set of bilinear operators \cite{zagierrc}, 
\begin{equation}
[ \cdot, \cdot]_n : M_* \times M_* \rightarrow M_{* + * + 2n} \;.
\end{equation}
The $0$th bracket makes $(M_*, [\cdot, \cdot]_0)$ into
a commutative and associative algebra. The $1$st bracket, on the other hand, makes
$(M_*, [\cdot, \cdot]_1)$ into
a graded Lie algebra, since the $1$st bracket, given by
\begin{equation}
[ f, g]_1 = - [g,f]_1 = k \, f \, g' - l \, f'  \, g \;,
\end{equation}
satisfies the Jacobi identity
\begin{equation}
[[f,g]_1, h]_1 + [[g,h]_1, f]_1 + [[h,f]_1, g]_1 =0 \;\;\;,\;\;\; \forall f, g, h \in M_* \;.
\end{equation}
The space $(M_*,  [\cdot, \cdot]_0, [\cdot, \cdot]_1)$ is a Poisson algebra in view of the Leibnitz rule relation
\begin{equation}
[[f,g]_0, h]_1 = [[f,h]_1, g]_0 + [[g,h]_1, f]_0  \;\;\;,\;\;\; \forall f, g, h \in M_* \;.
\label{poissonalg}
\end{equation}

We note the formal analogy between Rankin-Cohen brackets $ [\cdot, \cdot]_1$
and the adjoint representation of the Lie algebra $\mathfrak{g}$ of a linear group $G$. To this end,
let us recall the differential of the exponential map $\exp : \mathfrak{g}
\rightarrow G$ \cite{Stern}. Define a curve $a(t) = \exp(Z(t)) \subset G$ with $Z(t) =  X + t \, Y$ ($X, Y \in \mathfrak{g}$). Then
\begin{equation}
\exp(-X) \, \left(\frac{d}{dt} a(t) \right)_{\vert_{t=0}}= \exp(-X) \, \left(\frac{d}{dt} \exp(X + t \, Y) 
\right)_{ \vert_{t=0}} = \frac{1 - \exp \left(- ad (X) \right)}{ad(X)}
\, Y \, 
\label{diffexplie}
\end{equation}
where 
\begin{equation}
\frac{1 - \exp \left(- ad (X) \right)}{ad(X)}
\equiv
\sum_{k=0}^{\infty} \frac{ (-1)^k}{(k+1)!} \left(ad (X) \right)^k  \;,
\label{adk}
\end{equation}
and
\begin{eqnarray}
ad(X) : \mathfrak{g} &\rightarrow& \mathfrak{g} \;, \nonumber\\
Y & \mapsto & [X, Y] \;.
\end{eqnarray}

Now consider replacing the derivatives with respect to $\tau$ in \eqref{rcmod}
by the covariant derivative ${\cal D}_{\tau}$ given in \eqref{calDf2k}, to obtain
the $n$th Serre-Rankin-Cohen bracket \cite{cdmr} for $\Gamma_0(2)$
(we drop the normalization
factor $\frac{1}{(2 \pi i)^n} $),
\begin{equation}
SRC_n (f,g) (\tau)  = \sum_{r =0}^n (-)^r \, \begin{pmatrix}
k + n -1\\
n-r
\end{pmatrix} \, \begin{pmatrix}
l+ n -1\\
r
\end{pmatrix} \, 
{\cal D}_{\tau}^r f(\tau) \, {\cal D}_{\tau} ^{n-r} g(\tau) \;.
\label{rcmodcov}
\end{equation}
Then, the sequence $(SRC_n)_{n \in \mathbb{Z} \geq 0}$ of bilinear maps defines a 
 formal deformation 
\cite{cdmr}
of 
$M_* \equiv  \-
\oplus_{k \geq 0} \, {\cal M}_k (\Gamma_0(2) )$ through the non-commutative  Eholzer product $ \#$,
\begin{equation}
f \# g = \sum_{n \geq 0} SRC_n (f,g) \, \hbar^n \;,
\label{ehol}
\end{equation}
 which is associative \cite{zagierrc}, i.e. given any
modular forms $f, g, p$ (of weight $k, l, h \in \mathbb{N}$, respectively),
\begin{equation}
\sum_{r =0}^n SRC_{n-r} \left( SRC_r (f,g), p \right) = \sum_{r =0}^n SRC_{n-r} \left( f, SRC_r (g,p) \right) \;\;\;,\;\;\;  \forall n \in \mathbb{N}_0 \;.
\end{equation}
Note that the Eholzer product \eqref{ehol} is suggestive of a deformation quantization of the Poisson algebra \eqref{poissonalg}.

\section{Jacobi forms \label{sec:jacfor}}

In this section, we follow \cite{EiZag,Dabholkar:2012nd}.

Let $ \cal{H}$ denote the complex upper half plane. Let  $\tau \in \cal{H}$  and $z \in \mathbb{C}$.\\

\noindent
{\bf Definition:}
A 
Jacobi form of $SL(2, \mathbb{Z})$ is a holomorphic function
$\phi_{k,m}: \cal{H} \times \mathbb{C} \rightarrow \mathbb{C}$
that transforms as follows under the modular group $SL(2, \mathbb{Z})$,
\begin{equation}
\phi_{k,m} (\frac{a\tau+b}{c\tau+d} , \frac{z}{c\tau + d}) = (c\tau + d)^k \, e^{2 \pi i  \frac{m c z^2}{c\tau + d}} \;
\phi_{k,m}(\tau,z)  \;\;\;,\;\;\ \forall \;\;\; 
\begin{pmatrix}
a & b \\
c & d\\
\end{pmatrix}
\in SL(2, \mathbb{Z}) \;,
\label{phikmtransf}
\end{equation}
and under translations of $z$ by $\mathbb{Z} \tau + \mathbb{Z}$ as
\begin{equation}
\phi_{k,m}(\tau, z + \lambda \tau + \mu) = e^{- 2 \pi i m (\lambda^2 \tau + 2 \lambda z)} \;
 \phi_{k,m} (\tau,z) 
\;\;\;,\;\;\ \forall \;\;\; \lambda, \mu \in \mathbb{Z} \;.
\label{phikmelli}
\end{equation}
	Here, $k \in \mathbb{Z}$ is called the weight, and $m \in \mathbb{N}$
is called the index of the Jacobi form.  \\

\noindent
{\it Remark:}
$\tau$ is called modular parameter, while $z$ is called elliptic parameter.
\\

Due to the periodicities $\phi_{k,m} (\tau + 1, z) = \phi_{k,m} (\tau,z)$ and $ \phi_{k,m}(\tau, z+1) =
\phi_{k,m} (\tau,z)$, $\phi_{k,m}$ possesses a Fourier expansion,
\begin{equation}
\phi_{k,m} (\tau,z) = \sum_{n,r \in \mathbb{Z}} c(n,r) \, q^n \, y^r \;\;\;,\;\;\;
q = e^{2 \pi i \, \tau} \;\;\;,\;\; y = e^{2 \pi i \, z} \;.
\end{equation}
The transformation behaviour under elliptic transformations \eqref{phikmelli} implies
that
\begin{equation}
c(n,r) = C(\Delta, r) \;\;\;,\;\;\; \Delta \equiv  4 n m - r^2 ,
\end{equation}
where $C(\Delta, r) $ depends only on $ r \mod 2m$. 
\\

\noindent
{\bf Definition:}
$\phi_{k,m} $ is called a holomorphic Jacobi form of weight $k$ and index $m$ if
\begin{equation}
c(n, r) = 0 \;\;\;,\;\; {\rm for}\; \Delta < 0 \;,
\end{equation}
i.e.
\begin{equation}
\phi_{k,m} (\tau,z) = \sum_{n \geq 0,\; r \in \mathbb{Z}, \; 4 n m \geq r^2} c(n,r) \, q^n \, y^r \;.
\end{equation}
\\

\noindent
{\bf Definition:} 
$\phi_{k,m} $ is called a cuspidal holomorphic Jacobi form if
\begin{equation}
c(n, r) = 0 \;\;\;,\;\; {\rm for}\; \Delta \leq 0 \;,
\end{equation}
i.e.
\begin{equation}
\phi_{k,m} (\tau,z) = \sum_{n \geq 1,\; r \in \mathbb{Z}, \; 4 n m > r^2} c(n,r) \, q^n \, y^r \;.
\end{equation}
\\

\noindent
{\bf Definition:}
$\phi_{k,m} $ is called a weak holomorphic Jacobi form if
\begin{equation}
c(n, r) = 0 \;\;\;,\;\; {\rm for} \;  n < 0 \;,
\end{equation}
i.e. if 
\begin{equation}
\phi_{k,m} (\tau,z) = \sum_{n \geq 0,\; r \in \mathbb{Z}} c(n,r) \, q^n \, y^r \;.
\end{equation}
\\

\noindent
{\bf Definition:} 
$\phi_{k,m} $ is called a weakly holomorphic Jacobi form if 
\begin{equation}
c(n, r) = 0 \;\;\;,\;\; {\rm for} \;  n < - n_0 , \;, n_0 \in \mathbb{N} \;,
\end{equation}
i.e. if 
\begin{equation}
\phi_{k,m} (\tau,z) = \sum_{n \geq - n_0,\; r \in \mathbb{Z}} c(n,r) \, q^n \, y^r \;.
\end{equation}
\\

Depending on the nature of the Jacobi form, the coefficients $c(n,r)$ will have a certain growth
property \cite{Dabholkar:2012nd}.
\\

\noindent
{\it Remark:} The above generalizes to Jacobi forms of subgroups $\Gamma \subset SL(2, \mathbb{Z})$
\cite{EiZag}.
\\

\noindent
{\bf Example:}  Consider $\vartheta_2^8 (\tau,z)$, where  $\vartheta_2(\tau, z)$ denotes 
the Jacobi theta function defined by
\begin{equation}
\vartheta_2 (\tau,z) = q^{1/8} \, \sum_{n \in \mathbb{Z}} \, q^{\tfrac12 n(n+1)} \, e^{i \pi (2 n + 1) z} \;,
\end{equation}
which has the  following product representation, valid for $|q |<1$,
\begin{equation}
\vartheta_2 (\tau,z) = 2 \, q^{1/8} \, \cos \left( \pi z \right) \, \prod_{n=1}^{\infty}  (1-q^n) (1 + q^n \, y) \;
(1 + q^n \, y^{-1}) \;.
\end{equation}
For fixed $z$, $\vartheta_2 (\tau,z)$ has one branch point inside the unit circle of the complex $q$-plane, at $q=0$,
and a branch cut along the interval $(-1,0)$ on the real axis.
$\vartheta_2 (\tau,z)$
has zeroes at
 $z= \tfrac12 (2 m+1) + n \, \tau$ with $(m,n) \in \mathbb{Z}^2$.

$\vartheta_2^8 (\tau,z)$, on the other hand, is analytic in both $\tau \in {\cal H}$  and $z \in \mathbb{C}$,
and is an example of a modular form of weight $k=4$ and index $m=4$ 
with trivial multiplyer system under $\Gamma_0 (2)$-transformations.
\\

\noindent
{\it Remark:} $\vartheta_2$ is even with respect to $z$, 
\begin{equation}
\vartheta_2(\tau,- z) = \vartheta_2 (\tau,z) \;,
\label{theta2even}
\end{equation}
and it solves the heat equation
\begin{equation}
\frac{\partial}{\partial S} \vartheta_2(S, z) = \frac{1}{4 \pi} \, \frac{\partial^2}{\partial z^2} \vartheta_2(S, z) \;,
\label{heatvartheta2}
\end{equation}
with periodic boundary conditions ($z \rightarrow z + 1$) imposed in the $z$ direction. Here $\tau = i S$.

Also observe that when taking an odd number of derivatives with respect to $z$, we have
\begin{equation}
\frac{\partial^n}{\partial z^n} \vartheta_2 (S, z)\bigg\rvert_{z=0}=0\quad \forall n \in 2 \mathbb{N} + 1 
\label{odddtheta2}
\end{equation}
due to \eqref{theta2even}.
\\

\noindent
{\it Remark:}   Consider $f(S, z) = \ln \vartheta_2^8 (S, z)$.
We fix $S$,  and we view $\vartheta_2^8  (S, z)$ as a function of $z$, which is non-vanishing in a small open neighbourhood of $z=0$.
It is then possible to define a single-valued analytic branch of $\ln \vartheta_2^8 (S, z)$ in that neighbourhood.
We then Taylor expand around $z=0$,
retaining only the first few terms for illustrative purposes,
\begin{equation}
f(S,z)=  \ln \vartheta_2^8 (S)
+\frac{z^2}{2!}\partial_z^2f\rvert_{z=0}+\frac{z^4}{4!}\partial_z^4 f\rvert_{z=0} 
+\frac{z^6}{6!}\partial_z^6 f\rvert_{z=0} 
+\frac{z^8}{8!}\partial_z^8 f\rvert_{z=0} +
 {\cal O}(z^{10}) \;.
 \label{tayf}
\end{equation}
Using \eqref{theta2even} and \eqref{odddtheta2}  to convert an even number of derivatives with respect to $z$
into derivatives with respect to $S$, we obtain 
\begin{eqnarray}
f(S, z) &=& - \frac{2}{\gamma} \left( \omega(S) 
+ 4 \pi \,  \frac{z^2}{2 !} \, I_1 (S) + (4 \pi)^2 \,  \frac{z^4}{4 !} I_2 (S) + 
 (4 \pi)^3 \, \frac{z^6}{6 !} I_3 (S)  \right. \nonumber\\
&& \left.   \qquad + 
 (4 \pi)^4 \, \frac{z^8}{8 !} \left( I_4 (S) + \frac{5}{\gamma}  \, I_2^2 (S) \right)
 +
 {\cal O} (z^8)
\right) \;,
\label{fzexp}
\end{eqnarray}
where
\begin{equation}
\omega(S) = - \frac{\gamma}{2} \, \ln \vartheta_2^8 (S) \;\;\;,\;\;\; \gamma = - \frac{1}{256 \pi} \;,
\end{equation}
and where the $I_n$ are defined in \eqref{I1I2} and in \eqref{eq:DI-I}.

Thus, the expansion \eqref{fzexp} is indicative of an expansion in powers of $I_n(S)$, as follows:
\\

\noindent
{\bf Proposition:} 
\begin{eqnarray}
f(S, z) 
&=& - \frac{2}{ \gamma} \left[
\omega(S) + 4 \pi \,  \frac{z^2}{2 !} \, I_1 (S)
+
\sum_{n=2}^{\infty} (4 \pi)^n \, 
\frac{z^{2n}}{(2n)!} \, I_n (S) \right. \nonumber\\
&& \left. \qquad 
+
\sum_{m, n \geq 2} (4 \pi)^{m+n} \,f_{m,n} \,  
z^{2(m+n)} \, I_m(S) \,  I_n (S) + \dots
 \right] \;,
 \label{fImonbin}
\end{eqnarray}
where the dots stand for higher powers of $I_n$. 

\begin{proof}

 Let us first focus on the sector involving
single powers of $I_n$ only, and let us verify the expression for the monomials in $I_n$ given in 
\eqref{fImonbin}. Let us work in an open neighbourhood of $z = 0$.
First, observe that under $\Gamma_0(2)$-transformations \eqref{omSGam02}, the first two terms in \eqref{fImonbin}, 
$\omega(S) + 4 \pi \,  \frac{z^2}{2 !} \, I_1 (S)$,  transform precisely as $f(S,z)$.
Therefore, the Taylor series in \eqref{tayf} starting with the term $z^4$
has to be invariant under  $\Gamma_0(2)$-transformations.
Since each power of $z$ transforms as $z \mapsto z/  [\Delta (S)]$ under $\Gamma_0(2)$, it follows
that each summand in \eqref{tayf} starting with the term $z^4$ must be separately invariant under  $\Gamma_0(2)$, i. e.
$\partial^{2n}_z f\vert_{z=0}  \mapsto [ \Delta(S)]^{2n} \, \partial^{2n}_z f\vert_{z=0} $ for $n \geq 2$.
Using
\begin{equation} 
\partial^{2n}_z f\vert_{z=0} = 8 \, \frac{\partial_z^{2n} \vartheta_2 (S,z)}{\vartheta_2 (S, z)} \vert_{z=0} + \dots \;,
\end{equation}
where the dots stand for products of terms involving a lower number of $z$-derivatives of $\vartheta_2(S,z)$, 
 we infer, using  \eqref{heatvartheta2},
\begin{equation} 
\partial^{2n}_z f\vert_{z=0} = 8 \, (4 \pi)^n \,
 \frac{\partial_S^{n} \vartheta_2 (S)}{\vartheta_2 (S)} + \dots  =  (4 \pi)^n \,
\partial_S^n \ln \vartheta_2^8(S) + \cdots = - \frac{2 (4 \pi)^n }{\gamma} \, I_n (S) + \dots \;.
\label{der-2nIn}
\end{equation}
Covariance under $\Gamma_0(2)$-transformations dictates that the terms involving a lower number
of $z$-derivatives of $\vartheta_2(S,z)$ have to organize themselves into products of $I_k(S)$, 
so that 
the dots in \eqref{der-2nIn} stand  for higher powers of $I_k (S)$, such that under $\Gamma_0(2)$-transformations,
$\partial^{2n}_z f\vert_{z=0}  \mapsto [ \Delta(S)]^{2n} \, \partial^{2n}_z f\vert_{z=0} $. Thus we establish 
\eqref{fImonbin}.

\end{proof}

Another way of obtaining the same result is as follows.
We note that $f(S, z) = \ln \vartheta_2^8 (S,z) $, 
satisfies 
the non-linear PDE
\begin{equation}
\label{derivativerelation}
32 \pi \, \frac{\partial}{\partial S}f(S,z) = 
8 \, \frac{\partial^2}{\partial z^2}f(S,z) + \left(\frac{\partial}{\partial z}f(S,z)\right)^2,
\end{equation}
as  a consequence of \eqref{heatvartheta2}, with boundary conditions $f(S, 0) = \ln \vartheta_2^8(S), \; 
\partial_z  f \vert_{z =0} =0, \; 
\partial^2_z  f \vert_{z =0} = - 8 \pi \, I_1 (S)/\gamma$.
 Let us verify that \eqref{fImonbin} solves
\eqref{derivativerelation} when restricting to the monomial sector in \eqref{fImonbin}.
To distinguish between the various sectors, we rescale $\omega$ and each power of $I_n$ in \eqref{fImonbin} by a real constant $\lambda \in 
\mathbb{R}$, in which case  \eqref{fImonbin} becomes
\begin{eqnarray}
f(S, z, \lambda) 
= - \frac{2}{ \gamma} \left[\lambda \, 
\omega(S) + \lambda \, 4 \pi \,  \frac{z^2}{2 !} \, I_1 (S)
+ \lambda
\sum_{n=2}^{\infty} (4 \pi)^n \, 
\frac{z^{2n}}{(2n)!} \, I_n (S) + {\cal O} (\lambda^2) 
 \right] .
 \label{fImonbin2}
\end{eqnarray}
Inserting \eqref{fImonbin2} into \eqref{derivativerelation} 
(we verify below that we may differentiate the series 
\eqref{fImonbin2} term by term)
 and working to first
order in $\lambda$, we establish that \eqref{fImonbin2} solves \eqref{derivativerelation}  
at this order. Then, setting $\lambda =1$ yields the expression for the monomials given in 
\eqref{fImonbin}.

The uniqueness of solutions of the form \eqref{fImonbin2} can be established as follows.
Suppose that the differential equation \eqref{derivativerelation} admits a solution $f(S, z)$ that is expressed as a power series expansions in $z^2$,
\begin{eqnarray}
f(S, z) = \sum_{n\geq 0} f_n(S)z^{2n}\,,
\label{fzrec}
\end{eqnarray}
satisfies the boundary conditions given above, and that can be differentiated term by term. Then, inspection of
\eqref{derivativerelation} shows that 
$f_2(S) = - 4 \pi^2 I_2 (S)/(3 \gamma)$, $f_3 (S) = 2 \pi {\cal D}_S f_2(S)/15$, and that the functions $f_{n+1} (S)$ with $n \geq 3$ are determined recursively by
\begin{eqnarray}
(2 n + 2) (2n+1) \, f_{n+1}(S) = c_1 {\cal D}_S f_n (S) +  c_2 \, \sum_{k+l=n+1; \, k, l > 1}k \, l \, f_k(S)f_l(S)\,,
\end{eqnarray}
where $c_1, c_2$ are numerical constants. Hence, a solution of the form \eqref{fzrec} satisfying the conditions mentioned above is uniquely specified.

Thus, we have the following corollary:
\\

\noindent
{\bf Corollary:}  Let
\begin{equation}
g(S,z) \equiv 
\sum_{n=2}^{\infty} (4 \pi)^n \, 
\frac{z^{2n}}{(2n)!} \, I_n (S) \;.
\label{g2n}
\end{equation}
$g(S,z)$ is the unique $\Gamma_0(2)$-invariant solution, 
that is given by a Taylor series in even powers of $z$, 
to the PDE
\begin{equation}
4 \pi \,  \frac{\partial}{\partial S}g(S,z) = 
\frac{\partial^2}{\partial z^2}g(S,z) - \frac{2 \pi }{\gamma} \,  I_1(S) \,z  \frac{\partial }{\partial z } g(S,z) - \frac{(4 \pi)^2}{2} \, z^2 \, I_2(S) \;,
\label{pdeg}
\end{equation}
with boundary conditions $g \vert_{z =0} = 
\partial_z g \vert_{z =0} =  \partial^2_z g \vert_{z =0}  =  \partial^3_z g \vert_{z =0}  = 0$. 

\begin{proof}
Let 
\begin{equation}
g(S,z) \equiv 
\sum_{n=2}^{\infty} (4 \pi)^n \, 
\frac{z^{2n}}{(2n)!} \,  p_n (S) \;.
\label{p2n}
\end{equation}
Imposing $\Gamma_0(2)$ invariance shows that $p_n(S)$
are modular forms for $\Gamma_0(2)$ of weight $2n$.
Inserting \eqref{p2n} into \eqref{pdeg}
and differentiating term by term, which will be justified below, 
one infers $p_2 (S) = I_2 (S) $ and $p_{n+1} (S) = {\cal D}_S p_n (S)$ for $n \geq 2$.
Hence, $p_n = I_n$.

\end{proof}

In the main part of the paper, we encountered a closely related series, namely (c.f. \eqref{Inexp})
\begin{equation}
H(S, z) \equiv 
\sum_{n=2}^{\infty}    c^n \, \frac{z^{2n} }{n!} 
\, I_n(S) \;,
\label{sumnIn}
\end{equation}
with $c \in \mathbb{C}$ a constant. Note that the suppression factor is now $1/n!$, instead of $1/(2n)!$ in \eqref{g2n}.
The PDE which $H(S, z)$ satisfies will thus be different from \eqref{pdeg}, c.f. \eqref{PDEXi}.
For fixed $S$, this series is a power series in $z$, and hence
both absolutely convergent in a open neighbourhood $D(0, R)$ of $z=0$ as well as 
uniformly convergent on any compact subset contained in $D(0, R)$. The radius of convergence $R$ will depend on $S$.
We can  infer a lower bound on $R$, as follows. First, using \eqref{thet2}, we infer that for large ${\rm Re} \, S$
( i.e. $ |q| \rightarrow 0$), $\omega(S)$ behaves as $\ln q \sim S$,
and hence, $\partial \omega / \partial S$ is constant, up to exponentially suppressed corrections. 
Thus, at $q=0$, $I_k $ (with $k \geq 2$) is given by
\begin{equation}
I_k (q=0) = \frac{(k-1)!}{2 \gamma^{k-1}} \left( \frac{ \partial \omega}{\partial S} \right)^k \vert_{q=0} \;.
\end{equation}
On the other hand, evaluating \eqref{Ikeps2eps4} at $q=0$, we infer
\begin{equation}
\sum_{m + 2n = k, \, m \geq 0, n \geq 1} a_{m,n} = \frac{(k-1)!}{2 \gamma^{k}} \left( \frac{ \partial \omega}{\partial S} \right)^k \vert_{q=0} 
= \frac{(k-1)!}{2} \, \pi^k > 0
\;.
\end{equation}
Using \eqref{cE2qtil} and 
 \eqref{eisengam02}, we obtain (with $|q| < 1$)
\begin{eqnarray}
| {\tilde{\cal E}}_2 | & \leq & 1 + C_2 \, \sum_{n=1}^{\infty} n^2 \, |q|^n = 1 + C_2 \, \left( | q | \frac{d}{d |q|} \right)^2 \frac{|q|}{1 -|q|} \equiv f_2 (|q|)
\;\;\;,\;\;\; C_2 = 24 \;, \nonumber\\
| {\cal E}_4 | & \leq & 1 + C_4 \, \sum_{n=1}^{\infty} n^3 \, |q|^n = 1 + C_4 \, \left( | q | \frac{d}{d |q|} \right)^3 \frac{|q|}{1 -|q|}  \equiv f_4 (|q|)
\;\;\;,\;\;\; C_4 = 80 \, \zeta(3)  \;, \nonumber\\
\end{eqnarray}
and hence, using \eqref{Ikeps2eps4},
\begin{eqnarray}
|I_k (S) | \leq | \gamma| \,  \left| f^k_2 (|q|) \right| \;  \sum_{m + 2n = k, \, m \geq 0, n \geq 1}  a_{m,n}  \, | h^n(|q|) | \;,
\label{modIk}
\end{eqnarray}
where we recall that $a_{m,n}$ are positive, and where
\begin{equation}
h(|q|) \equiv \frac{f_4(|q|)}{f^2_2(|q|) } \;.
\end{equation}
If, for a fixed $q$, $|h(|q|)| \leq 1$, we may approximate \eqref{modIk} by
\begin{eqnarray}
|I_k (S) | \leq  |\gamma| \,  \left| f^k_2 (|q|) \right| \;  \frac{(k-1)!}{2} \, \pi^k \;, 
\end{eqnarray}
whereas if $|h(|q|)| > 1$, we have the bound $| h^n(|q|) | \leq | h(|q|) |^{[k/2]}$, and
hence we 
approximate \eqref{modIk} by
\begin{eqnarray}
|I_k (S) | \leq  |\gamma| \,  \left| f^k_2 (|q|) \right| \;  | h(|q|) |^{[k/2]} \;  \frac{(k-1)!}{2} \, \pi^k \;.
\end{eqnarray}
Therefore, we obtain the following lower bound for the radius of convergence $R$,
\begin{eqnarray}
\frac{1}{R} = \lim_{k \rightarrow \infty} ( |c|^k \,  | I_k(S)| / k!)^{1/k} \leq  \pi \, |c| \,  | f_2 (q) |  \, 
\left\{ 
\begin{matrix}
1 \;\;\; &,& \;\;\;  |h(|q|)| \leq 1 \\
\sqrt{  | h(|q|) | } \;\;\; &,&\;\;\; |h(|q|)| > 1 \;.
\end{matrix}
\right.
\end{eqnarray}
The functions $f_2 (|q|)$ and $f_4(|q|)$ are monotonically increasing
in the interval $0 \leq |q| < 1$, and they blow up at $|q| = 1$. 
The function $h(|q|)$ is bounded, $0 \leq h(|q|) < 1.5$.
Thus, so long as $|q| \leq r < 1$, we have $R > 0$. 

\begin{figure}
\centering
\begin{subfigure}[b]{0.4\linewidth}
\includegraphics[width=\linewidth] {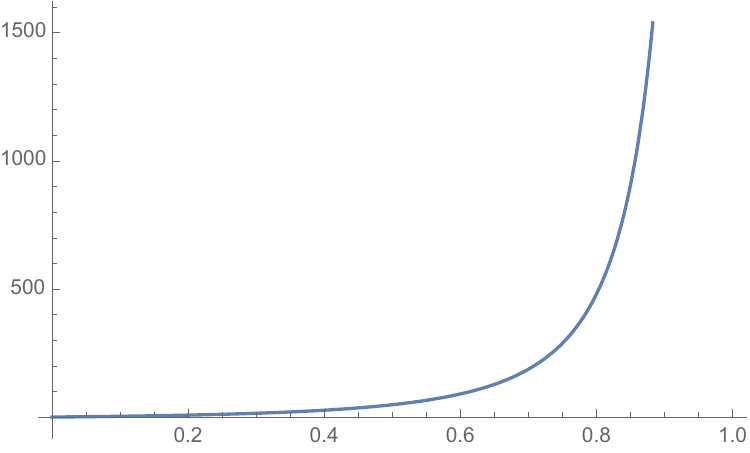}
\caption{Function $f_2(|q|)$, with  $0 \leq |q| < 1$}
\end{subfigure}
\begin{subfigure}[b]{0.4\linewidth}
\includegraphics[width=\linewidth] {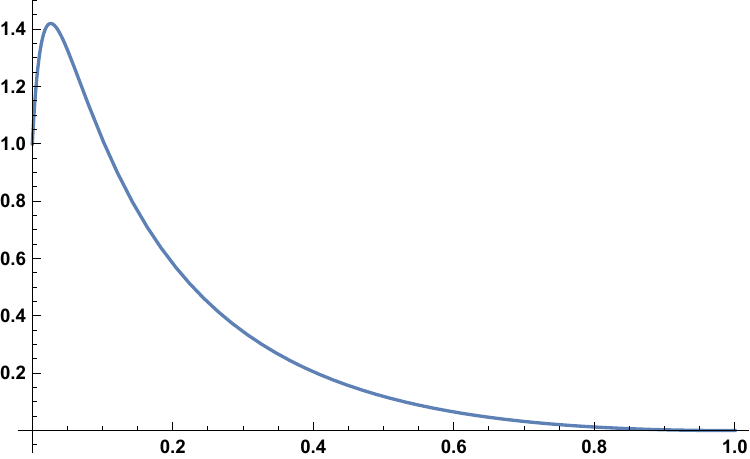}
\caption{Function $h(|q|)$, with  $0 \leq |q| < 1$ }
\end{subfigure}
\caption{Behaviour of the functions $f_2(|q|)$ and $h(|q|)$}
\end{figure}

Next, let us consider fixing $z$ in \eqref{sumnIn}. 
Then, using \eqref{modIk}, we infer
\begin{eqnarray}
|H(S, z)| \leq \tfrac12 |\gamma| \, \sum_{k=2}^{\infty} \frac{|\pi \, c \, z^2 \, f_2 (|q|) |^k}{k}
\left\{ 
\begin{matrix}
1 \;\;\;&,&\;\;\;  |h(|q|)| \leq 1 \\
 | h(|q|) |^{[k/2]} \;\;\; &,&\;\;\; |h(|q|)| > 1 \;.
\end{matrix}
\right.
\end{eqnarray} 
The right hand side is a convergent series for $|q| <1$, provided 
that $ |\pi \, c \, z^2 \, f_2 (|q|) | < 1$ and $ |\pi \, c \, z^2 \, f_2 (|q|) | \sqrt{| h(|q|)| } <1$.
Consider $|q| \leq r < 1$. Then, since $f_2 (|q|)$ and $f_4(|q|)$ are monotonically
increasing functions, we infer
 $ |\pi \, c \, z^2 \, f_2 (|q|) | \leq  |\pi \, c \, z^2 \, f_2 (r) | < 1$
 and 
  $ |\pi \, c \, z^2 \, f_2 (|q|) | \sqrt{| h(|q|)| } \leq
  |\pi \, c \, z^2 \, f_2 (r) | \sqrt{| h(r)| } <1$.
Thus,  by the Weierstrass M test, the function 
$H(S,z)$ converges both absolutely and uniformly for all
$|q| \leq r < 1$, provided $ |\pi \, c \, z^2 | < 1/M$, 
with $M = \max \{ f_2 (r ) ,  f_2 (r)  \sqrt{| h(r)|}  \}$.
In particular, it converges  absolutely and uniformly in an open neighbourhood
of $(q,z) = (0,0)$.

Next, let us consider the series of $S$-derivatives,
\begin{equation}
\sum_{n=2}^{\infty}    c^n \, \frac{z^{2n} }{n!} 
\, \partial_S I_n(S) \;.
\label{dersser}
\end{equation}
Proceeding as above we obtain the bound
\begin{eqnarray}
| \partial_S I_k(S) | \leq  \pi^{k+1}  |\gamma|  \, k! \, v(|q|) \, |f_2(|q|)|^k \, 
\left\{ 
\begin{matrix}
1 \;\;\;&,&\;\;\;  |h(|q|)| \leq 1 \\
 | h(|q|) |^{[k/2]} \;\;\; &,&\;\;\; |h(|q|)| > 1 \;,
\end{matrix}
\right.
\end{eqnarray} 
where the function $v(|q|)$ is given by
\begin{eqnarray}
v (|q|) =  24 f_2^{-1} (|q|) \, \left( | q | \frac{d}{d |q|} \right)^3 \frac{|q|}{1 -|q|} 
+ \tfrac12 C_4 f_4^{-1} (|q|) \, \left( | q | \frac{d}{d |q|} \right)^4 \frac{|q|}{1 -|q|}  \;.
\end{eqnarray}
The function $v(|q|)$ is monotonically increasing
in the interval $0 \leq |q| < 1$ and blows up at $|q| = 1$. 
Hence we obtain the estimate
\begin{eqnarray}
\sum_{k=2}^{\infty}   \vert c^k \, \frac{z^{2k} }{k!} 
\, \partial_S I_k(S) \vert \leq \pi |\gamma| \, v(|q|) \, 
\sum_{k=2}^{\infty}   \vert \pi c \, z^{2} \, f_2(|q|) \vert^k
\left\{ 
\begin{matrix}
1 &,& |h(|q|)| \leq 1 \\
 | h(|q|) |^{[k/2]} &,& |h(|q|)| > 1 \;. 
\end{matrix}
\right.
\end{eqnarray}
Thus, at fixed $z$,  this series converges both absolutely and uniformly for all
$|q| \leq r < 1$, provided $ |\pi \, c \, z^2 | < 1/M$, 
with $M = \max \{ f_2 (r ) ,  f_2 (r)  \sqrt{| h(r)|}  \}$, as above.
In particular, it converges  absolutely and uniformly in an open neighbourhood
of $(q,z) = (0,0)$.

Thus, at fixed $z$ satisfying $ |\pi \, c \, z^2 | < 1/M$, both \eqref{sumnIn} and \eqref{dersser} 
converge absolutely and uniformly for all
$|q| \leq r < 1$, and hence the $S$-derivative of \eqref{sumnIn} equals  \eqref{dersser}.

\begin{figure}
\centering
\includegraphics[width=0.4\linewidth] {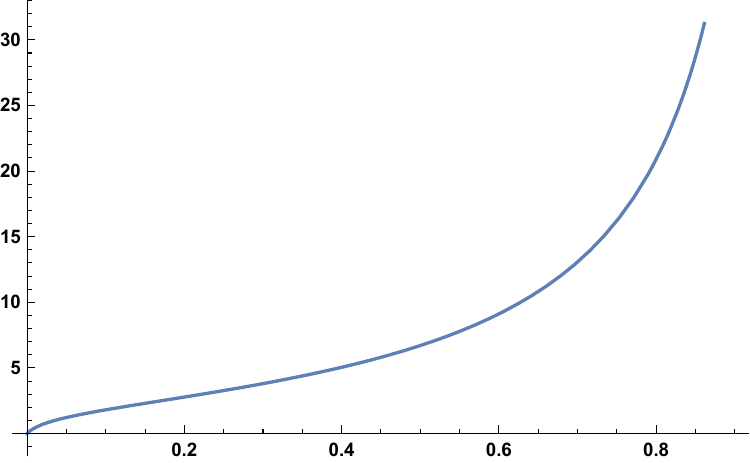}
\caption{Function $v(|q|)$, with  $0 \leq |q| < 1$}
\end{figure}

In the main part of the paper, we also encountered the series  (c.f. \eqref{Inexp})
\begin{equation}
\sum_{m,n\geq 2}  \, c^{m+n} \, \beta_{m,n}  \, z^{2(m+n)} \, I_m(S) \, I_n(S) \;,
\label{binseri}
\end{equation}
with $c \in \mathbb{C}$ a constant. For large $m,n$, the coefficients
    behave as $\beta_{m,n} \propto 1/ [ (m-1)!  \, (n-1)! ] $, and the series
\eqref{binseri} becomes
\begin{equation}
\left( z^2 \partial_{z^2} H (S, z) \right)^2 \;,
\end{equation}
with $H(S,z)$ given by \eqref{sumnIn}. Thus, the series \eqref{binseri}
has the same convergence properties as  $H(S,z)$.

\section{Siegel modular forms \label{sec:siegelH}}

In this section, we follow \cite{BGHZ,EiZag} as well as appendix A of 
\cite{Bossard:2018rlt}.
\\

\noindent
{\bf Definition:}  
Siegel's upper half plane ${\cal H}_2$ is defined by 
\begin{equation}{
\cal H}_2 = \left\{ \Omega \in {\rm Mat}(2 \times 2, \mathbb{C}): \Omega^T = \Omega, \; {\rm Im} \, \Omega > 0 
\right\} \;,
\end{equation} 
where ${\rm Im} \, \Omega > 0 $ means that ${\rm Im} \, \Omega$ is a positive definite matrix,
i. e. ${\cal H}_2$ consists of complex symmetric matrices 
\begin{eqnarray}
\Omega =
\begin{pmatrix}
\rho & v\\
v & \sigma
\end{pmatrix}
\label{Omrvs}
\end{eqnarray}
with $\rho_2 > 0, \; \sigma_2 > 0$ and $\det ( {\rm Im} \Omega) > 0$, i.e. $\rho_2 \, \sigma_2 - v_2^2 >0$,
where $\rho = \rho_1 + i \rho_2, \; \sigma = \sigma_1 + i \sigma_2, \; v=v_1 + i v_2$.
\\

The Siegel modular group  $\mathrm{Sp}(4,\mathbb{Z})$ acts on ${\cal H}_2$ as follows. An element $M \in \mathrm{Sp}(4,\mathbb{Z})$,
\begin{eqnarray}
M = \begin{pmatrix}
A&B \\C& D
\end{pmatrix} \;,
\end{eqnarray}
can be written in terms of four real two-by-two blocks $A$, $B$, $C$, and $D$ satisfying
$  A^T D -C^T B = D\,A^T - C\,B^T
  =\mathbb{I}_2$, \; $  A^T C= C^T A$, \; $B^T D= D^T B $. Then, $M \in \mathrm{Sp}(4,\mathbb{Z})$
  acts on ${\cal H}_2$ by
  \begin{equation}
  \label{eq:SP2}
   \Omega \mapsto \Omega' = (A\,\Omega + B)\,(C \,\Omega  + D)^{-1}\,.
\end{equation}
One then obtains $\Omega-\bar\Omega \mapsto (\Omega\,C^{T}
  +D^{T})^{-1} (\Omega-\bar\Omega)\, (C\,\bar\Omega + D)^{-1}$, which shows that $({\rm Im} \, \Omega)'$ is positive definite.
    \\

A standard fundamental domain for the action of $\mathrm{Sp}(4,\mathbb{Z})$ on ${\cal H}_2$ is
the set defined by \cite{jaber}
\begin{equation}
- \frac12 \leq \rho_1, \sigma_1, v_1 \leq  \frac12 , \;\;\; 0 < 2 v_2 \leq \rho_2 \leq \sigma_2 , \;\;\;
| \det (C \Omega + D) | \geq 1 \:.
\end{equation}
The condition $| \det (C \Omega + D) | \geq 1 $ applies to all $\mathrm{Sp}(4,\mathbb{Z})$ transformations of 
$\Omega$.\\

Next, we discuss the action of various subgroups of $\mathrm{Sp}(4,\mathbb{Z})$ on ${\cal H}_2$:

\begin{enumerate}

\item $SL(2, \mathbb{Z})_{\rho}$ (which leaves $\sigma_2 - (v_2)^2/\rho_2$ invariant): 
\begin{eqnarray}
\label{g1}
g_1(a,b,c,d) = \begin{pmatrix}
A&B \\C& D
\end{pmatrix} = 
\begin{pmatrix}
a & 0 & b & 0\\
0 & 1 &0 & 0\\
c & 0 & d & 0 \\
0 & 0 & 0 & 1 
\end{pmatrix} \;\;\;,\;\;\; (\rho, v, \sigma) \mapsto (\frac{a \rho + b}{c \rho + d}, \frac{v}{c \rho + d},
\sigma - \frac{c v^2}{c \rho + d} ) \;, \nonumber\\
\end{eqnarray}
where $a, b, c, d \in \mathbb{Z}, \;\;\;
ad - bc = 1$.

\item $SL(2, \mathbb{Z})_{\sigma}$ (which leaves $\rho_2 - (v_2)^2/\sigma_2$ invariant): 
\begin{eqnarray}
\begin{pmatrix}
A&B \\C& D
\end{pmatrix} = 
\begin{pmatrix}
1 & 0 & 0 & 0\\
0 & a &0 & b\\
0 & 0 & 1 & 0 \\
0 & c & 0 & d 
\end{pmatrix} \;\;\;,\;\;\; (\rho, v, \sigma) \mapsto (\rho - \frac{ c v^2}{c \sigma + d}, 
 \frac{v}{c \sigma + d},
\frac{a \sigma + b}{c \sigma + d} ) \;,
\end{eqnarray}
where $a, b, c, d \in \mathbb{Z}, \;\;\;
ad - bc = 1$.

\item Heis$_{\rho}$ (which leaves $\Omega - \bar \Omega$ invariant)
\begin{eqnarray}
\begin{pmatrix}
A&B \\C& D
\end{pmatrix} = 
\begin{pmatrix}
1 & 0 & 0 & \mu\\
\lambda & 1 & \mu & \kappa\\
0 & 0 & 1 & - \lambda \\
0 & 0 & 0 & 1 
\end{pmatrix} \;\;\;,\;\;\; (\rho, v, \sigma) \mapsto (\rho, v + \lambda \, \rho + \mu, \sigma + 
2 \lambda \, v + \lambda^2 \, \rho + \lambda \, \mu + \kappa) \:, \nonumber\\
\end{eqnarray}
where $\lambda, \mu, \kappa \in \mathbb{Z}$.

\item Heis$_{\sigma}$ (which leaves $\Omega - \bar \Omega$ invariant)
\begin{eqnarray}
\begin{pmatrix}
A&B \\C& D
\end{pmatrix} = 
\begin{pmatrix}
1 & \lambda & \kappa & \mu\\
0 & 1 & \mu & 0\\
0 & 0 & 1 & 0 \\
0 & 0 & -\lambda & 1 
\end{pmatrix} \;\;\;,\;\;\; (\rho, v, \sigma) \mapsto (\rho
+ 
2 \lambda \, v + \lambda^2 \, \sigma + \lambda \, \mu + \kappa, v + \lambda \, \sigma+ \mu, \sigma) \;,
\nonumber\\
\end{eqnarray}
where $\lambda, \mu, \kappa \in \mathbb{Z}$.

\item $\rho \leftrightarrow \sigma, \;\;\; v \rightarrow - v$:
\begin{eqnarray}
\label{g2}
g_2 = \begin{pmatrix}
A&B \\C& D
\end{pmatrix} = 
\begin{pmatrix}
0  & 1 & 0 & 0\\
-1 & 0 & 0 & 0\\
0 & 0 & 0 & 1 \\
0 & 0 & -1 & 0 
\end{pmatrix} \;\;\;,\;\;\; (\rho, v, \sigma) \mapsto (\sigma, -v, \rho) \;.
\end{eqnarray}

\end{enumerate}

In the main part of the paper, we are interested in the subgroup $H \subset \mathrm{Sp}(4,\mathbb{Z})$ 
consisting of elements $h \in H$ with $h = g_1(a,b,c,d) \, g_2 \, g_1(a,-b,-c,d) \, (g_2)^{-1}$
with the restriction $c = 0 \mod 2, \; a,d = 1 \mod 2$,
\begin{equation}
h = g_1(a, b, c, d) \, g_2 \, g_1 (a, -b, -c, d) \, (g_2)^{-1} = 
\begin{pmatrix}
A & B\\
C & D
\end{pmatrix} = 
\begin{pmatrix}
a &0 & b  & 0\\
0 & a & 0  & -b \\
c & 0 & d & 0 \\
0 & -c & 0 & d
\end{pmatrix} \;\;,\;\; \begin{pmatrix}
a & b\\
c & d
\end{pmatrix} \in \Gamma_0(2) \;.
\label{gggg}
\end{equation}
\\

\noindent
{\bf Definition:}
A Siegel modular form $\Phi_k$ of weight {\bf $k \in \mathbb{N}$} with respect
to the full Siegel modular group
$\mathrm{Sp}(4,\mathbb{Z})$
is a holomorphic function $\Phi_k: {\cal H}_2 \rightarrow \mathbb{C}$ 
that satisfies
\begin{equation}
\Phi_k( (A\,\Omega + B)\,(C \,\Omega  + D)^{-1} ) = ( \det( C \, \Omega + D) )^k \, \Phi_k (\Omega) \:\;\;,\;\;\; \forall \, M \in \mathrm{Sp}(4,\mathbb{Z}) \;.
\end{equation}
\\

\noindent
{\it Remark} (K\"ocher's principle):  A Siegel modular form $\Phi_k$
is bounded on any subset of ${\cal H}_2$ of the 
form $\{\Omega \in {\cal H}_2 | {\rm Im} (\Omega) \geq 
{\rm Im} (\Omega_0) \}$, for any ${\rm Im} (\Omega_0) > 0$.
\\

\noindent
{\bf Theorem:}
Let  $\Phi_k$ be a 
Siegel modular form of weight $k \in \mathbb{N}$. It has the Fourier development \cite{EiZag}
\begin{equation}
\Phi_k (\rho, \sigma, v) = \sum_{m\geq 0} \psi_{k,m}(\rho,v) \, e^{2 \pi i m \sigma} \;.
\label{sie-jac}
\end{equation}
For $m > 0$, the function $\psi_{k,m}$ is a holomorphic Jacobi form of weight $k \in \mathbb{N}$ and
index $m$. For $m=0$, the function $\psi_{k,0}$ transforms as a Jacobi form with $m=0$.
If the first coefficient $\psi_{k,0}$ is identically zero, the Siegel form is called 
Siegel cusp form.
\\

\noindent
{\it Remark:} 
The Fourier expansion of $\Phi_k$ is
\begin{equation}
\Phi_k (\rho, \sigma, v) = \sum_{n, m, r \in \mathbb{Z}; \, n,m,4mn-r^2 \geq 0}
A(n,m,r) \; e^{2 \pi i (n \rho + m \sigma + r v) }
\;.
\end{equation}
It is well defined on ${\cal H}_2$.
\\

\noindent
{\bf Theorem:} 
Let  $\psi_{k,1}$ be a  holomorphic Jacobi form of weight $k$ and index $1$. Then the functions $T_m \psi_{k,1}$, with $m \geq 1$, defined
in terms of the Hecke lift below, are the Fourier coefficients of a Siegel modular form of weight $k$ \cite{EiZag}.
\\

\noindent
{\it Remark:} These theorems also hold if $\mathrm{Sp}(4,\mathbb{Z})$ is replaced by a congruence subgroup 
$\Gamma \subset \mathrm{Sp}(4,\mathbb{Z})$.
We shall be interested in the congruence subgroup $\Gamma = \Gamma_{2,0} (2)$ of $\mathrm{Sp}(4,\mathbb{Z})$, where one restricts to 
$C = 0 \mod 2$. In particular, we shall be interested in the subgroup $H$
given by \eqref{gggg}. 
Then, the functions $\psi_{k,m}$ 
in the Fourier development are Jacobi forms for  $\Gamma_0(2) \subset
SL(2, \mathbb{Z})$.

\section{Hecke operators for $SL(2, \mathbb{Z})$ \label{sec:hecke}}
We follow \cite{BGHZ}.
Let $f \in {\cal M}_k (SL(2, \mathbb{Z}))$ with Fourier expansion $f(\tau) = \sum_{n \geq 0} a_n q^n$.
Let $m$ be an integer with $m \geq 1$.
\\

\noindent
{\bf Definition:} The $m$-th Hecke operator is  a linear operator $T_m$ that acts on modular
forms of weight $k$ by  
\begin{eqnarray}
T_m :  {\cal M}_k (SL(2, \mathbb{Z})) &\rightarrow & {\cal M}_k (SL(2, \mathbb{Z}) )\nonumber\\
f(\tau) &\mapsto& T_m f(\tau) =  m^{k-1} \, \sum_{\alpha \delta =m, \; \alpha, \delta
> 0} \delta^{-k} \sum_{0 \leq \beta < \delta}
f\left(\frac{\alpha \tau + \beta}{\delta}
\right) \;,
\label{Tf}
\end{eqnarray}
with $\alpha, \beta, \delta \in \mathbb{Z}$.
\\

\noindent
{\bf Proposition:} \cite{BGHZ}
\begin{equation}
 T_m f(\tau) =  \sum_{n \geq  0} \left( \sum_{r | (m,n) , \; r> 0} r^{k-1} \, a_{m n/r^2} \right) q^n \;.
 \label{Tmheckam}
 \end{equation}

\begin{proof}
Using
$ \sum_{0 \leq \beta < \delta}
 e^{2 \pi i \beta n /\delta}= \delta $ if $\delta |n$, and $0$ otherwise, we get
\begin{equation}
T_m f(\tau) =  m^{k-1} \, \sum_{n \geq 0} \;  \sum_{\alpha \delta =m, \; \alpha, \delta> 0} \delta^{1-k} \, a_n \, q^{\alpha n/\delta} \;.
\end{equation}
Since $\delta |n$, setting $n' = \alpha n/\delta \in \mathbb{Z}$ and using  $n' = \alpha n/\delta = \alpha^2 n /m \geq 0$, yields
\begin{equation}
T_m f(\tau) =  \sum_{n' \geq 0} \left(  \sum_{\alpha |(m,n') } \alpha^{k-1} \, a_{m n'/\alpha^2} \right) q^{n'} \;,
\label{tmft}
\end{equation}
where we used $m/\alpha = \delta$, hence $\alpha|m$, and $n = m n'/\alpha^2= (m/\alpha) (n'/\alpha)$, and hence $\alpha|n'$.

Note that $T_m f(\tau) $ is holomorphic at $\infty$.

\end{proof}

Now consider the case when $a_0=0$ and $a_1 = 1$. Suppose also that $f(\tau)$ is an eigenfunction
of all the Hecke operators, i.e. $T_m f = \lambda_m \, f$, with eigenvalues $\lambda_m$. $f$ is then called a Hecke eigenform.
The first term of $\lambda_m \, f$ is $\lambda_m \, a_1 = \lambda_m$, whereas the first term
of $T_m f$ (c.f. \eqref{tmft}) is the term with $n' = 1$ given by $a_m$. Hence, $\lambda_m = a_m$.
On the other hand,  $T_m f = \lambda_m \, f$ implies 
\begin{equation}
\sum_{\alpha|(m,n') } \alpha^{k-1} \, a_{m n'/\alpha^2} = \lambda_m \, a_{n'} \;,
\end{equation}
and hence \cite{BGHZ}
\begin{equation}
\sum_{\alpha|(m,n') } \alpha^{k-1} \, a_{m n'/\alpha^2} = a_m \, a_{n'} \;\;\;,\;\;\; m, n' \geq 1 \;.
\label{relaaa}
\end{equation}
Below we will encounter a similar relation when discussing the Hecke lift of a Jacobi form.

\section{Arithmetic lift of a particular Jacobi form for $\Gamma_0(2)$\label{sec:Heckelift}}

In the following, we focus on a particular cuspidal holomorphic Jacobi form  for $\Gamma_0(2)$ of weight $k=2$ and index $m=1$,
and we discuss its arithmetic lift to a Siegel modular form $\Phi_2$ of weight $2$.
The lift of this particular Jacobi form was briefly discussed in the context of black holes
in \cite{David:2006ru},
following the analysis given in \cite{Jatkar:2005bh}. Here, we first review this
construction, and then we use the technique of Hecke eigenforms of
the previous subsection to prove the decomposition \eqref{F2ff} in the limit $v =0$.

Jacobi forms $\phi_{2,1}$ of weight $2$ and index $m=1$ transform as follows under modular  and elliptic
transformations (c.f. \eqref{phikmtransf}),
\begin{eqnarray}
\label{modular1m}
\phi_{2,1} (\frac{a\tau+b}{c\tau+d} , \frac{z}{c\tau + d}) &= &
(c\tau + d)^2 \, e^{2 \pi i  \frac{c z^2}{c\tau + d}} \, \phi_{2,1}(\tau,z) \;, \nonumber\\
\phi_{2,1}(\tau, z + \lambda \tau + \mu) &=&  e^{- 2 \pi i (\lambda^2 \tau + 2 \lambda z)} \phi_{2,1} (\tau,z) \;\;\; \forall \;\;\; \lambda, \mu \in \mathbb{Z} \;.
\end{eqnarray}
Here, we restrict to $\Gamma_0(2)$ transformations, i.e. $c = 0 \mod 2, \;
a, d = 1 \mod 2$.

We will consider the following Jacobi form of $\Gamma_0(2)$ of weight $k=2$ and index $m=1$ (with trivial multiplyer system)
as a seed for the arithmetic lift, 
\begin{equation}
\phi_{2,1} (\tau, z) = \frac{\vartheta^2_1 (\tau, z)}{\eta^6(\tau)} \, f(\tau) \;\;\;,\;\;\; f(\tau) = 2^{-8} \, \vartheta_2^8(\tau) \;,
\label{seedjac}
\end{equation}
with $\vartheta_1 (\tau, z)$ one of the Jacobi theta functions.
Note that $f(\tau)$, which has weight $4$, is not a cusp form of $\Gamma_0(2)$ (the first cusp form of $\Gamma_0(2)$
arises at weight $8$; since $\dim {\cal S}_k (\Gamma_0(2)) = [ \frac{k}{4} ] -1$, the vector space
of cusp forms of weight $8$ has dimension one, and is spanned by
$\vartheta_2^8 (\tau) \, {\cal E}_4 (\tau)$ 
\cite{ACH}, which equals $[ \eta (\tau) \eta (2 \tau) ]^8$, up to a normalization constant,
as can be verified by using the relation given below \eqref{theta2eisen} and $\vartheta_2^4(\tau) = 16 \eta^8 (2 \tau)/\eta^4(\tau)$.
The form $[ \eta (\tau) \eta (2 \tau) ]^8$ is the standard cusp form of $\Gamma_0(2)$).

$\phi_{2,1}$ has a Fourier expansion in both $\tau$ and $z$ given by
\begin{equation}
\phi_{2,1} (\tau,z) = \sum_{l, r \in \mathbb{Z}, \; l\geq 1, \; 4l > r^2} \, c(4 l - r^2) \, e^{2 \pi i l \tau} 
\, e^{2 \pi i r z} \;,
\label{fourierphi21}
\end{equation}
with \cite{Jatkar:2005bh,David:2006ru}
\begin{equation}
c(\Delta) = (-1)^{\Delta} \, \sum_{s, n \in \mathbb{Z}, \; n \geq 1}   f_n \, \delta_{4 n + s^2-1, \Delta} \quad \;,
\end{equation}
where $f_n$ denote the Fourier coefficients of 
\begin{equation}
\frac{f(\tau)}{\eta^6 (\tau)} = \sum_{n \geq 1} f_n \, e^{2 \pi i \tau (n - \frac14)} \;.
\end{equation}
Note the relations $4 n + s^2 -1 >0, \; \Delta = 4l -r^2 > 0$, and hence $c(\Delta)=0$ for $\Delta \leq 0$. 
Hence, $\phi_{2,1}$ is a cuspidal holomorphic Jacobi form of $\Gamma_0(2)$.  Note that the coefficients $f_n$ are integers,
and so are the coefficients $c(\Delta)$. 

For latter use, we note that
in the limit  $z \rightarrow 0$ we have  $\vartheta_1  (\tau, z) = 2 \pi \, \eta^3(\tau) \, z + {\cal O} (z^3)$, and hence, we infer
\begin{equation}
\phi_{2,1} (\tau, z)  = 4 \pi^2 \, f (\tau) \, z^2 + {\cal O} (z^4) \;.
\label{p21z}
\end{equation}

Next, we construct Jacobi forms $\phi_{2,m}$ of $\Gamma_0(2)$ of weight $k=2$ and index $m \in \mathbb{N}$ by means of the Hecke operator $T_m$, which maps $\phi_{2,1}$ in
\eqref{fourierphi21} to $\phi_{2,m}$. 
The action of the Hecke operator $T_m$ on a Jacobi form $\phi_{k,1}$  is defined 
in a  manner similar to \eqref{Tf} \cite{EiZag}, 
\begin{equation}
\label{Hecke}
\phi_{2,m}(\tau,z)=T_m \phi_{2,1}(\tau,z) = m
\sum_{\alpha \delta =m, \; \alpha, \delta> 0, \; \alpha = 1 \, {\rm mod} \, 2} \delta^{-2} \sum_{0 \leq \beta < \delta}
\phi_{2,1} \left(\frac{\alpha \tau + \beta}{\delta}, \frac{m z}{\delta}
\right) \;.
\end{equation}
One then verifies \cite{Jatkar:2005bh}
that $\phi_{2,m}(\tau,z)$ transforms as a Jacobi form of weight $k=2$ and
index $m$ under modular and elliptic transformations \eqref{phikmtransf} and
\eqref{phikmelli}. 
$\phi_{2,m} (\tau,z) $ 
can be brought to the form
\begin{equation}
\phi_{2,m} (\tau,z) = 
\sum_{n, r \in \mathbb{Z}, \; n\geq 1, \; 4mn > r^2} \, a(n,m,r) \, e^{2 \pi i n \tau} 
\, e^{2 \pi i r z} \;,
\label{phi2ma}
\end{equation}
 with Fourier expansion coefficients \cite{Jatkar:2005bh,David:2006ru}
\begin{equation}
a(n,m,r) = \sum_{\alpha \in \mathbb{Z}, \; \alpha >0 , \, \alpha = 1 \; {\rm mod }\, 2, \; \alpha | (n, m, r)}
\, \alpha \,  \, c(\frac{4 m n  - r^2}{\alpha^2}) \;.
\label{expanmr}
\end{equation}
Note that $a(n,m,r) =0$ for $4 m n  - r^2 \leq 0$. The expression \eqref{expanmr} 
is obtained in a manner analogous to
\eqref{Tmheckam}. The Fourier coefficients $a(n,m,r)$ are integer
valued.

Next, define the Siegel cusp form $\Phi_2 (\rho, \sigma, v)$ by \cite{David:2006ru} 
\begin{equation}
\Phi_2 (\rho, \sigma, v)= \sum_{m \geq 1} \phi_{2,m} (\rho, v) \, e^{2 \pi i m \sigma} 
= \sum_{n, m, r \in \mathbb{Z}, \; n, m \geq 1, \; 4 m n - r^2 > 0} \, a(n, m, r) \, 
e^{2 \pi i (n \rho + m \sigma + r v) }
\;,
\label{def_siegel}
\end{equation}
where we have relabelled $\tau \rightarrow \rho, z \rightarrow v$.
By construction, $\Phi_2 (\rho, \sigma, v)$ is invariant under the exchange $\rho \leftrightarrow \sigma$, since the coefficients $a(n, m, r)$ are invariant under the exchange $n \leftrightarrow m$.
$\Phi_2 (\rho, \sigma, v)$ transforms as  a modular form under a certain subgroup $G$ of $\mathrm{Sp(4,
\mathbb{Z})}$ \cite{David:2006ru}. Here, we will focus on the following subgroup $H$ of
$G$ that consists of elements $h \in H$ given by \eqref{gggg}.
Under $H\subset \mathrm{Sp}(4,
\mathbb{Z}) $, $\Phi_2$ transforms as \cite{Jatkar:2005bh}
\begin{equation}
\Phi_2 ( (A \Omega + B) (C \Omega + D)^{-1}) = \det(C \Omega + D )^2 \, \Phi_2 (\Omega) \;,
\end{equation}
with $\Omega$ given in \eqref{Omrvs}.
Introducing $d^3 \Omega = d\rho \, d \sigma \, d v$, we note that
\begin{equation}
\frac{d^3 \Omega}{ (2 v - \rho - \sigma)^5 \, \Phi_2}
\label{inv_comb}
\end{equation}
is invariant under $H$-transformations. The combination \eqref{inv_comb}
is one of the building blocks of the microstate proposal. 
Since $\Phi_2 (\rho, \sigma, v)$ has integer valued Fourier coefficients
(c.f. \eqref{def_siegel}), so does $1/\Phi_2 (\rho, \sigma, v)$ when expanded
in powers of $e^{2 \pi i (n \rho + m \sigma + r v) }$.

Finally, as $v \rightarrow 0$, we obtain from \eqref{def_siegel}, using  \eqref{Hecke} and \eqref{p21z},
\begin{equation}
\Phi_2 (\rho, \sigma, v) = v^2 \, F_2(\rho, \sigma) + {\cal O} (v^4) \;.
\end{equation}
with
\begin{equation}
F_2(\rho, \sigma)  = 4 \pi^2 \, 
\sum_{m \geq 1}
m^3 \, e^{2 \pi i m \sigma}
\sum_{\alpha \delta =m, \; \alpha, \delta> 0, \; \alpha = 1 \, {\rm mod} \, 2} 
\, \delta^{-4} \,  \sum_{0 \leq \beta < \delta}
 f\left(\frac{\alpha
\rho +\beta}{\delta}\right) \;.
\label{F2}
\end{equation}
Although not apparent, $F_2(\rho, \sigma)$ is symmetric under the exchange $\rho \leftrightarrow
\sigma$, and given by:\\

\noindent
{\bf Proposition:}
\begin{equation}
F_2(\rho, \sigma)  = 4 \pi^2 \, f(\rho) \, f(\sigma) \;.
\label{F2ff}
\end{equation}

\begin{proof}

Using the Fourier expansion of $f(\rho)$,
\begin{equation}
f(\rho) = \sum_{n \geq 1} a_n \, e^{2 \pi i n \rho} \;,
\end{equation}
and similarly to \eqref{Tmheckam}, we rewrite \eqref{F2} into
\begin{equation}
F_2(\rho, \sigma)  = 4 \pi^2 \, 
\sum_{m \geq 1, n' \geq 1}
\, e^{2 \pi i m \sigma} \, e^{2 \pi i n' \rho}
 \sum_{\alpha \in \mathbb{Z}, \; \alpha > 0, \; \alpha = 1 {\rm mod} \, 2, \; \alpha |(m, n')
} \, \alpha^3 \, a_{n' m /\alpha^2} \;.
\label{F2amn}
\end{equation}
Now we use that $f(\rho)$ can be expressed as in \eqref{theta2eisen},
\begin{equation}
f(\rho) = 2^{-8} \, \vartheta_2^8 (\rho) 
= \frac{1}{240} \left( E_4 (\rho) - E_4 (2 \rho) \right) \;.
\end{equation}
Inspection of \eqref{Tf}, \eqref{F2} and \eqref{F2amn} shows that
\begin{eqnarray}
T_m f(\rho) =
 \sum_{n' \geq 1} e^{2 \pi i n' \rho}
 \sum_{\alpha \in \mathbb{Z}, \; \alpha > 0, \; \alpha = 1\, {\rm mod} 2, \; \alpha |(m, n')
} \, \alpha^3 \, a_{n' m /\alpha^2} \;.
\label{Tmf}
\end{eqnarray}
Now we use that $T_m f(\rho)$ is a modular form for $\Gamma_0(2)$ of weight $4$.
The vector space 
$ {\cal M}_4 (\Gamma_0(2))$ is two-dimensional and 
spanned by $\tilde{\cal E}_2^2$ and ${\cal E}_4$
\cite{ACH}. Since (by inspection of \eqref{Tmf}) $ T_m f(\rho)$ does not
have a constant term in its Fourier expansion, $ T_m f$ must be proportional
to the combination $\tilde{\cal E}_2^2 - {\cal E}_4$, i.e.
\begin{equation}
 T_m f(\rho) = \lambda_m \left( E_4 (\rho) - E_4 (2 \rho) \right) \;,
\end{equation}
and hence $f$ is a Hecke eigenform,
\begin{equation}
 T_m f(\rho) = 240 \, \lambda_m f(\rho) \;, \forall \; m \geq 1 \;.
\end{equation}
Using $a_1 =1$ and repeating the argument that led to \eqref{relaaa}, we obtain
\begin{equation}
a_m \, a_{n'} = \sum_{\alpha \in \mathbb{Z}, \; \alpha > 0, \; \alpha = 1 \,{\rm mod} \, 2, \; \alpha |(m, n')
} \, \alpha^3 \, a_{n' m /\alpha^2} \;,
\label{absrel}
\end{equation}
which ensures \eqref{F2ff}.

\end{proof}

We perform numerical checks on the validity of \eqref{absrel}. We take $f(\rho) = \left( \tfrac{1}{2} \right)^8 \, \vartheta_2^8 (\rho)$,
and obtain for the first coefficients of the Fourier expansion, 
\begin{eqnarray}
f(\rho) = q + 8 q^2 + 28 q^3 + 64 q^4 + 126 q^5 + 224 q^6 + 344 q^7 + 512 q^8 + 
 757 q^9 + \dots \nonumber\\
\end{eqnarray}
We then test \eqref{absrel} by taking $m=n' =1$ ($\alpha =1 \rightarrow a_1^2 = a_1$), 
$m=1, n' =2$ ($\alpha =1 \rightarrow a_1 a_2 = a_2$),
$m=n'=2$ ($\alpha =1 \rightarrow  a_2^2 = a_4$), $m=n'=3$ ($\alpha=1,3 \rightarrow a_3^2 =
a_9 + 27 a_1)$, to find perfect agreement.

\providecommand{\href}[2]{#2}\begingroup\raggedright\endgroup

\end{document}